# Modified Causal Forests
## for Estimating Heterogeneous Causal Effects

Michael Lechner[*]

Swiss Institute for
Empirical Economic Research

University of St.Gallen

**Abstract:** Uncovering the heterogeneity of causal effects of policies and business decisions at various levels of granularity provides substantial value to decision makers. This paper develops new estimation and inference procedures for multiple treatment models in a selection-on-observables framework by modifying the Causal Forest approach suggested by Wager and Athey (2018) in several dimensions. The new estimators have desirable theoretical, computational and practical properties for various aggregation levels of the causal effects. While an Empirical Monte Carlo study suggests that they outperform previously suggested estimators, an application to the evaluation of an active labour market programme shows the value of the new methods for applied research.

**Keywords**: Causal machine learning, statistical learning, conditional average treatment effects, individualized treatment effects, multiple treatments, selection-on-observables.

**JEL classification:** C21, J68.

---

[*] I am also affiliated with CEPR, London, CESIfo, Munich, IAB, Nuremberg, IZA, Bonn, and RWI, Essen. I thank Michael Knaus, Anthony Strittmatter and Michael Zimmert for many important comments and discussions on this and related topics. This paper also greatly benefited from the data preparation done for previous papers using the same data. I am indebted to Martin Huber, Giovanni Mellace, Michael Knaus and Anthony Strittmatter for this. I also thank Jonathan Chassot, Daniel Goller, and Gabriel Okasa for help with optimizing the code. A previous version of the paper was presented at research seminars at the Universities of St. Gallen (2018) and Gent (2019), at Amazon, Seattle (2018), and at the annual meeting of the Swiss Society of Economics and Statistics in Geneva (2019). I thank participants for helpful comments and suggestions. The usual disclaimer applies. The programmes are coded in Gauss 18.1.5 and 19.1.0. They are freely available at *[www.michael.lechner.eu/software](www.michael.lechner.eu/software)* and at my *ResearchGate* homepage.

# 1 Introduction

Although academia and the public celebrated the amazing predictive power of the new machine learning (ML) methods, many researchers experience some unease, simply because prediction does not imply causation. The ability to uncover causal relations is, however, at the core of most questions concerning the *effects* of particular policies, medical treatments, marketing campaigns, business decisions, etc. (see Athey, 2017, for a recent discussion).

The currently rapidly expanding *causal* ML literature holds great promise for the improved estimation of causal effects by merging the statistics and econometrics literature on causality with the supervised ML literature focussing on prediction. On the one hand, the causality literature clarifies the conditions needed to identify and estimate causal effects. It also shows how to transform a counterfactual causal problem into specific prediction problems (e.g., Imbens and Wooldridge, 2009). On the other hand, the literature on ML provides tools that can be highly effective in solving prediction problems (e.g. Hastie, Tibshirani, and Friedman, 2009). Bringing those two literatures together can lead to more precise, less biased, and thus more reliable estimators of average causal effects. Furthermore, it appears now possible to uncover systematically their heterogeneity (for an overview, see Athey and Imbens, 2017).

In many important applications of causal methods in economics, e.g. in evaluation studies of an active labour market programmes (e.g. Card, Kluve, and Weber, 2018), identification is likely to be based on a selection-on-observables assumption within a multiple-programme setting (for this so-called 'multiple treatment' case, see Imbens, 2000, and Lechner, 2001). In such studies, researchers (and policy makers) are interested to learn the effects of policies at different levels of granularity. For example, average programme (treatment) effects might serve as inputs into a cost-benefit analysis. Uncovering the heterogeneity of the programme effects at the finest possible level of granularity is of interest as well. It not only allows a better understanding of the causal mechanisms at work, but it also sheds some light on distributional aspects of the policy by, e.g., identifying groups who win and groups who lose. Furthermore, such disaggregated results will also hint at potential improvements that could be obtained by different allocations of, in this example, unemployed to different programmes. However, frequently the intermediate aggregation level, e.g. based on several easily understandable groups, is also of interest. One example is that such findings are much easier to communicate to, and understood by decision makers. Of course, for all these estimated policy parameters, measures of estimation uncertainty help to gauge the confidence one should have in the various point estimates. Furthermore, overall estimation costs should not be excessive.

The literature on estimating heterogeneous causal effects with the help of ML methods is quickly expanding. Knaus, Lechner, and Strittmatter (2018, KLS18 henceforth) characterized this literature by three types of (not necessarily exclusive) approaches used. The first two approaches use standard tools of ML but modify the data such as to obtain causal effects instead of predictions. The options are either to modify the outcome (first proposed by Signorovitch, 2007) or to modify the covariates[1] (first proposed by Tian, Alizadeh, Gentles, and Tibshirani, 2014) in different ways. The third generic approach is to change the ML algorithm directly (early Regression Tree based examples are Sue, Tsai, Wang, Nickerson, Li, 2009, and Athey and Imbens, 2016). This is also the path taken in this paper.

This paper provides a new estimator that allows to obtain estimation and inference results for the above-mentioned different levels of aggregation at reasonable computational costs. To be more precise, the Causal Forest estimators proposed by Wager and Athey (2018, WA18 henceforth) and Athey, Tibshirani, and Wager (2019, ATW19 henceforth) form the basis of the proposed methods. Their outcome-based version is extended in two important directions while still keeping most of its theoretical guarantees.

---

[1] The terms *covariates* or *features* are used interchangeably below.

The first extension concerns the splitting rule used for growing the Causal Trees that form the Causal Forest of WA18. They propose to base it on maximising estimated treatment effect heterogeneity. The rational is that this is an unbiased estimator of the MSE of the estimated causal effects. However, this is only true if selection bias is not relevant. Since this is usually not the case for the early splits of the tree, we propose an alternative splitting criteria (for the binary as well as the multiple treatment case). It is based on the observation that the estimation of a causal effect at the lowest level of granularity in a selection-of-observables setting corresponds to predicting the *difference* of two outcome regressions conditional on the covariates. The complication is that the observations used to estimate these regressions are observed in different subsamples defined by treatment status. We propose to establish the necessary link by a matching step conducted prior to building the Causal Forest. Extensive simulation results show that there could be indeed considerable improvements when there is selection-bias. As expected, in the case of experimental data the differences of the rule of WA18 and the new splitting criteria lead to similar results. The simulations also show that adding an additional component to the splitting criteria that favours splits leading to larger heterogeneity of treatment probabilities (propensity scores) across leaves may drastically reduce (finite-sample) selection bias.

The second modification addresses the desire to obtain point estimates and inference for many aggregation levels with limited additional computational costs. In particular, it will be computationally inconvenient to use separate ML estimators for the different aggregation levels (by using one of the many aggregation-level-specific estimators suggested in the literature). Therefore, this paper proposes a procedure that uses a single Causal Forest for estimation and inference for all these parameters. This procedure exploits the fact that predictions of Random Forests can be expressed explicitly as weighted sums of the outcome variables. These weights obtained at the lowest aggregation level can subsequently be aggregated to obtain estimators and inference results at higher aggregation levels. The simulation results show that this approach leads to rather accurate inference for higher aggregation levels, and conservative inference for the lowest level. It also turned out that using in addition a method proposed by ATW19, local centering, which is based on subtracting the estimated conditional-on-covariates but unconditional-on-treatment expectation of the outcome from the actual outcome, essentially removes the 'conservativeness' of the estimated standard errors while not changing much the point estimates. However, this 'one-stop' general weights-based estimator may come at the price of a potential efficiency loss due to the need for sample splitting, which is required for the implemented weights-based inference approach. However, again, the simulation results suggests that such a loss is small, if existent at all.

As a further contribution, the paper contains a large-scale simulation analysis of the properties of various 'one-stop' estimators addressing the behaviour of point estimators and their inference for various aggregation levels. The simulations come from an Empirical Monte Carlo Study (EMCS) that is based on an actual labour market programme evaluation. While KLS18 contains a very extensive EMCS-based analysis of point-estimators, there is to the best of our knowledge so far no large-scale simulation evidence for inference procedures for disaggregated treatment effects estimated by causal ML algorithms.

In the next section, we introduce the related literature. In Section 3, we discuss the parameters of interest and their identification. Estimation and inference is discussed in Section 4. Section 5 contains the Empirical Monte Carlo study. Finally, Section 6 presents an empirical application and Section 7 concludes. Appendices A and B contain more details on the estimators and the simulation study. Appendix C contains the results for several other DGPs relevant in observational studies to document the robustness and sensitivities of the results. Appendix D covers the simulation results for the case without selection bias (experiment). Free computer code is provided in terms of Gauss programme files and is downloadable from www.michael-lechner.eu/statistical-software and www.researchgate.net/project/Causal-Machine-Learning.

## 2  Literature

There is a considerable and rapidly increasing literature related to the estimation of effect heterogeneity by ML methods in observational studies within a selection-on-observables research design. Therefore, we will discuss the main papers only briefly and refer the reader to the much more in-depth and systematic discussion of KLS18.

We start this review with the methodological literature concerned with estimating heterogeneous causal or treatment effects by ML methods followed by applications in economics and comparative studies.[2] Contributions to this literature come from various fields, like epidemiology, econometrics, statistics, and informatics. Proposed estimators are based on regression trees (Su, Tsai, Wang, Nickerson, and Li, 2009; Athey and Imbens, 2016), Random Forests (Athey, Tibshirani, and Wager, 2019; Friedberg, Tibshirani, Athey, and Wager, 2018; Oprescu, Syrgkanis and Wu, 2018; Seibold, Zeileis, Hothorn, 2017; Wager and Athey, 2018), bagging nearest neighbour estimators (Fan, Lv and Wang, 2018), the least absolute shrinkage and selection operator (LASSO, Qian and Murphy, 2011; Tian, Alizadeh, Gentles, and Tibshirani, 2014), support vector machines (Imai and Ratkovic, 2013), boosting (Powers, Qian, Jung, Schuler, Shah, Hastie, and Tibshirani, 2018), neural networks (Ramachandra, 2018; Schwab, Linhardt, and Karlen, 2018; Shalit, Johansson and Sontag, 2017), and Bayesian ML methods (Hill, 2011; Wang, and Rudin, 2015; Taddy, Gardner, Chen, and Draper, 2016). Finally, Chen, Tian, Cai, and Yu (2017), Künzel, Sekhon, Bickel, and Yu (2018), and Nie and Wager (2018) propose general estimation approaches that are not specific to any particular ML method.

---

[2]  Flexible **C**onditional **A**verage **T**reatment **E**ffect (CATE) estimation has also been discussed using non-machine learning methods. These are usually multi-step procedures based on a first step estimation of the propensity score (and possibly the expectation of the outcome given treatment and confounders) and a second non- or semi-parametric step to obtain a low-dimensional CATE function. Finally, this function is used for predicting CATEs (or aggregated versions thereof) in and out-of-sample. For example, Xie, Brand and B. Jann (2012) base their estimator on propensity score stratification and regression, while Abrevaya, Hsu and Lieli (2015) use propensity score weighting, and Lee, Okui, and Whang (2017) use a doubly robust estimation approach.

While there is a large number of proposed methods, in economics only few studies used these methods so far. Ascarza (2018) investigates retention campaigns for customers. Bertrand, Crépon, Marguerie, and Premand (2017) analyse active labour market programmes in a developing country. Davis and Heller (2017) investigate summer jobs in the US. Strittmatter (2018) reinvestigates a US welfare programme. Finally, Knaus, Lechner, and Strittmatter (2017) evaluate the heterogeneous effects of a Swiss job search programme for unemployed workers.

There are also a few simulation studies investigating the properties of the suggested methods. Of course, almost every methodological paper contains a simulation study. However, these studies tend to be very specific, and usually conclude that the estimator proposed in the particular paper performs very well. However, there appear to be four studies that compare a larger number of estimators: Three of them have an epidemiological background. Two of them are using data generating processes that are only to a limited extent informed by real data (Zhao, Runfold, and Kemper, 2017; Powers, Qian, Jung, Schuler, Shah, Hastie, and Tibshirani, 2018). The third study uses large medical databases to inform their simulation designs (Wendling, Callahan, Schuler, Shah, and Gallego, 2018). Since these data generating processes are very specific to biometrics, it appears to be hard to draw strong lessons for many applications in social sciences. The fourth study by KLS18 uses the same data and similar simulation designs as this paper. KLS18 investigate various Random Forest and LASSO based estimators for causal heterogeneity in a selection-on-observables setting. Generally, they conclude that the Forest based versions, in particular the Generalized Forest by Athey, Tibshirani, and Wager (2019) belong to the best performing estimators if explicitly adjusted to take account of confounding. This adjustment is done by a pre-estimation ML step that purges the outcomes from some of their dependence on the covariates by subtracting their estimated conditional-on-covariates mean (*local centering*).[3] Perhaps somewhat surprisingly, the estimators that predict the conditional mean of the outcomes among the treated and among the controls by standard Random Forests and subsequently take the difference perform often similar to the more sophisticated estimators explicitly optimized for causal estimation.

# 3 Causal framework

## 3.1 The potential outcome model

We use Rubin's (1974) potential outcome language to describe a multiple treatment model under unconfoundedness, or conditional independence (Imbens, 2000, Lechner, 2001). Let *D* denote the treatment that may take a known number of *M* different integer values from *0* to *M-1*. The (potential) outcome of interest that realises under treatment *d* is denoted by $Y^d$. For each observation, we observe only the particular potential outcome that is related to the treatment status the observation is observed to be in, $y_i = \sum_{d=0}^{M-1} 1(d_i = d) y_i^d$ ($1(\cdot)$ denotes the indicator function, which is one if its argument is true).[4] There are two groups of variables to condition on, $\tilde{X}$ and *Z*. $\tilde{X}$ contains those covariates that are needed to correct for selection bias (confounders), while *Z* contains variables that define (groups of) population members for which an average causal effect estimate is desired. For identification, $\tilde{X}$ and *Z* may be discrete, continuous, or both (for estimation, we will consider discrete *Z* only). They may overlap in any way. In line with the ML literature, we call them 'features' from now on. Denote the union of the two groups of variables by *X*, $X = \{\tilde{X}, Z\}$, $\dim(X) = p$.[5]

---

[3] Oprescu, Syrgkanis, and Wu (2018) suggested a related, computationally more intensive adjustment procedure.

[4] If not obvious otherwise, capital letters denote random variables, and small letter their values. Small values subscripted by '*i*' denote the value of the respective variable of observation '*i*'.

[5] To avoid complications, we assume *p* to be finite (although it may be very large).

Below, we investigate the following average causal effects:

$$IATE(m,l;x,\Delta) = E(Y^m - Y^l \mid X = x, D \in \Delta),$$

$$GATE(m,l;z,\Delta) = E(Y^m - Y^l \mid Z = z, D \in \Delta) = \int IATE(m,l;x,\Delta) f_{X \mid Z=z, D \in \Delta}(x) dx,$$

$$ATE(m,l;\Delta) = E(Y^m - Y^l \mid D \in \Delta) = \int IATE(m,l;x,\Delta) f_{X \mid D \in \Delta}(x) dx.$$

The **I**ndividualized **A**verage **T**reatment **E**ffects (IATEs), $IATE(m,l;x,\Delta)$ measure the mean impact of treatment $m$ compared to treatment $l$ for units with features $x$ that belong to treatment groups $\Delta$, where $\Delta$ denotes all treatments of interest. The IATEs represent the causal parameters at the finest aggregation level of the features available. On the other extreme, the **A**verage **T**reatment **E**ffects (ATEs) represent the population averages. If $\Delta$ relates to the population with $D=m$, then this is the **A**verage **T**reatment **E**ffect on the **T**reated (ATET) for treatment $m$. The ATE and ATET are the classical parameters investigated in many econometric causal studies. The **G**roup **A**verage **T**reatment **E**ffect (GATE) parameters are in-between those two extremes with respect to their aggregation levels.[6] The IATEs and the GATEs are special cases of the so-called **C**onditional **A**verage **T**reatment **E**ffects (CATEs).

## 3.2 Identifying assumptions

The classical set of unconfoundedness assumptions consists of the following parts (see Imbens, 2000, Lechner 2001):[7]

---

[6] Note that we presume that the analyst selects the variables $Z$ prior to estimation. They are not assumed to be determined in a data driven way, e.g., by statistical variable selection procedures. However, the estimated IATE may be analysed by such methods to describe their dependence on certain features. See Section 6 for more details. Note that Abrevaya, Hsu and Lieli (2015) and Lee, Okui, and Whang (2017) introduce similar aggregated parameters that depend on a reduced conditioning set and discuss inference in the specific settings of their papers.

[7] To simplify the notation, we take the strongest form of these assumptions. Some parameters are identified under weaker conditions as well (for details, see Lechner, 2001, or Imbens, 2000, 2004).

$$\{Y^0,...,Y^m,...,Y^{M-1}\} \amalg D \mid X = x, \qquad \forall x \in \chi; \qquad (CIA)$$

$$0 < P(D = d \mid X = x) = p_d(x), \qquad \forall x \in \chi, \forall d \in \{0,...,M-1\}; \ (CS)$$

$$Y = \sum_{j=0}^{M-1} \mathbf{1}(D = j) Y^j; \qquad (SUTVA)$$

The conditional independence assumption (CIA) implies that there are no features other than $X$ that jointly influence treatment and potential outcomes (for the values of $X$ that are in the support of interest, $\chi$). The common support (CS) assumption stipulates that for each value in $\chi$, there must be the possibility to observe all treatments. The stable-unit-treatment-value assumption (SUTVA) implies that the observed value of the treatment does not depend on the treatment allocation of the other population members (ruling out spillover and treatment size effects). Usually, in order to have an interesting interpretation of the effects, it is required that $X$ is not influenced by the treatment (exogeneity). In addition to these *identifying* assumptions, assume that a large random sample of size $N$ from the random variables $Y, D, X$, $\{y_i, d_i, x_i\}$, $i = 1,...,N$, is available and that these random variables have at least first and second moments.[8]

If this set of assumption holds, then all IATEs are identified in the sense that they can be uniquely deduced from expectations of variables that have observable sample realisations (see Hurwicz, 1950):

$$\begin{aligned} IATE(m,l;x,\Delta) &= E(Y^m - Y^l \mid X = x, D \in \Delta) \\ &= E(Y^m - Y^l \mid X = x) \\ &= E(Y^m \mid X = x, D = m) - E(Y^l \mid X = x, D = l) \\ &= E(Y \mid X = x, D = m) - E(Y \mid X = x, D = l) \\ &= IATE(m,l;x); \qquad \forall x \in \chi, \forall m \neq l \in \{0,...,M-1\}. \end{aligned}$$

Note that IATE does not depend on the conditioning treatment set, $\Delta$. Since the distributions used for aggregation, $f_{X \mid Z=z, D \in \Delta}(x)$ and $f_{X \mid D \in \Delta}(x)$, relate to observable variables ($X, Z, D$) only, they are identified as well (under standard regularity conditions). This in turn implies that the GATE and ATE parameters are identified (their dependence on Δ remains, if the distribution of the features depends on Δ).

---

[8] The identification results will also hold under weights-based and dependent sampling (if the dependence is not too large and certain additional regularity conditions are imposed), but for simplicity we stick to the i.i.d. case. Second moments are not needed for identification, but for the inference part below.

# 4 Estimation and inference

In this section, we discuss estimation and inference of the proposed Causal Forest based estimators. The first subsection introduces the modified splitting criteria for Causal Trees that form the Causal Forest. The following subsection briefly reviews the theoretical guarantees and properties such estimators have, followed by an implementation of weights-based inference as a computationally convenient tool to conduct inference for all desired aggregation levels. Subsection 4 considers the aggregation step required for the ATE and GATE parameters explicitly. The final subsection considers several issues related to the practical implementation of the estimator, including local centering.

## 4.1 IATE: Towards a MSE minimal estimator

Denoting the conditional expectations of $Y$ given $X$ in the subpopulation $D = d$ by $\mu_d(x)$, i.e. $\mu_d(x) = E[Y|X = x, D = d]$ leads to the following expression of $IATE(m,l;x)$ as a difference of $\mu_m(x)$ and $\mu_l(x)$:

$$IATE(m,l;x) = \mu_m(x) - \mu_l(x); \quad \forall x \in \chi, \forall m \neq l \in \{0,...,M-1\}.$$

This estimation task is different to standard ML problems because the two conditional expectations have to be estimated in different, treatment-specific subsamples.[9] Thus, the ML prediction of the difference cannot be directly validated in a holdout sample. This observation is indeed the starting point of the current causal ML literature. The papers then differ on how to tackle this issue.

---

[9] This is implied by the fact that the causal effect is a hypothetical construct that is per se unobservable. Thus, in the words of Athey and Imbens (2016), the 'ground truth' is unobservable in causal analysis.

An easy-to-implement estimator consists in estimating the two conditional expectations separately by standard ML tools, and taking a difference. Below, we denote this estimator as *Basic*, $\widehat{IATE}^{basic}(m,l;x) = \hat{\mu}_m^{basic}(x) - \hat{\mu}_l^{basic}(x)$. This approach has the disadvantage that standard ML methods attempt to maximise out-of-sample predictive power of the two estimators *separately*. More concretely, if a Random Forest is used, the difference of the predictions of the two different estimated forests (one estimated in subpopulation *m*, the other one estimated in subpopulation *l*) may suggest a variability of the IATE that is just estimation error due to (random) differences in the estimated forests. This problem can be particularly pronounced when the features are highly predictive for *Y*, but the IATEs are rather constant. Another example is the case of very unequal treatment shares. As (standard) tree-building uses a stopping rule defined in terms of minimum leaf size, there will be many more observations in treatment *m* compared to treatment *l*, even for similar values of *x*. Thus, the forest estimated for treatment *m* will be finer than the one estimated for treatment *l*. Again, this may lead to spurious effect heterogeneity. However, despite these methodological drawbacks, the large-scale EMCS of KLS18 finds that the Random Forest based *Basic* estimator may perform well compared to technically more sophisticated approaches, in particular when the IATEs are large and vary strongly with the features.

An alternative approach is to use the same trees in both subsamples in which $\mu_m(x)$ and $\mu_l(x)$ are estimated by $\hat{\mu}_m(x)$ and $\hat{\mu}_l(x)$, respectively. Of course, the key is then how to obtain a plausible splitting rule for this 'joint' forest. The dominant approach in the literature so far seems to consider the analogy to a classical random forest regression problem in which the 'ground truth', i.e. the individual treatment effect, would be observable. In this case, the tree estimate of $IATE(m,l;x)$ would be equal to the mean of the 'observed' individual treatment effects in each leaf. For such a case, some algebra reveals that minimising the mean squared expected error of the prediction and maximising the variance of the predicted treatment effects leads to the same sample splits. Therefore, Athey and Imbens (2016) suggest for their causal CARTs to split the parent leaf such as to maximise the heterogeneity of the estimated effects (subject to some adjustments for overfitting). This criterion is also used in one of the approaches of Wager and Athey (2018) and in Oprescu, Syrgkanis, and Wu (2018). However, in case of causal estimation, when the individual treatment effect is unobservable, there is no guarantee that the difference of the outcome means of treated and controls within all leaves equals the means of the true effects within all leaves. Without this condition, the maximisation of treatment effect heterogeneity is not equivalent to MSE minimisation of treatment effects prediction. The reason for the difference to the standard predictive case is due to potential selection bias. Intuitively, if selection bias does not matter, such an equality holds in expectation and this criterion should be a good approximation to minimizing the MSE of the estimated and true individual treatment effect. However, if selection bias is relevant (as it is likely to be, particularly in the early splits of the tree), then the quality of this approximation may be questionable.

An alternative approach, which is our first modification of the approach by WA18, is to derive a splitting rule that considers the mean square error of the particular estimation problem directly:

$$\begin{aligned}
MSE\left[\widehat{IATE}(m,l;x)\right] &= E\left\{\left[\widehat{IATE}(m,l;x) - IATE(m,l;x)\right]^2\right\} \\
&= E\left[\hat{\mu}_m(x) - \mu_m(x)\right]^2 + E\left[\hat{\mu}_l(x) - \mu_l(x)\right]^2 - 2E\left[\hat{\mu}_m(x) - \mu_m(x)\right]\left[\hat{\mu}_l(x) - \mu_l(x)\right] \\
&= MSE\left[\hat{\mu}_m(x)\right] + MSE\left[\hat{\mu}_l(x)\right] - 2\underbrace{E\left[\hat{\mu}_m(x) - \mu_m(x)\right]\left[\hat{\mu}_l(x) - \mu_l(x)\right]}_{MCE\left[\hat{\mu}_m(x), \hat{\mu}_l(x)\right]} \\
&= MSE\left[\hat{\mu}_m(x)\right] + MSE\left[\hat{\mu}_l(x)\right] - 2MCE\left[\hat{\mu}_m(x), \hat{\mu}_l(x)\right].
\end{aligned}$$

This derivation of the mean square error is instructive.[10] It shows that the *Basic* estimator fails to take into account that estimation errors may be correlated, conditional on the features. Thus, there may be a substantial advantage of tying the estimators together in a way such that the correlation of their estimation errors becomes positive (and cancels to some extent). The complication here is that the **M**ean **C**orrelated **E**rror (MCE) is difficult to estimate.

For constructing estimators based on this criterion, the MSE of $\hat{\mu}_d(x)$ has to be estimated. This is straightforward, as the MSEs of all $M$ functions $\hat{\mu}_d(x)$ can be computed in the respective treatment subsamples in the usual way. Denote by $N_{S_x}^d$ the number of observations with treatment value $d$ in a certain stratum (leaf) $S_x$, which is defined by the values of the features $x$. Then, the following estimator is a 'natural' choice:

$$\widehat{MSE}_{S_x}\left[\hat{\mu}_d(x)\right] = \frac{1}{N_{S_x}^d} \sum_{i=1}^{N} \underline{1}(x_i \in S_x)\underline{1}(d_i \in d)\left[\hat{\mu}_d(x_i) - y_i\right]^2.$$

Note that the overall MSE in $S_x$ is the sum of the MSEs in the treatment specific subsamples of $S_x$, where each subsample receives the same weight (independent of the number of observations in that subsample), as implied by the above MSE formula for causal effect estimation.

In order to compute the correlation of the estimation errors, we need a proxy for cases when there are no observations with exactly the same values of $x$ in all treatment states (as is always true for continuous features). In this case, we propose using the closest neighbour available instead.[11]

---

[10] Note that in Random Forests, the expectation is taken with respect to the distribution of $X$ in the training data.

[11] Closeness is based on a simplified Mahalanobis metric as in Abadie and Imbens (2006). This simplified version has the inverse of the variances of the features on the main diagonal. Off-diagonal elements are zero. The simplification avoids computational complications when inverting the variance-covariance matrix of potentially large-dimensional features at the cost of ignoring correlations between covariates.

$$\widehat{MCE}(m,l;S_x) = \frac{1}{N_{S_x}^l + N_{S_x}^m} \sum_{i=1}^{N} \underline{1}(x_i \in S_x)\left[\underline{1}(d_i = m) + \underline{1}(d_i = l)\right]\left[\hat{\mu}_m(x_i) - \tilde{y}_{(i,m)}\right]\left[\hat{\mu}_l(x_i) - \tilde{y}_{(i,l)}\right],$$

$$\tilde{y}_{(i,m)} = \begin{cases} y_i & \text{if} \quad d_i = m \\ y_{(i,m)} & \quad d_i \neq m \end{cases}.$$

The splitting rule that minimizes $\widehat{MSE}[\widehat{IATE}(m,l,x)]$ is motivated by maximising the predictive power of the estimator. However, in causal analysis inference is important. Thus, if the MSE-minimal estimator has a substantial bias, this is problematic. In causal studies, a substantial source of bias is a non-random allocation of treatment (*selection bias*). This is captured by the propensity score, $P(D=d\,|\,X=x)$, which thus has a certain role to play in many proposed estimators of IATEs. This is usually tackled by a first stage estimation of $P(D=d\,|\,X=x)$ and / or $E(Y\,|\,D=d, X=x)$ and treating it as a nuisance parameter in the Random Forest estimation of $\widehat{IATE}(m,l;x)$.[12] Here, we like to avoid computer-time-consuming additional estimation steps, but still improve on the robustness of the estimator with respect to selection bias, in particular in smaller samples where the Random Forests may not be automatically fine enough to remove all selection biases.

These considerations lead us to suggest a further modification of the splitting rules. Denote by *leaf(x')* and *leaf(x'')* the values of the features in the daughter leaves resulting from splitting some parent leaf. We propose to add a penalty term to $\widehat{MSE}\left[\widehat{IATE}(m,l;x)\right]$ that penalizes possible splits where the treatment probabilities in the resulting daughter leaves are similar (splits leading to leaves with similar treatment shares will not be able to remove much selection bias, while if they are very different in this respect, they approximate differences in $P(D = d | X = x)$ well). In other words, the modified criterion prefers splits with high propensity score heterogeneity and puts explicit emphasis on tackling selection bias. In the simulations below, the following penalty function is added to the splitting criteria of (some) estimators considered:

$$penalty(x',x'') = \lambda \left\{ 1 - \frac{1}{M} \sum_{d=0}^{M-1} [P(D = d | X \in leaf(x')) - P(D = d | X \in leaf(x''))]^2 \right\}.$$

---

[12] See for example Oprescu, Syrgkanis, and Wu (2018). There is also a substantial literature on how to exploit so-called double-robustness properties when estimating the causal effects at higher aggregation level, see, e.g., Belloni, Chernozhukov, Fernández-Val, and Hansen (2017) and the references therein.

This penalty term is zero if the split leads to a perfect prediction of the probabilities in the daughter leaves. It reaches its maximum value, $\lambda$, when all probabilities are equal. Thus, the algorithm prefers a split that is not only predictive for *y* but also for *D*. Of course, the choice of the exact form of this penalty function is arbitrary. Furthermore, there is the issue of how to choose $\lambda$ (*without* expensive additional computations) which is taken up again in Section 4.5.

So far, the intuitive exposition focused on the comparison of two treatments. When there more than two treatments, this algorithm can be implemented in at least two different ways. Based on the 'sample reduction properties' in Lechner (2001), the first version estimates all parameters in pair-wise comparisons independent of each other. This may become computationally cumbersome when there are many treatments. Thus, an alternative second version is to build the same trees for all treatments based on summing up the MSE's of the estimated IATEs (as well as the respective values of the penalty functions) over all binary comparisons of interest. If some effects are more important to a researcher than others, such a summation may use some pre-specified weights. If this is not the case, it appears to be 'natural' to weight all MSE's equally.

## 4.2 Properties

Theorem 1 of WA18 shows that under certain conditions predictions from Random Forest based estimators are normally distributed. The assumptions necessary for achieving this asymptotic distribution in the case of i.i.d. sampling require (i) Lipschitz continuity of lower and existence of some higher order moments of the outcome variable conditional on the features, (ii) using subsampling to form the training data for tree building (subsamples should increase with $N$, slower than $N$, but not too slow), (iii) conditions on the features (independent, continuous with bounded support, $p$ is fixed, i.e. low-dimensional), as well as (iv) some conditions on how to build the trees. The latter conditions require (a) that building a tree is independent from computing its predictions (honesty), (b) that every feature has a some positive probability to be used for tree splitting, (c) that trees are fully grown up to a minimum leaf size, (d) that the minimum share of observations required at least to end up in the smaller daughter-leaf in each split is not more than 20% of the observations in the parent leaf, and (e) that the tree prediction is independent of the indexing order of the features.

The proposed estimators (see below) fulfil these conditions on tree building. However, some of these conditions are very specific and sometimes difficult to match (like covariates being independent), or impossible to verify (like the regularity conditions on the conditional outcome expectations). Nevertheless, the simulation results show that the predictions from all Random Forest based estimators appear to be very close to be normally distributed even for the smallest sample size investigated (*N=1000*).

## 4.3 Inference

There are several suggestions in the literature on how to conduct inference and how to compute standard errors of Random Forest based predictions (e.g., Wager, Hastie, and Efron, 2014; Wager and Athey, 2018; and the references therein). Although these methods can be used to conduct inference on the IATE, it is yet unexplored how these methods could be readily generalized to take account of the aggregation steps needed for the GATE and ATE parameters.

Therefore, we suggest an alternative inference method useful for estimators that have a representation as weighted averages of the outcomes. This perspective is attractive for Random Forest based estimators (e.g. Athey, Tibshirani, and Wager, 2019) as they consist of trees that first stratify the data (when building a tree), and subsequently average over these strata (when building the forest). Thus, we exploit the weights-based representation explicitly for inference (see also Lechner, 2002, and Abadie and Imbens, 2006, for related approaches).

Let us start with a general weights-based estimator. Denote by $\hat{w}_i$ the weight that the dependent variable $y_i$ receives in the desired estimator, $\hat{\theta}$ (which could be one of the IATEs, or GATEs, or ATEs).

$$\hat{\theta} = \frac{1}{N}\sum_{i=1}^{N}\hat{w}_i y_i; \qquad Var(\hat{\theta}) = Var\left(\frac{1}{N}\sum_{i=1}^{N}\hat{w}_i y_i\right).$$

Next, we apply the law of total probability to the variance:[13]

$$Var\left(\frac{1}{N}\sum_{i=1}^{N}\hat{w}_i y_i\right) = E_{\hat{W}} Var\left(\frac{1}{N}\sum_{i=1}^{N}\hat{w}_i y_i \mid \hat{w}_1,...,\hat{w}_N\right) + Var_{\hat{W}} E\left(\frac{1}{N}\sum_{i=1}^{N}\hat{w}_i y_i \mid \hat{w}_1,...,\hat{w}_N\right)$$

$$= E_{\hat{W}}\left(\frac{1}{N^2}\sum_{i=1}^{N}\hat{w}_i^2 Var(y_i \mid \hat{w}_1,...,\hat{w}_N)\right)$$

$$+ Var_{\hat{W}}\left(\frac{1}{N}\sum_{i=1}^{N}\hat{w}_i E(y_i \mid \hat{w}_1,...,\hat{w}_N)\right).$$

However, the large conditioning sets of $E(y_i \mid \hat{w}_1,...,\hat{w}_N)$ and $Var(y_i \mid \hat{w}_1,...,\hat{w}_N)$ makes it impossible to estimate these terms precisely without further assumptions. The conditioning sets can be drastically reduced, though, if observation 'i' is not used to build the forest,[14] and the data used for the computations of the conditional mean and variance are an i.i.d. sample. To see this, recall that Random Forest weights are computed as functions of $\vec{X}^T = (x_1,...,x_{N_T})$ and $\vec{Y}^T = (y_1,...,y_{N_T})$ in the training sample with $N_T$ training observations. These weights are then assigned to observation *i* based on the value of $x_i$ only. Thus, the weights are functions of $x_i$ and the training data, $\hat{w}_i = \hat{w}(x_i, \vec{X}^T, \vec{Y}^T)$. If observation 'i' does not belong to the training data and there is i.i.d. sampling, $y_i$ and $\hat{w}_j = \hat{w}(x_j, \vec{X}^T, \vec{Y}^T)$ are independent. Thus, we obtain $E(y_i | \hat{w}_1,...,\hat{w}_N) = E(y_i | \hat{w}_i) = \mu_{Y|\hat{W}}(\hat{w}_i)$ and $Var(y_i | \hat{w}_1,...,\hat{w}_N) = Var(y_i | \hat{w}_i) = \sigma^2_{Y|\hat{W}}(\hat{w}_i)$. This leads to the following expression of the variance of the proposed estimators:

$$Var\left(\frac{1}{N}\sum_{i=1}^{N}\hat{w}_i y_i\right) = E_{\hat{W}}\left(\frac{1}{N^2}\sum_{i=1}^{N}\hat{w}_i^2 \sigma^2_{Y|\hat{W}}(\hat{w}_i)\right) + Var_{\hat{W}}\left(\frac{1}{N}\sum_{i=1}^{N}\hat{w}_i \mu_{Y|\hat{W}}(\hat{w}_i)\right).$$

---

[13] Letting $A$ and $B$ be two random variables, then $Var(A) = E_B Var(A \mid B) + Var_B E(A \mid B)$.

The above expression suggests using the following estimator:

$$\widehat{Var(\hat{\theta})} = \frac{1}{N^2}\sum_{i=1}^{N}\hat{w}_i^2 \hat{\sigma}^2_{Y|\hat{W}}(\hat{w}_i) + \frac{1}{N(N-1)}\sum_{i=1}^{N}\left[\hat{w}_i \hat{\mu}_{Y|\hat{W}}(\hat{w}_i) - \frac{1}{N}\sum_{i=1}^{N}\hat{w}_i \hat{\mu}_{Y|\hat{W}}(\hat{w}_i)\right]^2.$$

The conditional expectations and variances may be computed by standard non-parametric or ML methods, as this is a one-dimensional problem for which many well-established estimators exist. Bodory, Camponovo, Huber, and Lechner (2018) investigate *k*-nearest neighbour estimators to obtain estimates for these quantities. They found good results in a binary treatment setting for the ATET. The same method is used here.[15] Finally, note that because of the weighting representation, this approach can also be readily used to account for, e.g., clustering, which is a common feature in economic data, or to conduct joint tests of several linear hypotheses, such that several groups have the same or no effect, leading to Wald-type statistics. It is, however, beyond the scope of this paper to analyse rigorously the exact statistical conditions needed for this estimator to lead to valid inference.

---

[14] Note that this condition goes beyond 'honesty' (i.e. continuously switching the role of observations used for tree building and effect estimation in Random Forests), because even if honesty is used, each weight may still depend on many observations. Clearly, the price to pay for sample splitting is a loss of precision (as so-called 'cross-fitting' does also not appear to work in this set-up without further adjustments).

[15] They also found a considerable robustness on how exactly to compute the conditional means and variances. Note that since their results relate to aggregate results, their generalisability to the level of IATE's is unclear.

## 4.4 GATE and ATE

Estimates for GATEs and ATE are most easily obtained by averaging the IATEs in the respective subsamples defined by $z$ (assuming discrete $Z$) and $\Delta$. Although estimating ATEs and GATEs directly instead of aggregating IATEs may lead to more efficient and more robust estimators (e.g. Belloni, Chernozhukov, Fernández-Val, and Hansen, 2017), the computational burden would also be higher, in particular if the number of GATEs of interest is large, as is common in many empirical studies. Therefore, letting $\widehat{IATE(m,l;x)}$ be an estimator of $IATE(m,l;x)$, we suggest to estimate the GATEs and ATEs as appropriate averages of $\widehat{IATE(m,l;x)}$'s:

$$\widehat{GATE}(m,l;z,\Delta) = \frac{1}{N^{z,\Delta}} \sum_{i=1}^{N} 1(z_i = z, d_i \in \Delta) \widehat{IATE}(m,l;x_i)$$

$$= \frac{1}{N^{z,\Delta}} \sum_{i=1}^{N} 1(z_i = z, d_i \in \Delta) \frac{1}{N} \sum_{j=1}^{N} \hat{w}_j^{IATE(m,l;x_i)} y_j$$

$$= \frac{1}{N} \sum_{i=1}^{N} \left( \frac{1}{N^{z,\Delta}} \sum_{j=1}^{N} 1(z_j = z, d_j \in \Delta) \hat{w}_i^{IATE(m,l;x_j)} \right) y_i$$

$$= \frac{1}{N} \sum_{i=1}^{N} \hat{w}_i^{GATE(m,l;z,\Delta)} y_i;$$

$$\hat{w}_i^{GATE(m,l;z,\Delta)} = \frac{1}{N^{z,\Delta}} \sum_{j=1}^{N} 1(z_j = z, d_j \in \Delta) \hat{w}_j^{IATE(m,l;x_i)}; \qquad N^{z,\Delta} = \sum_{i=1}^{N} 1(z_i = z, d_i \in \Delta).$$

$$\widehat{ATE}(m,l;\Delta) = \frac{1}{N^z} \sum_{i=1}^{N} 1(d_i \in \Delta) \widehat{IATE}(m,l;x_i)$$

$$= \frac{1}{N} \sum_{i=1}^{N} \hat{w}_i^{ATE(m,l;\Delta)} y_i;$$

$$\hat{w}_i^{ATE(m,l;z,\Delta)} = \frac{1}{N^{\Delta}} \sum_{j=1}^{N} 1(d_j \in \Delta) \hat{w}_j^{IATE(m,l;x_i)}; \qquad N^{\Delta} = \sum_{i=1}^{N} 1(d_i \in \Delta).$$

From this expression, it is clear that ATEs and GATEs have the same type of weights-based representation as the IATEs. Hence, inference can be conducted in the same way as for the IATEs, just using different weights.

## 4.5 Implementation

Beyond the *Basic* estimator, we investigate seven different estimators (and some additional variants), all based on using the same single forest for all treatment subsamples. They differ in their complexity. *OneF* ignores the MCE. It constructs the tree solely based on the sum of the treatment specific MSEs of estimating $\hat{\mu}_d(x)$. *OneF.MCE* estimates the MCE in a computationally not too expensive way by using nearest neighbours. *OneF.VarT* is the Causal Forest estimator of Wager and Athey (2018) based on maximising treatment effect heterogeneity. *OneF.MCE.Penalty* and *OneF.VarT.Penalty* use the same splitting criteria as *OneF.MCE* and *OneF.VarT* but add an additional penalty term to the splitting rule to reduce potential selection biases. *OneF.MCE.LCk* and *OneF.MCE.Penalty.LCk* recenter the outcome variables before using the *OneF.MCE* and the *OneF.MCE.Penalty* algorithms.

The main elements of the algorithms used for estimating the results in the simulation and application parts are the following:

1) Split the estimation sample randomly into two parts of equal size (sample A and sample B)

2) Estimate the trees that define the respective random forest in sample A.

   a. *Basic*: Estimate Random Forests for each treatment state in the subsamples defined by treatment state. The splitting rule consists of independently minimizing the mean squared prediction error within each subsample.

   b. *OneF*: Estimate the same forest for all treatment states jointly. The splitting rule is based on minimising the sum of the MSE's for all treatment state specific outcome predictions (MCEs are set to 0).

c. *OneF.MCE*: Same as b), but before building the trees, for each observation in each treatment state, find a close 'neighbour' in every other treatment state and save its outcome (to estimate MCE). The splitting rule is based on minimising the overall MSEs, taking account of all MCEs.

d. *OneF.MCE.LCk*: Same as c) but with recentered outcomes (based on *k* folds, see Appendix A.2 for details).

e. *OneF.VarT*: Same as b), but splitting is based on maximising estimated treatment effect heterogeneity.

f. *OneF.MCE.Penalty*: Same as c) but a penalty term penalizing propensity score homogeneity is added.[16]

g. *OneF.MCE.Penalty.LCk*: Same as f) but with recentered outcomes.

h. *OneF.VarT.Penalty*: Same as e) but a penalty term penalizing propensity score homogeneity is added.[17]

3) Apply the sample splits obtained in sample A to all subsamples (by treatment state) of sample B and take the mean of the outcomes in the respective leaf as the prediction that comes with this Forest.

4) Obtain the weights from the estimated Forest by counting how many times an observation in sample B is used to predict IATE for a particular value of *x*.

---

[16] In the simulations below, setting λ to *Var(Y)* works well. *Var(Y)* corresponds to the MSE when the effects are estimated by the sample mean without any splits. Thus, it provides some benchmark for plausible values of λ. In small-scale experiments with values smaller and larger than *Var(Y)* the MSE shows little sensitivity for values half as well as twice the size of *Var(Y)* (available on request). Generally, decreasing the penalty increases biases and reduces variances, et vice versa. As will be seen below in the simulations, biases are more likely to occur when selection is strong. Thus, if a priori knowledge about the importance of selectivity is available, then the researcher might increase (strong selectivity) or decrease (weak selectivity or experiment) the penalty term accordingly.

[17] Setting λ to the square of the sum of the differences of the treatment means corresponds to the intuition used for *OneF.MCE.Penalty*. However, in the simulations below it appeared that such a value is far too small to reduce biases significantly when there is selectivity (available on request). Therefore, a value corresponding to 100 x that value is used below.

5) Aggregate the IATEs to GATEs by taking the average over observations in sample B that have the same value of *z* and treatment group Δ. Do the same aggregation with the weights to obtain the new weights valid for the GATEs.

6) Do the same steps as in 5) to obtain the ATEs, but average over all observations in treatment group Δ.

7) Compute weights-based standard errors as described above. Use the estimated standard errors together with the quantiles from the normal distribution to obtain critical values (p-values).

While Appendix A details the implementation further, at least three points merit some more discussion. The efficiency loss inherent in the two-sample approach could be avoided by cross-fitting, i.e. by repeating the estimation with exchanged roles of the two samples and averaging the two (or more) estimates.[18] However, in such a case it is unclear on how to compute the weights-based inference for the averaged estimator as the two components of this average will be correlated. A second issue concerns the fact of forming the neighbours by simplified Mahalanobis matching, which has the issue of being potentially large-dimensional. A lower dimensional alternative might be to estimate a prognostic score, $[\hat{\mu}_0(x),....,\hat{\mu}_{M-1}(x)]$, by ML methods and then use this score instead.[19] While this is a viable alternative, it requires again (costly) nuisance parameter estimation which we want to avoid with the suggested estimator. The third note concerns *k*-fold cross-fitting as suggested by ATW19 and implemented in the application of Athey and Wager (2019). Appendix A.2 details the algorithm performing the recentering that is modified (compared to ATW19) to take into account the special structure of the two-sample approach used here in order to obtain weights-based inference.

---

[18] However, note that the simulations below comparing the two-sample estimator with a one-sample-with-honesty strategy suggest that the efficiency loss is minor, if existent at all.

[19] Note that propensity scores will not be helpful as the intended correction is not directly related to selection bias.

# 5 Simulation results of an Empirical Monte Carlo Study

## 5.1 Data base, simulation concept, and data generating processes

It is a general problem of simulation results that they depend on the particular design of the data generating process (DGP) chosen by the researcher, which might reduce their generalizability to specific empirical applications. While this may be innocuous if simulations are used to investigate specific theoretical properties of estimators or test statistics only (e.g. analysing what happens if the correlation of features increases), it is more problematic when simulations are used to investigate the suitability of estimators for particular applications. In this case, a simulation environment that closely mimics real applications is advantageous. In that way, the results generalize more easily to applications with a similar data structure.

Huber, Lechner, and Wunsch (2013) and Lechner and Wunsch (2013) proposed a specific data driven method for simulation analyses which they called Empirical Monte Carlo Study (EMCS). The main idea of an EMCS is to have a large real data set from which to draw random subsamples using much information from the real data. In the simulations below, we follow this approach (see Appendix B for the exact algorithm used).[20]

We base the simulations on Swiss social security data previously used to evaluate active labour market policies. More precisely, as in Huber, Lechner, and Mellace (2017) for analysing a mediation framework, and in KLS18 for comparing different estimators of the IATE, we consider the effects of a job search programme on employment outcomes. This data is well suited for this type of analysis, as it is long and wide (about 95'000 observations, more than 50 base covariates). Furthermore, the programme is one for which the literature argues that a selection-on-observable assumption is plausible when rich social security data are available (see, e.g.

---

[20] Note the similarity of the concept of EMCS with the very recently proposed method of Synth Validation (Schuler, Jung, Tibshirani, Hastie, and Shah, 2017). Although the latter method is intended to select methods for a particular datasets (in the ML spirit of comparing predictions with observed variables), both methods could be used for method selection as well as for method validation. Advani, Kitagawa, and Słoczyński (2018) point to the potential limits of generalizing results from such simulation exercises.

Lechner and Gerfin, 2002, for the Swiss case, and the survey by Card, Kluve, and Weber, 2018).[21]

The main steps are the following: Using the initial data, we estimate the propensity score, *p(x)*, using the same specification as KLS18. As such, *p(x)* depends on 77 features that enter a logit model estimated by maximum likelihood.[22] This estimated propensity score plays the role of the true selection process in the following steps. Next, the treated are removed from the data. Thus, for all remaining observations (approx. 84'000) we observe $Y^0$, the non-treatment outcome (measured as number of months in employment in the next 33 months after the programme starts), the features, and *p(x)*. This information is used to simulate the individual treatment effects (ITE), to compute the IATEs, and to compute $Y^1$ as the sum of $Y^0$ and the ITE. In the next step, we draw randomly a validation data set with 5'000 observations and remove it from the main data. From the remaining data, we draw random samples of size *N = 1'000*, *N = 4'000,* and *N = 8'000*, simulate a treatment status for each observation using the 'true' propensity score (shifted such that the expected treatment share is about 50%), and take the potential outcome that corresponds to the simulated treatment as observed outcome, *Y*. These random samples are the training data for the algorithms, while their performance is measured out-of-sample on the 5'000 observations of the validation sample.

What remains is the specification of the ITEs, which we base on different IATEs variations. Here we specify four different IATEs such that they reflect different correlations with the propensity score (as disentangling selectivity from effect heterogeneity is a key challenge for all estimators in this setting) and different strengths and variability of the effects. To be specific, the first specification sets the ITE to zero for all individuals. In the next two specifications, the IATE is a non-linear function of the propensity score given through a non-linear deterministic component $\xi(x)$:

$$\xi(x) = \sin\left(1.25\pi \frac{p(x)}{\max_{i=1:N} p(x_i)}\right),$$

$$IATE(x) = \alpha \frac{\xi(x) - \bar{\xi}}{SD(\xi)}; \qquad \bar{\xi} = \frac{1}{N}\sum_{i=1}^{N} \xi(x_i), \quad SD(\xi) = \sqrt{\frac{1}{N}\sum_{i=1}^{N}\left[\xi(x_i) - \bar{\xi}\right]^2}.$$

---

[21] Our implementation follows closely KLS18. Therefore, for the sake of brevity, we do not repeat their extensive documentation of all steps that lead to the final sample and their descriptions of the estimation sample. The reader interested in more details is referred to KLS18.

[22] The number is larger than the number of base covariates due to the addition of some transformations of the base covariates, such as dummy variables.

The parameter α determines the variability of the IATE. In the simulations, we consider two values of α (i) $\alpha = 2$ ('normal' heterogeneity); and (ii) $\alpha = 8$ ('strong' heterogeneity). Due to the non-linear way in which the features enter the IATEs via the propensity score, it is a difficult task for every estimator not to confuse selection effects with heterogeneous treatment effects. Finally, we also consider an 'easier' case in which $\xi(x)$ depends linearly on the *insured earnings* of the unemployed. The latter is an officially defined pre-unemployment earnings measure used to compute unemployment benefits. Although this variable is related to the selection into programmes as well, the link to the IATEs is much weaker.

Adding two independent random components to the IATE leads to the ITE. The first random term is a (minus) Poisson (1) variate adjusted to have zero expectation. The second random term ensures that the ITE, and thus $Y^1$, keeps its character as integer (month) in a way that the rounding 'error' is independent of the IATEs:

$$ITE(x) = IATE(x) + (1-u) + v;$$
$$u \sim Poisson(1);$$
$$v^* \sim Uniform[0,1];$$
$$v^{diff} = IATE(x) + u - floor(IATE(x) + u);$$
$$v = \underline{1}(v^* > v^{diff})(-v^{diff}) + \underline{1}(v^* \leq v^{diff})(1 - v^{diff}).$$

*Floor* denotes the integer part of *IATE(x)*.

The non-linear ('normal') IATE with *α = 2* has a standard deviation of about 1.7 (*Corr(p(x),IATE)=0.73*), while the 'strong' (*α = 8*) IATE has a standard deviation of about 6.8 (*Corr(p(x),IATE)=0.85*). The standard deviation of the earnings related ('earnings') IATE is also about 1.7 (*Corr(p(x),IATE) = 0.24*). These numbers should be related to the standard deviation of the outcome of about 12.9. The IATEs are well predictable (if observed). For example, plain-vanilla Random Forests have out-of-sample $R^2$'s above 99%. Naturally, the predictability of the ITE is smaller with out-of-sample $R^2$'s of about 76% ('normal'), 92% ('strong'), and 94% ('earnings') respectively.

These IATEs are also used to compute the ATE and the GATEs. The true ATE is taken as average of the IATEs in the validation sample, while the true GATEs are taken as averages of the IATEs in the respective groups. There are two types of GATEs considered: The first type consists of two GATEs, one for man (56%) and one for women (44%). The second type of GATEs considers 32 yearly age categories (24-55). As younger individuals are more likely to become unemployed in Switzerland, the largest, i.e. youngest, age group has about three times as many observations as the smallest, i.e. oldest, age group.

In addition to varying the IATEs and the corresponding aggregate effects in four scenarios, we also vary the assignment process by considering random assignment of the treatment as in a randomized control trial (RCT) as well. Furthermore, we vary the sample size. Overall, this leads to 24 different DGPs in total (4 IATEs x 3 sample sizes x 2 selection processes). In addition to the estimators introduced in Section 4, we consider *Basic* in a version without the a priori sample splitting but with honest trees instead.[23] Finally, for the smaller sample, *N = 1'000*, estimation and inference is repeated on 1000 independent random training samples (replications, *R*). Since computation time is a constraint and because the larger sample produces estimates with lower variability, only 250 replications are used for *N = 4'000,* and only 125 replications for *N=8'000*.

---

[23] So that for each random subsample used for a particular tree the data is split randomly into two parts: Half the data is used to build the tree, and half the data is used to estimate the effects given the tree.

## 5.2 Results

### 5.2.1 General remarks

For each replication, we obtain 5'035 results in the validation sample, consisting of 5'000 estimates of different IATEs, 32 and 2 estimates of the GATEs, and 1 estimate of the ATE. For each of these parameters, we compute the usual measures for point estimators, like bias (col. (1) of the following table), standard deviation (col. (8)), mean squared error (MSE, col. (4)), skewness (col. 5), and kurtosis (col. 6). The Jarque-Bera statistic (JB, col. (7)) summarizes these third and fourth moments. These three measures are used to check whether the estimators are normally distributed (JB is $\chi^2(2)$ distributed when the estimators are normally distributed, and shifts to the right when they are not; the 5% and 1%-critical values are 6 and 9.2, respectively). Finally, we compute the bias of the estimated standard errors (col. 9) and the coverage probability of the 90% confidence interval (col. 10).[24] For the MSE, we also compute measures of its variability across replications to assess the simulation error (in the footnote of the tables).

It is nonsensical to report these measures for all 5035 parameters. While we report the measures for the ATE directly, we aggregate the measures for the IATEs and the GATEs by taking their average across groups. Note that due to the way the estimators are constructed, the bias of the ATE and the average biases of the IATEs are identical. Since this type of cancellation of biases for the IATEs is undesirable in a quality measure (as a negative bias is as undesirable as a positive one), we report their average absolute bias (col. 1). Finally, we report the standard deviation of all the true (2) and estimated GATEs and IATEs (3) to see how the estimators capture the cross-sectional variability of the true effect heterogeneity.

---

[24] This is the share of replications for which the true value was included in the 90% confidence interval (computed with the point estimate and the variance estimate under a normality assumption). 90% instead of the more common 95% is used, because the number of replications may be too small (in particular for the larger sample) to estimate the more extreme tail probability precisely. Since all estimators are normally distributed, the particular tail quantile should hardly affect the conclusions.

5.2.2 Discussion of detailed results

We begin the discussion with the main simulation results for the DGP with $N=8'000$ and a 'normal' IATE ($\alpha = 2$) presented in Table 1. Further simulation results are referred to various appendices. Appendix C.1 contains the results for the two smaller sample sizes, while Appendix C.2 considers the other three specifications of the IATEs, and Appendix C.3 shows the results for the additional estimators discussed in Section 4.5. Finally, Appendix D contains the results for the experimental case.

Table 1 contains the results for the estimators *Basic*, *OneF.VarT*, and *OneF.MCE* and its locally centred version based on two folds, *OneF.MCE.LC-2*. Both MCE based estimators are also considered in their penalized form (*OneF.MCE.Pen, OneF.MCE.Pen.LC-2)*. To see the impact of the (computationally more expensive) increase in folds for the locally centred estimator, we consider also a version based on 5-folds (*OneF.MCE.Pen.LC-5*).

*Basic* is substantially biased, but it captures the cross-sectional variation of the heterogeneity well. Furthermore, *Basic* (as well as all other estimators considered) appears to be normally distributed. The estimated weights-based standard errors are somewhat too large. The main problem for inference is, however, the substantial bias leading to too small coverage probabilities for all effects, in particular for ATE and the GATEs.

Estimating only one forest instead of two as in *Basic* and using the splitting rule of WA18 (*OneF.VarT*) improves the MSE somewhat for IATEs (but increases them for ATE and GATEs), but biases get even larger. This leads to lower coverage probabilities than the already too low ones of *Basic*.

*Table 1: Simulation results for N=8'000, main DGP, and main estimators*

| | | | True & estimated effects | | | Estimation error of effects (averages) | | | | | Estimation of std. error | |
|---|---|---|---|---|---|---|---|---|---|---|---|---|
| | Groups | Est. | Avg. bias | X-sectional std. dev. | | MSE | Skewness | Kurtosis | JB-Stat. | Std. err. | Avg. bias | CovP (90) in % |
| | # | | | true | est. | | | | | | | |
| | (1) | | (2) | (3) | (4) | (5) | (6) | (7) | (8) | (9) | (10) | (11) |
| ATE | 1 | Basic | 1.25 | - | - | 1.75 | 0.0 | 2.9 | 0.1 | 0.43 | 0.04 | 15 |
| GATE | 2 | | 1.27 | - | - | 1.78 | -0.1 | 2.8 | 0.7 | 0.45 | 0.05 | 18 |
| GATE | 32 | | 1.21 | 0.17 | 0.14 | 1.81 | 0.0 | 2.8 | 1.6 | 0.56 | 0.08 | 38 |
| IATE | 5000 | | 1.38 | 1.72 | 1.42 | 4.89 | 0.0 | 2.9 | 1.9 | 1.45 | 0.22 | 78 |
| ATE | 1 | OneF. | 1.71 | - | - | 3.12 | 0.3 | 3.1 | 1.5 | 0.51 | 0.06 | 3 |
| GATE | 2 | VarT | 1.71 | - | - | 3.14 | 0.2 | 3.1 | 2.2 | 0.54 | 0.07 | 4 |
| GATE | 32 | | 1.68 | 0.17 | 0.11 | 3.07 | 0.2 | 3.0 | 1.0 | 0.58 | 0.10 | 7 |
| IATE | 5000 | | 1.73 | 1.72 | 1.33 | 4.64 | 0.0 | 3.0 | 2.4 | 1.50 | 0.50 | 71 |
| ATE | 1 | OneF. | 1.29 | - | - | 1.86 | -0.1 | 3.3 | 0.4 | 0.46 | 0.04 | 14 |
| GATE | 2 | MCE | 1.28 | - | - | 1.86 | -0.1 | 3.2 | 0.4 | 0.46 | 0.05 | 17 |
| GATE | 32 | | 1.27 | 0.17 | 0.08 | 1.87 | 0.0 | 3.1 | 1.1 | 0.50 | 0.14 | 34 |
| IATE | 5000 | | 1.34 | 1.72 | 0.88 | 3.92 | 0.0 | 2.9 | 2.0 | 1.00 | 0.46 | 75 |
| ATE | 1 | OneF. | 0.90 | - | - | 1.06 | 0.1 | 2.3 | 2.5 | 0.50 | 0.02 | 44 |
| GATE | 2 | MCE. | 0.90 | - | - | 1.06 | 0.1 | 2.5 | 1.9 | 0.51 | 0.02 | 48 |
| GATE | 32 | LC-2 | 0.88 | 0.17 | 0.06 | 1.10 | 0.1 | 2.5 | 1.9 | 0.56 | 0.03 | 55 |
| IATE | 5000 | | 1.12 | 1.72 | 0.70 | 3.36 | 0.0 | 3.0 | 2.3 | 1.09 | 0.08 | 70 |
| ATE | 1 | OneF. | 0.21 | - | - | 0.27 | 0.1 | 3.3 | 0.8 | 0.49 | 0.28 | 97 |
| GATE | 2 | MCE. | 0.20 | - | - | 0.28 | 0.1 | 3.2 | 0.7 | 0.49 | 0.29 | 98 |
| GATE | 32 | Pen | 0.20 | 0.17 | 0.16 | 0.36 | 0.1 | 3.2 | 1.6 | 0.56 | 0.31 | 97 |
| IATE | 5000 | | 0.32 | 1.72 | 1.72 | 2.29 | 0.1 | 3.0 | 2.7 | 1.44 | 0.78 | 98 |
| ATE | 1 | OneF. | 0.23 | - | - | 0.49 | -0.2 | 3.3 | 1.5 | 0.66 | 0.01 | 90 |
| GATE | 2 | MCE. | 0.24 | - | - | 0.50 | -0.2 | 3.3 | 1.5 | 0.67 | 0.01 | 90 |
| GATE | 32 | Pen | 0.23 | 0.16 | 0.15 | 0.53 | -0.2 | 3.2 | 1.7 | 0.69 | 0.02 | 89 |
| IATE | 5000 | LC-2 | 0.39 | 1.72 | 1.65 | 2.39 | 0.0 | 3.0 | 1.5 | 1.45 | 0.10 | 90 |
| ATE | 1 | OneF. | 0.21 | - | - | 0.35 | 0.3 | 3.1 | 1.5 | 0.55 | 0.02 | 86 |
| GATE | 2 | MCE. | 0.21 | - | - | 0.35 | 0.3 | 3.0 | 1.5 | 0.56 | 0.03 | 88 |
| GATE | 32 | Pen | 0.21 | 0.16 | 0.15 | 0.37 | 0.2 | 3.1 | 1.7 | 0.57 | 0.04 | 88 |
| IATE | 5000 | LC-5 | 0.37 | 1.72 | 1.57 | 1.83 | 0.1 | 2.9 | 2.0 | 1.27 | 0.07 | 90 |

Note: For GATE and IATE the *average bias* is the absolute value of the bias for the specific group (GATE) / observation (IATE) averaged over all groups / observation (each group / observation receives the same weight). *CovP (90%)* denotes the (average) probability that the true value is part of the 90% confidence interval. The simulation errors of the mean MSEs are around 0.08.

Using the MCE splitting rule proposed in this paper (*OneF.MCE*) leads to substantially lower MSEs than for *Basic* (for IATEs) and *OneF.VarT* (for all parameters). Part of this gain in MSEs comes from a reduction of biases (compared to *OneF.VarT*) and another part comes from a reduction of the variances of the IATEs. However, due to the still substantial bias, coverage probabilities are still too small to be useful for inference. Adding local centering to the MCE criterion (*OneF.MCE.LC-2)* reduces the bias for all parameters and improves the MSE. However, despite a rather accurate estimation of the standard errors, coverage probabilities are still too low, because the bias is still relevant.

Adding the penalty terms to the splitting rule drastically reduces the remaining biases for the MCE based estimators (same for *OneF.VarT*, see Appendix C.3) and leads to substantially smaller MSEs. The bias reductions come at the cost of some additional variance. Interestingly, once the penalty is used, local centering does not affect the bias much, but improves the estimation of the weights-based standard errors. It is a general feature of all simulations that for estimators not using local centering estimated standard errors particularly for the IATEs are always too large (leading to conservative inference). However, estimated standard errors for the locally centred estimators are fairly accurate. The comparison of the estimators using two-fold and five-fold local centering shows that the additional variance reduction coming from using more than two folds can be relevant. This leads to the conclusion that across all parameters, and considering point estimates and inference jointly, either *OneF.MCE.Pen* (if having conservative inference is no problem) or *OneF.MCE.Pen.LC-5* are the preferred estimators.

A similar conclusion also holds for the case of no effect heterogeneity (Table C.5) as well as for the case of earnings dependent heterogeneity (Table C.11). However, the gain of local centering for estimating the IATEs is less clear (in fact, the IATEs are best estimated by *OneF.MCE*; for ATE and the GATEs, the penalty function is important, though). The most challenging case for estimation is the one with strong effects strongly linked to the propensity score ($\alpha=8$, Table C.8). In this case, the clear winner of all estimators is the uncentred, penalised, MCE based estimator (*One.MCE.Pen*). Thus, if one cannot rule out strong heterogeneity strongly linked to selection, *One.MCE.Pen* is the estimator of choice because it is the only one that performs well in all scenarios. The cost of this robustness is that for IATE estimated standard errors are too large leading to conservative inference. Otherwise, as mentioned before, its locally centred version may perform substantially better with respect to inference.

Considering the results for the smaller samples based on the same DGP as in Table 1 (Tables C.1 and C.2), we arrive at very similar conclusion favouring *One.MCE.Pen.LC-5*. Using the three sample sizes of 1'000, 4'000, and 8'000 allows to consider approximate convergence rates of the estimators (of course, based on 3 data points only). For example, for the ATE and *One.MCE.Pen.LC-5* the biases (standard errors) fall with sample size as 0.79 – 0.39 – 0.21 (1.34 – 0.73 – 0.55) which seems to be roughly in line with $\sqrt{N}$-convergence. The absolute biases (standard errors) of the IATE fall slower at 0.79 – 0.53 – 0.37 (1.96 – 1.45 – 1.27). A similar picture appears for the other specification of the IATEs. The results for the strong-effect-highly-correlated-with-selection case (Tables C.6 and C.7) point to the need of a large enough sample when such a phenomenon occurs. For *N = 1'000* none of the estimators performs well, while for *N=4'000* their performances improve, but even the best estimator for this DGP (*OneF.MCE.Pen*) has a bias large enough to lead to too small coverage probabilities. This problem disappears for *N=8'000* (Table C.8).

Considering now further estimators (Appendix C.3), it turns out that *OneF* performs well in the case when there is no effect. This is expected, because in this case the dependence of the conditional expectation of the outcome on covariates is the same among treated and controls. In all other cases, one of the other estimators always dominates it. Adding a penalty term to *OneF.VarT* leading to *OneF.VarT.Pen* also reduces its bias, but this is not enough to become competitive with estimators performing well in these scenarios. Finally, comparing *Basic* in its one-sample (*Basic.OneSam*) and two-sample versions shows that the one-sample-with-honesty version might have a slightly lower MSE (if at all). However, its bias tends to be somewhat larger and, as expected, the estimated standard errors are too small. This indicates that the costs of sample splitting needed for inference, seem to be low and unavoidable.

The simulations with random assignment (see Appendix D) show the importance of the selection process for the differential performance of the estimators. The first finding for the experimental case is that the other estimators considered dominate *Basic* (with the exception of the strong-effect DGP and the small sample). The second finding is in fact a non-finding in the sense that it is very difficult to rank the other single-forest estimators among themselves as they perform very similarly. This is as expected, as the splitting rule of WA18 seems to be well justified if there is no selectivity. Another good news of this finding is that the penalty term does not inflict much harm when used in cases in which it is redundant (as there is no selection bias in these DGPs). Finally, since the bias of all estimators are small in the experimental case, coverage probabilities substantially improve.

The conclusion of the simulation exercise is the following: If there is some a priori knowledge that the propensity score and the effect heterogeneity are not too important and tied too closely together, then OneF.MCE.Pen.LC-5 is the preferred estimator. It has low MSE and good coverage. However, OneF.MCE.Pen is more robust and works in this difficult case in medium sized samples. However, this robustness comes at the price of the inference being too conservative. If effects are very precisely estimated, because of their large size or of large samples, such costs may not matter much and may be outweighed by the additional robustness.

# 6  Application to the evaluation of a job search programme

This section shows how the estimator that came out best in the Empirical Monte Carlo study across all scenarios, i.e. *OneF.MCE.Penalty*, can be productively applied in empirical studies. To this extent, we rely on the data set that formed the basis of the Empirical Monte Carlo study above, but we use a more homogeneous subsample of men living in cantons where German is the dominant language (about 38'000 observations). As before, we investigate the effect of participating in a job search programme (share of participants is 8%) on months of employment accumulated over 9 months, as well as over 3 years, respectively. Based on a previous reanalysis of the effects of this programme in Knaus, Lechner, and Strittmatter (2017), we expect to find essentially zero effects with very limited heterogeneity over three years, but substantial heterogeneity over the first 9-month period, the so-called lock-in period. On top of this, we expect selective programme participation (see below). Thus, this is a challenging setting.[25]

The estimation with *OneF.MCE.Penalty* is based on 1'000 subsampling replications (50% subsampling share). The minimum leaf size $N^{min} = (13, 50)$ and the number of coefficients used for leaf splitting $M = (4, 10, 27)$ (out of 31 ordered and 8 unordered variables;[26] 6 variables have been deleted a priori as their unconditional correlation with both the outcome and programme participation is below 1%) are treated as tuning parameters. $N^{min} = 13$ and $M = 27$ are chosen by out-of-bag minimisation of the optimization criterion of this estimator.

The matching literature has shown that it may be important to ensure that common support actually holds post-estimation (e.g. Imbens, 2004). This issue has not yet been discussed in the context of ML methods of the type proposed here.[27] The key issue is that for every combination of values of *X* that is relevant for estimation, there should be enough treated and non-treated observations for reliable estimation of the effects. While this has been shown to be relevant for the (conventional) estimation of *ATE* and related aggregated parameters (e.g. Lechner and Strittmatter, 2019), it is even more important when estimating *IATE(x)*, which will be particularly sensitive to support violations close to their evaluation points. Our proposal is to predict the propensity scores using the trees of the already estimated (outcome) forest in a first step. In a second step, all values of *X* related to very few predicted treated or controls are deleted. Here, we deleted all values of *X* with propensity scores outside the [5%, 95%]-interval. This led to discarding about 0.5% (9 months) and 1.1% (3 years) of the observations. In the discarded group, almost 80% of the observations have a native language that is different from any of the main Swiss official languages (German, French, and Italian).

---

[25] This section is merely a demonstration of possible findings. Therefore, for the sake of brevity, it is written very densely. For more details on programmes, institutional details, data, the reader is referred to the previous papers using this setting, e.g. Huber, Lechner, and Mellace (2017), Knaus, Lechner, and Strittmatter (2017, 2018) and the references therein.

[26] *k-means-clustering* is used to deal with unordered variables (as proposed in Chou, 1991, and Hastie, Tibshirani, and Friedman, 2009).

[27] For a general discussion on the implications of the overlap assumption in a high-dimensional covariate space, see D'Amour, Ding, Feller, Lei, and Sekhon (2018).

From the point of view of eliminating selection bias, any good estimator in this setting should ensure covariate balance. Table 2 therefore shows the means and standardized differences for selected covariates (mostly directly related to the pre-specified heterogeneity shown in Table 3 below) prior to the estimation. To see how the different forests related to the two respective outcome variables change the balancing, the same covariates are predicted using the estimated forests (and their implied weights; columns headed by 'post-estimation'). In both cases, the balancing improves considerably.

*Table 2: Pre- and post-estimation balancing of covariates*

| | Pre-estimation | | | | Post-estimation | |
|---|---|---|---|---|---|---|
| **Features** | Mean $d=1$ | Mean $d=0$ | Difference | Stand. diff. (%) | Difference | |
| **Outcome: Months of employment in** | | | | | 9 month | 3 years |
| # of pre UE employment episodes (last 2 years) | 0.09 | 0.12 | -0.027 | 20 | -0.019 | -0.016 |
| Foreign native language | 0.24 | 0.32 | -0.08 | 19 | -0.05 | -0.05 |
| Employability (1, 2, 3) | 2.01 | 1.93 | 0.09 | 18 | 0.02 | 0.05 |
| Pre-UE monthly earnings (in CHF) | 5435 | 4899 | 537 | 26 | 50 | 26 |

Note: The standardized difference (stand. diff.) is defined as $Stand.diff = \dfrac{|\bar{x}^1 - \bar{x}^0|}{\sqrt{[Var(x^1) + Var(x^0)]/2}}$, where the superscripts 1 and 0 denote the subpopulations of treated and controls. Post-estimation results are obtained by treating the variables as outcomes and running them through the estimated forests. Since these forests differ for the 9 and the 36 month outcomes, the post-estimation results differ with respect to the outcome variable used to train the forests.

Tables 3 and 4 contain the results of the estimation for the ATE (upper panel), the various GATEs that appeared to be of a priori interest (in the middle of the table), as well as summary statistics for the IATEs (lower panel). The GATEs relate to effect heterogeneity in the number of unemployment episodes in the 2 years prior to the unemployment spell analysed (12 categories), the native language being a 'Swiss' one or a different one (2 cat.), the employability index (3 cat.), and the economic sector of the last occupation (16 cat.). As mentioned before, all effects have the interpretation of additional months of regular employment due to programme participation.

*Table 3: Average, group effects, and individualized effects: 9 months*

| | Average potential outcomes | | | | Average effects | | |
|---|---|---|---|---|---|---|---|
| | Treated | | Controls | | | | |
| | Expectation | Stand. err. | Expectation | Stand. err. | ATE | Stand. err. | p-val. in % |
| | 1.59 | 0.06 | 2.09 | 0.02 | -0.52 | 0.06 | 0.0 |
| | **Group average treatment effects (GATEs)** | | | | | | |
| **Group** | | Expect. | Est. std. | p-val. in % | **Group** | Expect. | Est. std. |
| # E spells | 0 | -0.52 | 0.06 | 93 | Sector 1 | -0.56 | 0.07 |
| # E spells | 1 | -0.51 | 0.06 | 37 | Sector 2 | -0.64 | 0.07 |
| # E spells | 2 | -0.50 | 0.06 | 93 | Sector 3 | -0.51 | 0.06 |
| # E spells | 3 | -0.50 | 0.07 | 43 | Sector 4 | -0.41 | 0.09 |
| # E spells | 4 | -0.50 | 0.07 | 55 | Sector 5 | -0.44 | 0.07 |
| # E spells | 5 | -0.49 | 0.07 | 87 | Sector 6 | -0.45 | 0.07 |
| # E spells | 6 | -0.49 | 0.08 | 39 | Sector 7 | -0.50 | 0.07 |
| # E spells | 7 | -0.51 | 0.07 | 30 | Sector 8 | -0.50 | 0.06 |
| # E spells | 8 | -0.48 | 0.08 | 59 | Sector 9 | -0.53 | 0.06 |
| # E spells | 9 | -0.51 | 0.08 | 76 | Sector 10 | -0.50 | 0.06 |
| # E spells | 10 | 0.53 | 0.10 | - | Sector 11 | -0.64 | 0.09 |
| # E spells joint test (p-val in %) | | | | 94 | Sector 12 | -0.49 | 0.07 |
| **Swiss lang. native** | | -0.51 | 0.06 | 68 | Sector 13 | -0.54 | 0.07 |
| **Swiss lang. not nat.** | | -0.53 | 0.06 | - | Sector 14 | -0.45 | 0.07 |
| **Swiss joint test** | | | | 68 | Sector 15 | -0.46 | 0.07 |
| **Employability good** | | -0.49 | 0.06 | 11 | Sector 16 | -0.47 | 0.07 |
| **Employabil. middle** | | -0.52 | 0.06 | 52 | **Sector** joint test (p-val in %) | | 0.1 |
| **Employability bad** | | -0.54 | 0.07 | - | | | |
| **Employability** joint test (p-val in %) | | | | 17 | | | |
| | **Individualized average treatment effects (IATEs)** | | | | | | |
| **Mean** | Standard deviation | Share < 0 in % | Share > 0 in % | Average standard error of estimate | Share significant at 5% level in % | | |
| **-0.52** | 0.15 | 100 | 0 | 0.15 | 92 | | |

Note: *Expect.* and *est. std.* is defined for potential outcomes, effects, and their respective estimation errors. The p-values relate to hypothesis tests based on the (asymptotic) t- or Wald-type. For the ATE, it is the *t*-test that the effect is zero. For the GATEs, it is the *t*-test that the adjacent values are identical (therefore it is not given for the sectors that are not in any natural order) as well as the Wald test that all GATEs relating to this variables are identical. Therefore, this test has (# of groups-1) degrees of freedom.

For the average treatment effect, the first four columns show the mean of the potential outcomes as well as their estimated standard errors. The remaining columns show the average effects, their estimation error, and the p-value for the hypothesis that the average effects are zero. The results imply that over nine months (3 years), these individuals work 1.6 (16.5) month if they participate in the programme and 2.1 (16.8) months if not. This negative short-run effect is statistically significant at conventional levels (despite inference being likely conservative), while the smaller 3-year effect is not.

The middle part of the tables contains the group-mean effects and their standard errors for the four different GATEs relating to the variables discussed above. The p-values shown, however, do not relate to the hypothesis that the particular effects are zero but to the hypothesis that the differences of the effects minus the ones in the next row are zero. Such a test contains relevant information only if the heterogeneity variable is ordered. Since this is not true for the economic sectors, the p-values are not shown for those GATEs. Finally, for each GATE the last row shows the p-value of a Wald-test (again, based on the weighted representation of the estimator, which allows uncovering the full variance-covariance matrix of the estimated GATEs) with degrees of freedom equal to the number of different categories for the particular GATE for the hypothesis that all GATEs relating to this variable are equal.

No significant heterogeneity appears for the 3-year outcome, neither in a statistical nor in a substantive sense. However, for the 9-month employment outcome there appears to be heterogeneity with respect to the economic sector. Here, the Wald test clearly rejects homogeneity and the various negative effects differ substantially.[28]

While the estimated GATEs are presented in total since they are not based on too many different groups, this is impossible for the estimated IATEs. Thus, for a better description we may want to condense their information further. The first part of this exercise is reported in the lower panel of Table 3 and 4. The first two columns of these tables show the average and their standard deviation. Next, there is the share of IATEs with positive and negative values. Finally, we report the average of their estimated standard errors and the share of IATEs that are significantly different from zero (at the 5% level). For the 9-month outcome, all effects are negative and more than 90% significantly so (in a statistical sense). Their standard deviation across individuals is about 4 days. For the 3-year outcome, their standard deviation is larger, but this is essentially so because the potential outcomes are about 10 x larger. About two thirds of the IATEs are smaller than zero, and one third is larger. However, only 2% of them are statistically significantly different from zero.

---

[28] Given the space constraints of this paper, we refrain from analysing and interpreting this heterogeneity further.

*Table 4: Average, group effects, and individualized effects: 36 months*

| | Average potential outcomes | | | | Average effects | | |
|---|---|---|---|---|---|---|---|
| | **Treated** | | Controls | | | | |
| **Expectation** | Stand. err. | Expectation | Stand. err. | ATE | Stand. err. | p-val. in % |
| **16.50** | 0.33 | 16.79 | 0.099 | -0.30 | 0.34 | 39 |
| **Group average treatment effects (GATEs)** | | | | | | | |
| Group | | Expect. | Est. std. | p-val. in % of diff | Group | Expect. | Est. std. |
| # E spells | 0 | -0.55 | 0.37 | 12 | Sector 1 | 0.19 | 0.47 |
| # E spells | 1 | -0.19 | 0.36 | 12 | Sector 2 | -0.62 | 0.42 |
| # E spells | 2 | 0.04 | 0.40 | 16 | Sector 3 | -0.30 | 0.37 |
| # E spells | 3 | 0.16 | 0.44 | 43 | Sector 4 | 0.77 | 0.67 |
| # E spells | 4 | 0.22 | 0.50 | 9 | Sector 5 | 0.04 | 0.42 |
| # E spells | 5 | 0.34 | 0.55 | 96 | Sector 6 | -0.13 | 0.40 |
| # E spells | 6 | 0.34 | 0.50 | 37 | Sector 7 | -0.27 | 0.46 |
| # E spells | 7 | 0.15 | 0.50 | 65 | Sector 8 | -0.25 | 0.39 |
| # E spells | 8 | 0.27 | 0.55 | 92 | Sector 9 | -0.44 | 0.44 |
| # E spells | 9 | 0.24 | 0.60 | 37 | Sector 10 | -0.32 | 0.37 |
| # E spells | 10 | -0.16 | 0.60 | - | Sector 11 | -0.45 | 0.58 |
| # E spells  joint test | | | | 28 | Sector 12 | -0.25 | 0.42 |
| Swiss lang. native | | -0.22 | 0.40 | 59 | Sector 13 | -0.42 | 0.37 |
| Swiss lang. not nat. | | -0.33 | 0.36 | 35 | Sector 14 | -0.17 | 0.44 |
| Swiss joint test | | | | 66 | Sector 15 | -0.29 | 0.38 |
| Employability good | | -0.09 | 0.38 | 81 | Sector 16 | -0.19 | 0.40 |
| Employabil. middle | | -0.33 | 0.35 | 35 | Sector joint test (p-val in %) | | 41 |
| Employability bad | | -0.44 | 0.38 | 25 | | | |
| Employability  joint test (p-val in %) | | | | 34 | | | |
| **Individualized average treatment effects (IATEs)** | | | | | | | |
| **Mean** | Standard deviation | Share < 0 in % | Share > 0 in % | Average standard error of estimate | Share significant at 5% level in % | | |
| **-0.30** | 0.86 | 64 | 36 | 1.10 | 0.02 | | |

Note: *Expect.* and *est. std.* is defined as potential outcomes and effects and their estimation errors. The p-values relate all to the hypothesis test based on the (asymptotic) t- or Wald-type statistic. For the ATE, it is the *t*-test that the effect is zero. For the GATEs, it is the *t*-test that the adjacent values are identical (therefore it is not given for the sectors that are not in any natural ordered) as well as the Wald test that all GATEs relating to this variables are identical. Therefore, this test has (# of groups-1) degrees of freedom.

The final step consists in a post-estimation analysis of the IATEs in order to be able to *describe* the correlation of the estimated IATEs with exogenous variables. In principle, all standard descriptive tools could be used for this exercise. In a multivariate analysis, there is the difficulty that the covariate space is very large. Thus, we use a dimension reduction tool, here the LASSO, and perform OLS estimation with the covariates that had a non-zero coefficient in the LASSO estimation (this is also called Post-LASSO; for theoretical properties of the Post-LASSO and its advantages over LASSO see, e.g., Belloni, Chernozhukov and Wei, 2016).[29]

---
[29] This approach is in spirit very similar to Zhao, Small, and Ertefaie (2017).

Since model selection (LASSO) and coefficient estimation are performed on two random, non-overlapping subsamples (50% each), the OLS standard errors remain valid given the selected model. However, of course they do not explicitly account for the fact that the dependent variable is an estimated quantity.

*Table 5: Selected coefficients of post-lasso OLS estimation of IATE*

| Variable | Coefficient | Standard error | t-value | p-value in % | Uncond. correlation (if > 10%) in % |
|---|---|---|---|---|---|
| **Outcome: Months of employment in first 6 months** | | | | | |
| \|Age case worker – age UE\| in years | -0.002 | 0.001 | -2.0 | 4.7 | -24 |
| \|Age case worker – age UE\| < 5 | not | selected | - | - | - |
| Age | 0.45 | 0.22 | 2.0 | 4.6 | 27 |
| # E spells | 0.10 | 0.05 | 2.0 | 4.6 | - |
| Time in employment before (share) | -0.10 | 0.05 | -2.0 | 4.5 | -16 |
| Earnings before | -0.04 | 0.02 | -1.9 | 5.2 | - |
| Employability | -0.004 | 0.002 | 1.7 | 9.5 | - |
| **Outcome: Months of employment in first 3 years** | | | | | |
| \|Age case worker – age UE\| in years | not | selected | - | - | 11 |
| \|Age case worker – age UE\| < 5 | -0.12 | 0.03 | -4.1 | < 1% | - |
| Age | -0.01 | 0.001 | -7.8 | < 1% | -20 |
| # E spells | 0.12 | 0.01 | 14 | <1% | 26 |
| Time in employment before (share) | -1.20 | 0.06 | -21 | <1% | -32 |
| Earnings before | -0.008 | 0.0006 | -14 | <1% | - |
| Employability | -0.17 | 0.03 | -7.3 | <1% | - |

Note: OLS regression. Dependent variable is the estimated IATE. Independent variables are those with non-zero coefficients in the LASSO estimation. OLS and LASSO were estimated on randomly splitted subsamples (to facilitate the use of ordinary, robust OLS standard errors). The penalty term of the LASSO estimation was chosen by minimization of the 10-fold cross-validated mean square prediction error and the one-standard error rule. $R^2$ = 71% (9 months), 20% (3 years). *N*=8990. A constant term and further variables capturing caseworker, local and individual characteristics are included as well.

Table 5 shows a selection of variables only, to serve as examples. Here, heterogeneity appears for both outcome variables, but related to variables not specified in the GATEs, like age difference of case worker and unemployed, age, number of employment spells, pre-unemployment time in employment and earnings, as well as the employability according to the judgement of the case worker. While these are highly significant predictors for the 3-year outcome, their 9-month effects are significant at approximately the 5% level.

In summary, this section gave a first indication about the usefulness of the new methods in applications and provides some suggestions on how to deal with some practical issues (i.e.

common support, balancing of covariates, describing the IATEs) that arise when applying such ML tools in a causal framework.

# 7 Conclusion

In this paper, we develop and apply new estimators to estimate heterogeneous treatment effects for multiple treatments in a selection-on-observables setting. We compare them to existing estimation approaches in an empirically informed simulation study, and apply the best performing estimator to an empirical programme evaluation study. The new estimators are an extension of the Causal Forest approach proposed by Wager and Athey (2018).

The new estimators deviate from the original Causal Forest in that a new splitting criterion is proposed to build the trees that form the forest. The new splitting criterion has two components: First, we approximate the mean square error of the causal effects by combining the causal problem with the particular identification strategy directly. Second, a penalty term is added that favours splits that are heterogeneous with respect to the implied treatment propensity to reduce selection bias directly. It turned out that both changes worked as intended and that they can improve the performance of the Causal Forest approach in observational studies substantially.

An additional advantage of the Causal Forest approach is that the common estimators have a representation as weighted means of the outcome variables. This makes them particularly amendable to using a common inference framework for estimation effects at various aggregation levels, like for the average treatment effect (ATE), the group treatment effect (GATE), or the individualized treatment effect (IATE). The advantage is that ATE and GATEs (and their inference) can be obtained directly from aggregating the estimated IATEs without the need for different additional ML estimations at the various aggregation levels of interest.

These changes together with the good performance of the proposed methods in the simulation study and the application make this new estimator (even more) attractive to use in empirical studies. However, several issues are still unresolved and deserve further inquiry in the future: One open issue is how to choose the value of the penalty term in an optimal way at low computational costs. Another topic is to explore alternative methods to estimate the mean correlated errors, which is the main innovation in the new splitting rule. Three other issues concern the proposed inference methods: First, it would be desirable to have a more solid theoretical justification why the inference methods work well, e.g. by deriving explicitly the regularity conditions that the weights have to fulfil to lead to a consistent standard error estimator. Second, this paper explored just one way on how to implement this weights-based inference. It seems worthwhile to investigate alternatives, also with the goal of tackling the issue that the current standard errors for the IATEs in particular seem to be too large without local centering. Third, it would be useful to understand better the relation of the many other tuning parameters that come with these types of Random or Causal Forests (e.g., number of coefficients to be randomly chosen, minimum leaf size, subsampling size) to the quality of inference (and point estimation).

The Gauss code of the estimator is available at the websites of the author at *ResearchGate* and at [www.michael-lechner.eu/statistical-software](www.michael-lechner.eu/statistical-software). A release of the code in other computer languages (R/Python) is planned.

# Appendix A *(Online)*: More details on estimation and inference

Here, we present some more details of the Causal Forest algorithms used in the EMCS.

## Appendix A.1: Tuning parameters

The number of randomly chosen variables to form the next split in a tree (*V*) is either chosen from a *min(58,1+Poisson(5))* process or from a *min(58,1+Poisson(38))* process (58 variables in total). The means of the two Poisson processes are considered as the only tuning parameters. They are chosen based on the out-of-bag value of the objective function of the particular estimator. The motivation for using a random number of features is to foster the independence of the trees that appear in the Random Forest.[30]

The minimum leaf size in each tree equals 5. The number of trees contained in any forest equals 1000. Trees are formed on random subsamples drawn without replacement (subsampling) with a sample size of 50% of the size of sample A.

The nonparametric regressions that enter the estimation of the standard errors are based on *k*-NN estimation with number of neighbours equal to *2 sqrt(N)*.

## Appendix A.2: Local centering

Recentering is implemented in the following way:

1) Estimate the trees that define the random forest for $E(Y \mid X = x)$ in sample A.

2) Recentering of outcomes in sample A: Split sample A randomly into *K* equally sized parts, *A-1* to *A-K*. Use the outcomes in the union of the *K-1*-folds *A-1* to *A-(K-1)* to obtain the random forest predictions given the forest estimated in step 1). Use these prediction to predict $E(Y | X = x)$ in fold *A-K*. Do this for all-possible combinations of folds (cross-fitting as used in k-fold cross-validation). Subtract the predictions from the actual values of *Y*.

---

[30] This has also been suggested by Denil, Matheson, and de Freitas (2014) for regression forests.

3) Redo step 2 in sample B using the estimated forests of sample A.

Concerning the specifics of this algorithm, there are a couple of points worth mentioning.

First, in order to avoid overfitting, the outcomes of observation 'i' are not used to predict itself. Therefore, the implementation is chosen similar to cross-validation.

Second, weights-based inference requires avoiding a dependence of the weights in sample B on outcomes of sample A. However, since recentering uses outcome variables independent of the treatment state, this could induce a correlation between the recentered outcomes in different treatment states. This finite sample correlation will be ignored here (as in Athey, Tibshirani, and Wager, 2019).

Third, the number of folds is a tuning parameter that influences the noise added to the recentered outcome by subtracting an estimated quantity. The simulation results indicate that the computationally most attractive choice of *K=2* may be too small in medium sized samples and that a somewhat larger number of folds may be needed to avoid much additional noise to the estimators.

## Appendix B *(Online)*: Protocol of EMCS

The Empirical Monte Carlo study follows almost exactly the one used in KLS18. The main differences are in how the effects are generated, as explained in the main body of the text. In this appendix, we repeat their protocol for completeness, with some small adjustments.

1) Take the full sample and estimate the propensity score, $p^f(x)$. We use the specification of Huber, Lechner, and Mellace (2017) which is based on a logit model.

2) Remove all treated from the dataset and keep only the $N^{nt}$ non-treated observations ($Y^0$).

3) Specify the true ITE. Add them to $Y^0$ to obtain $Y^1$ for all observations.

4) Remove $N^v=5000$ observations from the main sample to form the validation sample.

5) Calculate the true GATEs and ATE by aggregating the true IATEs in the validation sample.

6) Modify $p^f(x)$ such that it equals approximately 50% and compute its value for all $N$ units.

7) Draw random samples with $N=1'000$, $N=4'000$, and $N=8'000$ from the ($N^{nt} - N^v$) observations.

8) Assign a treatment state based on the outcome of a draw in the Bernoulli distribution with this modified probability as parameter.

9) Depending on the assigned treatment state, use the value $Y^0$ or $Y^1$ as observable outcome variable $Y$.

10) Use these $N$ observations as training sample to compute all effects.

11) Predict all effects in the validation sample.

12) Predict the quality measures in the validation sample for each parameter to estimate.

13) Repeat steps 7 to 12 1'000 ($N=1'000$), 250 ($N=4'000$), or 125 ($N=8'000$) times.

14) Calculate quality measures by aggregating over all estimated effects.

# Appendix C *(Online)*: Further simulation results for the case with selection bias

The tables in this appendix follow exactly the structure of those in Section 5 of the main body of the text.

# Appendix C.1: Main specification with smaller sample sizes

This appendix contains the results of the main specification for the smaller samples of $N=1'000$ and $N=4'000$.

*Table C.1: Simulation results for N=1'000, main DGP, and main estimators*

| | Groups # | Est. | True & estimated effects | | | Estimation error of effects (averages) | | | | | Estimation of std. error | |
|---|---|---|---|---|---|---|---|---|---|---|---|---|
| | | | Avg. bias | X-sectional std. dev. | | MSE | Skewness | Kurtosis | JB-Stat. | Std. err. | Avg. bias | CovP (90) in % |
| | | | | true | est. | | | | | | | |
| | (1) | | (2) | (3) | (4) | (5) | (6) | (7) | (8) | (9) | (10) | (11) |
| **ATE** | *1* | Basic | 1.90 | - | - | 4.92 | -0.1 | 2.7 | 5.4 | 1.14 | 0.08 | 53 |
| **GATE** | *2* | | 1.90 | - | - | 4.96 | -0.1 | 2.7 | 4.7 | 1.16 | 0.09 | 54 |
| **GATE** | *32* | | 1.87 | 0.17 | 0.09 | 4.98 | -0.1 | 2.8 | 4.2 | 1.22 | 0.13 | 60 |
| **IATE** | *5000* | | 1.91 | 1.72 | 0.96 | 8.47 | -0.1 | 3.1 | 3.4 | 1.86 | 0.46 | 79 |
| **ATE** | *1* | OneF. | 2.09 | - | - | 5.64 | -0.1 | 3.0 | 0.7 | 1.14 | 0.11 | 48 |
| **GATE** | *2* | VarT | 2.09 | - | - | 5.67 | -0.1 | 3.0 | 0.7 | 1.15 | 0.12 | 50 |
| **GATE** | *32* | | 2.06 | 0.17 | 0.07 | 5.60 | -0.1 | 3.0 | 0.8 | 1.16 | 0.15 | 53 |
| **IATE** | *5000* | | 2.09 | 1.72 | 0.75 | 8.01 | 0.0 | 3.0 | 3.0 | 1.55 | 0.54 | 74 |
| **ATE** | *1* | OneF. | 1.97 | - | - | 5.28 | -0.1 | 3.1 | 1.6 | 1.18 | 0.04 | 50 |
| **GATE** | *2* | MCE | 1.97 | - | - | 5.31 | -0.1 | 3.1 | 2.2 | 1.19 | 0.06 | 52 |
| **GATE** | *32* | | 1.94 | 0.17 | 0.06 | 5.32 | -0.1 | 3.0 | 2.0 | 1.24 | 0.11 | 56 |
| **IATE** | *5000* | | 1.99 | 1.72 | 0.55 | 8.43 | 0.0 | 3.0 | 2.0 | 1.63 | 0.56 | 75 |
| **ATE** | *1* | OneF. | 1.39 | - | - | 3.56 | 0.0 | 3.1 | 0.2 | 1.28 | 0.11 | 76 |
| **GATE** | *2* | MCE. | 1.38 | - | - | 3.60 | 0.0 | 3.1 | 0.3 | 1.30 | 0.11 | 77 |
| **GATE** | *32* | LC-2 | 1.36 | 0.17 | 0.04 | 3.64 | 0.0 | 3.1 | 0.4 | 1.33 | 0.12 | 78 |
| **IATE** | *5000* | | 1.52 | 1.72 | 0.43 | 7.02 | 0.0 | 3.1 | 2.1 | 1.75 | 0.20 | 76 |
| **ATE** | *1* | OneF. | 1.11 | - | - | 3.35 | -0.1 | 2.9 | 2.6 | 1.46 | 0.10 | 81 |
| **GATE** | *2* | MCE. | 1.11 | - | - | 3.36 | -0.1 | 3.1 | 3.1 | 1.46 | 0.11 | 82 |
| **GATE** | *32* | Pen | 1.10 | 0.17 | 0.13 | 3.37 | -0.1 | 2.9 | 2.7 | 1.47 | 0.13 | 83 |
| **IATE** | *5000* | | 1.12 | 1.72 | 1.50 | 5.87 | -0.2 | 3.1 | 2.0 | 2.05 | 0.56 | 90 |
| **ATE** | *1* | OneF. | 0.97 | - | - | 3.42 | -0.1 | 3.1 | 1.6 | 1.57 | 0.03 | 84 |
| **GATE** | *2* | MCE. | 0.97 | - | - | 3.44 | -0.1 | 3.1 | 1.7 | 1.58 | 0.03 | 84 |
| **GATE** | *32* | Pen | 0.95 | 0.17 | 0.10 | 3.46 | -0.1 | 3.1 | 1.7 | 1.60 | 0.03 | 84 |
| **IATE** | *5000* | LC-2 | 1.00 | 1.72 | 1.24 | 6.52 | -0.1 | 3.2 | 9.5 | 2.25 | 0.12 | 86 |
| **ATE** | *1* | OneF. | 0.79 | - | - | 2.41 | -0.1 | 2.8 | 1.3 | 1.34 | 0.06 | 85 |
| **GATE** | *2* | MCE. | 0.79 | - | - | 2.43 | -0.1 | 2.8 | 1.3 | 1.34 | 0.06 | 85 |
| **GATE** | *32* | Pen | 0.78 | 0.17 | 0.10 | 2.44 | -0.1 | 2.8 | 1.4 | 1.35 | 0.06 | 85 |
| **IATE** | *5000* | LC-5 | 0.89 | 1.72 | 1.20 | 5.03 | -0.1 | 3.1 | 2.0 | 1.96 | 0.08 | 86 |

Note: For GATE and IATE the *average bias* is the absolute value of the bias for the specific group (GATE) / observation (IATE) averaged over all groups / observation (each group / observation receives the same weight). *CovP (90%)* denotes the (average) probability that the true value is part of the 90% confidence interval. The simulation errors of the mean MSEs are around 0.1.

*Table C.2: Simulation results for N=4'000, main DGP, and main estimators*

| | | | True & estimated effects | | | Estimation error of effects (averages) | | | | | Estimation of std. error | |
|---|---|---|---|---|---|---|---|---|---|---|---|---|
| | Groups # | Est. | Avg. bias | X-sectional std. dev. true | est. | MSE | Skew ness | Kurt- osis | JB- Stat. | Std. err. | Avg. bias | CovP (90) in % |
| | (1) | | (2) | (3) | (4) | (5) | (6) | (7) | (8) | (9) | (10) | (11) |
| ATE | 1 | Basic | 1.39 | - | - | 2.30 | 0.3 | 3.0 | 2.9 | 0.62 | 0.02 | 31 |
| GATE | 2 | | 1.39 | - | - | 2.35 | 0.2 | 3.1 | 2.4 | 0.64 | 0.03 | 35 |
| GATE | 32 | | 1.35 | 0.17 | 0.12 | 2.40 | 0.2 | 2.9 | 3.3 | 0.74 | 0.06 | 48 |
| IATE | 5000 | | 1.48 | 1.72 | 1.30 | 5.71 | 0.0 | 3.0 | 2.3 | 1.58 | 0.24 | 78 |
| ATE | 1 | OneF. | 1.78 | - | - | 3.57 | -0.2 | 3.0 | 1.0 | 0.63 | 0.05 | 17 |
| GATE | 2 | VarT | 1.78 | - | - | 3.58 | -0.1 | 3.0 | 1.4 | 0.65 | 0.06 | 18 |
| GATE | 32 | | 1.75 | 0.17 | 0.09 | 3.52 | -0.1 | 3.0 | 1.0 | 0.66 | 0.08 | 22 |
| IATE | 5000 | | 1.79 | 1.72 | 1.13 | 5.31 | -0.1 | 3.1 | 5.7 | 1.14 | 0.50 | 73 |
| ATE | 1 | OneF. | 1.46 | - | - | 2.50 | -0.6 | 3.7 | 20.1 | 0.61 | 0.04 | 22 |
| GATE | 2 | MCE | 1.45 | - | - | 2.50 | -0.6 | 3.6 | 17.4 | 0.63 | 0.06 | 27 |
| GATE | 32 | | 1.43 | 0.17 | 0.07 | 2.53 | -0.5 | 3.4 | 11.9 | 0.67 | 0.12 | 40 |
| IATE | 5000 | | 1.50 | 1.72 | 0.79 | 4.87 | -0.1 | 3.0 | 3.2 | 1.16 | 0.50 | 76 |
| ATE | 1 | OneF. | 1.05 | - | - | 1.54 | -0.2 | 2.9 | 1.9 | 0.71 | 0.04 | 56 |
| GATE | 2 | MCE. | 1.05 | - | - | 1.56 | -0.2 | 2.9 | 1.6 | 0.73 | 0.05 | 59 |
| GATE | 32 | LC-2 | 1.02 | 0.17 | 0.05 | 1.59 | 0.2 | 2.9 | 2.1 | 0.78 | 0.05 | 62 |
| IATE | 5000 | | 1.24 | 1.72 | 0.63 | 4.14 | 0.0 | 3.0 | 3.6 | 1.36 | 0.13 | 72 |
| ATE | 1 | OneF. | 0.39 | - | - | 0.89 | -0.1 | 2.9 | 0.9 | 0.86 | 0.10 | 90 |
| GATE | 2 | MCE. | 0.39 | - | - | 0.90 | -0.1 | 2.9 | 0.8 | 0.87 | 0.11 | 90 |
| GATE | 32 | Pen | 0.38 | 0.17 | 0.16 | 0.93 | -0.1 | 2.8 | 1.1 | 0.89 | 0.14 | 91 |
| IATE | 5000 | | 0.51 | 1.72 | 1.89 | 2.83 | -0.1 | 3.0 | 2.3 | 1.53 | 0.53 | 94 |
| ATE | 1 | OneF. | 0.39 | - | - | 1.02 | 0.1 | 3.0 | 0.1 | 0.90 | -0.03 | 86 |
| GATE | 2 | MCE. | 0.39 | - | - | 1.04 | 0.1 | 3.0 | 0.2 | 0.91 | -0.03 | 86 |
| GATE | 32 | Pen | 0.39 | 0.17 | 0.14 | 1.07 | 0.0 | 2.9 | 0.3 | 0.94 | -0.02 | 87 |
| IATE | 5000 | LC-2 | 0.51 | 1.72 | 1.59 | 3.38 | 0.0 | 3.1 | 2.9 | 1.80 | 0.10 | 89 |
| ATE | 1 | OneF. | 0.41 | - | - | 0.70 | -0.2 | 3.1 | 1.5 | 0.73 | 0.04 | 87 |
| GATE | 2 | MCE. | 0.41 | - | - | 0.71 | -0.2 | 3.1 | 1.5 | 0.74 | 0.05 | 87 |
| GATE | 32 | Pen | 0.40 | 0.17 | 0.13 | 0.75 | -0.2 | 3.1 | 1.9 | 0.76 | 0.05 | 87 |
| IATE | 5000 | LC-5 | 0.53 | 1.72 | 1.44 | 2.57 | -0.1 | 3.0 | 3.5 | 1.45 | 0.09 | 89 |

Note: For GATE and IATE the *average bias* is the absolute value of the bias for the specific group (GATE) / observation (IATE) averaged over all groups / observation (each group / observation receives the same weight). *CovP (90%)* denotes the (average) probability that the true value is part of the 90% confidence interval. The simulation errors of the mean MSEs are around 0.1.

## Appendix C.2: Alternative specifications of IATEs for main estimators

In this section, we show the results for alternative specifications of the individualized effects for all sample sizes.

Appendix C.2.1: No individual effects (ITE = 0)

*Table C.3: Simulation results for N=1'000, no effect, and main estimators*

| | Groups # | Est. | True & estimated effects | | | Estimation error of effects (averages) | | | | | Estimation of std. error | |
|---|---|---|---|---|---|---|---|---|---|---|---|---|
| | | | Avg. bias | X-sectional std. dev. | | MSE | Skewness | Kurtosis | JB-Stat. | Std. err. | Avg. bias | CovP (90) in % |
| | | | | true | est. | | | | | | | |
| | (1) | | (2) | (3) | (4) | (5) | (6) | (7) | (8) | (9) | (10) | (11) |
| ATE | 1 | Basic | 1.13 | - | - | 2.64 | 0.1 | 3.3 | 3-4 | 1.17 | 0.04 | 77 |
| GATE | 2 | | 1.13 | - | - | 2.71 | 0.1 | 3.2 | 3.7 | 1.19 | 0.05 | 78 |
| GATE | 32 | | 1.09 | 0 | 0.12 | 2.78 | 0.1 | 3.3 | 3.9 | 1.25 | 0.08 | 80 |
| IATE | 5000 | | 1.14 | 0 | 0.60 | 5.11 | 0.0 | 3.1 | 2.4 | 1.86 | 0.42 | 90 |
| ATE | 1 | OneF. | 1.23 | - | - | 2.80 | -0.1 | 3.0 | 0.8 | 1.12 | 0.11 | 76 |
| GATE | 2 | VarT | 1.24 | - | - | 2.84 | -0.1 | 3.0 | 0.9 | 1.14 | 0.13 | 76 |
| GATE | 32 | | 1.23 | 0 | 0.04 | 2.85 | -0.1 | 3.0 | 0.8 | 1.15 | 0.15 | 78 |
| IATE | 5000 | | 1.24 | 0 | 0.39 | 3.98 | 0.0 | 3.0 | 1.9 | 1.51 | 0.53 | 91 |
| ATE | 1 | OneF. | 1.11 | - | - | 2.53 | -0.1 | 2.9 | 2.0 | 1.14 | 0.06 | 77 |
| GATE | 2 | MCE | 1.11 | - | - | 2.58 | -0.1 | 2.9 | 2.1 | 1.15 | 0.08 | 78 |
| GATE | 32 | | 1.10 | 0 | 0.06 | 2.64 | -0.1 | 2.9 | 1.7 | 1.20 | 0.13 | 81 |
| IATE | 5000 | | 1.11 | 0 | 0.33 | 3.87 | 0.0 | 2.9 | 2.0 | 1.59 | 0.56 | 93 |
| ATE | 1 | OneF. | 0.81 | - | - | 2.31 | 0.1 | 3.0 | 1.0 | 1.29 | 0.08 | 85 |
| GATE | 2 | MCE. | 0.81 | - | - | 2.36 | 0.1 | 3.0 | 1.2 | 1.31 | 0.09 | 85 |
| GATE | 32 | LC-2 | 0.80 | 0 | 0.04 | 2.44 | 0.1 | 3.0 | 1.2 | 1.34 | 0.09 | 86 |
| IATE | 5000 | | 0.81 | 0 | 0.23 | 3.79 | 0.0 | 3.1 | 1.6 | 1.76 | 0.19 | 90 |
| ATE | 1 | OneF. | 0.63 | - | - | 2.26 | -0.1 | 3.1 | 0.9 | 1.37 | 0.16 | 89 |
| GATE | 2 | MCE. | 0.63 | - | - | 2.29 | -0.1 | 3.1 | 1.0 | 1.38 | 0.17 | 90 |
| GATE | 32 | Pen | 0.63 | 0 | 0.06 | 2.34 | -0.1 | 3.1 | 0.9 | 1.39 | 0.18 | 90 |
| IATE | 5000 | | 0.88 | 0 | 0.73 | 4.70 | -0.1 | 3.2 | 9.5 | 1.93 | 0.63 | 93 |
| ATE | 1 | OneF. | 0.50 | - | - | 2.59 | -0.1 | 3.1 | 1.8 | 1.53 | 0.05 | 89 |
| GATE | 2 | MCE. | 0.51 | - | - | 2.62 | -0.1 | 3.1 | 1.7 | 1.54 | 0.05 | 89 |
| GATE | 32 | Pen | 0.51 | 0 | 0.04 | 2.68 | -0.1 | 3.1 | 1.5 | 1.56 | 0.06 | 89 |
| IATE | 5000 | LC-2 | 0.64 | 0 | 0.50 | 5.43 | -0.1 | 3.3 | 6.3 | 2.34 | 0.14 | 90 |
| ATE | 1 | OneF. | 0.41 | - | - | 1.95 | -0.2 | 3.1 | 10.0 | 1.34 | 0.05 | 90 |
| GATE | 2 | MCE. | 0.42 | - | - | 1.98 | -0.2 | 3.1 | 9.8 | 1.34 | 0.05 | 90 |
| GATE | 32 | Pen | 0.42 | 0 | 0.03 | 2.01 | -0.2 | 3.1 | 9.4 | 1.35 | 0.06 | 90 |
| IATE | 5000 | LC-5 | 0.58 | 0 | 0.48 | 4.03 | -0.2 | 3.2 | 10.0 | 1.90 | 0.13 | 90 |

Note: For GATE and IATE the *average bias* is the absolute value of the bias for the specific group (GATE) / observation (IATE) averaged over all groups / observation (each group / observation receives the same weight). *CovP (90%)* denotes the (average) probability that the true value is part of the 90% confidence interval. The simulation errors of the mean MSEs are around 0.15.

*Table C.4: Simulation results for N=4'000, no effect, and main estimators*

| | | | True & estimated effects | | | Estimation error of effects (averages) | | | | | Estimation of std. error | |
|---|---|---|---|---|---|---|---|---|---|---|---|---|
| | Groups # | Est. | Avg. bias | X-sectional std. dev. true | est. | MSE | Skewness | Kurtosis | JB-Stat. | Std. err. | Avg. bias | CovP (90) in % |
| | (1) | | (2) | (3) | (4) | (5) | (6) | (7) | (8) | (9) | (10) | (11) |
| ATE | 1 | Basic | 0.84 | - | - | 1.09 | 0.0 | 2.8 | 0.2 | 0.63 | 0.00 | 60 |
| GATE | 2 | | 0.85 | - | - | 1.15 | 0.0 | 2.8 | 0.9 | 0.65 | 0.01 | 63 |
| GATE | 32 | | 0.78 | 0 | 0.17 | 1.18 | -0.1 | 2.9 | 1.1 | 0.73 | 0.06 | 73 |
| IATE | 5000 | | 1.00 | 0 | 0.86 | 3.84 | 0.0 | 3.0 | 2.4 | 1.54 | 0.24 | 87 |
| ATE | 1 | OneF. | 1.05 | - | - | 1.43 | 0.1 | 2.7 | 1.5 | 0.58 | 0.09 | 55 |
| GATE | 2 | VarT | 1.05 | - | - | 1.47 | 0.1 | 2.7 | 1.6 | 0.60 | 0.11 | 58 |
| GATE | 32 | | 1.03 | 0 | 0.06 | 1.44 | 0.1 | 2.8 | 1.4 | 0.61 | 0.12 | 61 |
| IATE | 5000 | | 1.06 | 0 | 0.57 | 2.60 | 0.0 | 2.9 | 1.7 | 1.08 | 0.53 | 89 |
| ATE | 1 | OneF. | 0.86 | - | - | 1.06 | -0.1 | 2.8 | 0.6 | 0.57 | 0.05 | 59 |
| GATE | 2 | MCE | 0.87 | - | - | 1.08 | -0.1 | 2.8 | 0.7 | 0.58 | 0.07 | 63 |
| GATE | 32 | | 0.84 | 0 | 0.08 | 1.13 | -0.1 | 2.9 | 1.1 | 0.64 | 0.14 | 75 |
| IATE | 5000 | | 0.87 | 0 | 0.43 | 2.18 | 0.0 | 3.0 | 2.4 | 1.12 | 0.52 | 94 |
| ATE | 1 | OneF. | 0.72 | - | - | 0.88 | 0.1 | 2.9 | 0.6 | 0.60 | 0.09 | 76 |
| GATE | 2 | MCE. | 0.72 | - | - | 0.91 | 0.1 | 2.9 | 0.7 | 0.62 | 0.09 | 76 |
| GATE | 32 | LC-2 | 0.71 | 0 | 0.05 | 0.95 | 0.1 | 3.0 | 1.5 | 0.67 | 0.10 | 79 |
| IATE | 5000 | | 0.72 | 0 | 0.29 | 1.95 | 0.0 | 3.0 | 2.2 | 1.16 | 0.18 | 89 |
| ATE | 1 | OneF. | 0.25 | - | - | 0.66 | -0.1 | 2.8 | 1.0 | 0.77 | 0.20 | 94 |
| GATE | 2 | MCE. | 0.25 | - | - | 0.67 | -0.1 | 2.8 | 0.9 | 0.78 | 0.20 | 95 |
| GATE | 32 | Pen | 0.26 | 0 | 0.05 | 0.70 | -0.1 | 2.9 | 1.6 | 0.80 | 0.22 | 95 |
| IATE | 5000 | | 0.62 | 0 | 0.59 | 2.42 | 0.0 | 2.9 | 1.6 | 1.41 | 0.66 | 96 |
| ATE | 1 | OneF. | 0.38 | - | - | 0.84 | 0.1 | 2.6 | 1.4 | 0.84 | 0.06 | 89 |
| GATE | 2 | MCE. | 0.39 | - | - | 0.86 | 0.1 | 2.7 | 1.4 | 0.85 | 0.06 | 89 |
| GATE | 32 | Pen | 0.38 | 0 | 0.04 | 0.89 | 0.0 | 2.7 | 1.4 | 0.86 | 0.07 | 89 |
| IATE | 5000 | LC-2 | 0.56 | 0 | 0.47 | 2.87 | 0.0 | 3.0 | 2.8 | 1.58 | 0.21 | 91 |
| ATE | 1 | OneF. | 0.17 | - | - | 0.55 | -0.4 | 3.5 | 10.2 | 0.73 | 0.06 | 91 |
| GATE | 2 | MCE. | 0.17 | - | - | 0.56 | -0.4 | 3.4 | 8.9 | 0.73 | 0.06 | 91 |
| GATE | 32 | Pen | 0.18 | 0 | 0.03 | 0.60 | -0.4 | 3.5 | 9.8 | 0.76 | 0.06 | 92 |
| IATE | 5000 | LC-5 | 0.36 | 0 | 0.34 | 2.22 | -0.1 | 3.1 | 2.3 | 1.43 | 0.12 | 91 |

Note: For GATE and IATE the *average bias* is the absolute value of the bias for the specific group (GATE) / observation (IATE) averaged over all groups / observation (each group / observation receives the same weight). *CovP (90%)* denotes the (average) probability that the true value is part of the 90% confidence interval. The simulation errors of the mean MSEs are around 0.1.

*Table C.5: Simulation results for N=8'000, no effect, and main estimators*

| | Groups # | Est. | True & estimated effects | | | Estimation error of effects (averages) | | | | | Estimation of std. error | |
|---|---|---|---|---|---|---|---|---|---|---|---|---|
| | | | Avg. bias | X-sectional std. dev. | | MSE | Skewness | Kurtosis | JB-Stat. | Std. err. | Avg. bias | CovP (90) in % |
| | | | | true | est. | | | | | | | |
| | (1) | | (2) | (3) | (4) | (5) | (6) | (7) | (8) | (9) | (10) | (11) |
| ATE | 1 | Basic | 0.77 | - | - | 0.77 | 0.1 | 2.9 | 0.4 | 0.40 | 0.05 | 46 |
| GATE | 2 | | 0.79 | - | - | 0.81 | 0.2 | 2.8 | 1.1 | 0.42 | 0.06 | 52 |
| GATE | 32 | | 0.70 | 0 | 0.25 | 0.85 | 0.0 | 2.7 | 1.6 | 0.54 | 0.10 | 69 |
| IATE | 5000 | | 0.99 | 0 | 0.91 | 3.39 | 0.0 | 2.9 | 1.7 | 1.39 | 0.25 | 86 |
| ATE | 1 | OneF. | 1.06 | - | - | 1.28 | -0.1 | 2.8 | 0.3 | 0.41 | 0.09 | 28 |
| GATE | 2 | VarT | 1.06 | - | - | 1.30 | -0.1 | 2.9 | 0.4 | 0.43 | 0.54 | 34 |
| GATE | 32 | | 1.04 | 0 | 0.07 | 1.29 | -0.1 | 2.8 | 0.6 | 0.45 | 0.58 | 40 |
| IATE | 5000 | | 1.09 | 0 | 0.64 | 2.48 | 0.0 | 3.0 | 1.9 | 0.97 | 1.47 | 86 |
| ATE | 1 | OneF. | 0.77 | - | - | 0.79 | -0.3 | 2.9 | 1.7 | 0.43 | 0.02 | 44 |
| GATE | 2 | MCE | 0.78 | - | - | 0.81 | -0.2 | 2.8 | 2.8 | 0.45 | 0.03 | 50 |
| GATE | 32 | | 0.75 | 0 | 0.09 | 0.84 | -0.2 | 2.8 | 2.8 | 0.52 | 0.11 | 66 |
| IATE | 5000 | | 0.79 | 0 | 0.42 | 1.74 | -0.1 | 3.0 | 3.0 | 0.98 | 0.48 | 94 |
| ATE | 1 | OneF. | 0.53 | - | - | 0.48 | -0.3 | 3.0 | 2.0 | 0.44 | 0.05 | 71 |
| GATE | 2 | MCE. | 0.54 | - | - | 0.50 | -0.3 | 2.9 | 1.9 | 0.45 | 0.05 | 74 |
| GATE | 32 | LC-2 | 0.52 | 0 | 0.06 | 0.54 | -0.2 | 2.9 | 1.7 | 0.51 | 0.06 | 76 |
| IATE | 5000 | | 0.55 | 0 | 0.32 | 1.40 | 0.0 | 2.9 | 1.0 | 1.00 | 0.13 | 89 |
| ATE | 1 | OneF. | 0.09 | - | - | 0.42 | -0.3 | 4.5 | 12.4 | 0.64 | 0.14 | 95 |
| GATE | 2 | MCE. | 0.09 | - | - | 0.43 | -0.3 | 4.3 | 11.0 | 0.65 | 0.14 | 95 |
| GATE | 32 | Pen | 0.08 | 0 | 0.03 | 0.47 | -0.2 | 4.1 | 10.4 | 0.68 | 0.20 | 95 |
| IATE | 5000 | | 0.34 | 0 | 0.36 | 2.40 | -0.1 | 3.0 | 2.3 | 1.49 | 0.77 | 98 |
| ATE | 1 | OneF. | 0.08 | - | - | 0.49 | -0.2 | 3.4 | 1.5 | 0.68 | -0.01 | 89 |
| GATE | 2 | MCE. | 0.09 | - | - | 0.49 | -0.2 | 3.4 | 1.3 | 0.69 | -0.01 | 89 |
| GATE | 32 | Pen | 0.09 | 0 | 0.02 | 0.52 | -0.2 | 3.3 | 1.6 | 0.72 | 0.00 | 90 |
| IATE | 5000 | LC-2 | 0.32 | 0 | 0.34 | 2.13 | 0.0 | 3.0 | 2.5 | 1.57 | 0.17 | 92 |
| ATE | 1 | OneF. | 0.08 | - | - | 0.34 | -0.4 | 3.9 | 7.5 | 0.58 | 0.02 | 90 |
| GATE | 2 | MCE. | 0.08 | - | - | 0.35 | -0.4 | 3.8 | 6.6 | 0.58 | 0.02 | 91 |
| GATE | 32 | Pen | 0.09 | 0 | 0.03 | 0.38 | -0.3 | 3.8 | 6.2 | 0.61 | 0.02 | 90 |
| IATE | 5000 | LC-5 | 0.28 | 0 | 0.29 | 1.72 | 0.0 | 3.1 | 1.9 | 1.27 | 0.10 | 91 |

Note: For GATE and IATE the *average bias* is the absolute value of the bias for the specific group (GATE) / observation (IATE) averaged over all groups / observation (each group / observation receives the same weight). *CovP (90%)* denotes the (average) probability that the true value is part of the 90% confidence interval. The simulation errors of the mean MSEs are around 0.1.

## Appendix C.2.2: Strong individual effects (α = 8)

*Table C.6: Simulation results for N=1'000, strong effect, and main estimators*

|  | Groups # | Est. | True & estimated effects | | | Estimation error of effects (averages) | | | | | Estimation of std. error | |
|---|---|---|---|---|---|---|---|---|---|---|---|---|
|  |  |  | Avg. bias | X-sectional std. dev. true | X-sectional std. dev. est. | MSE | Skewness | Kurtosis | JB-Stat. | Std. err. | Avg. bias | CovP (90) in % |
|  | (1) |  | (2) | (3) | (4) | (5) | (6) | (7) | (8) | (9) | (10) | (11) |
| ATE | 1 | Basic | 3.22 | - | - | 12.01 | 0.1 | 2.7 | 4.0 | 1.26 | 0.01 | 20 |
| GATE | 2 |  | 3.22 | - | - | 11.99 | 0.1 | 2.7 | 3.8 | 1.28 | 0.03 | 22 |
| GATE | 32 |  | 3.18 | 0.66 | 0.39 | 11.98 | 0.1 | 2.8 | 2.5 | 1.34 | 0.06 | 27 |
| IATE | 5000 |  | 3.33 | 6.87 | 3.91 | 28.63 | 0.0 | 2.9 | 2.9 | 2.24 | 0.35 | 65 |
| ATE | 1 | OneF. | 6.47 | - | - | 15.62 | -0.1 | 2.9 | 2.8 | 1.46 | -0.08 | 18 |
| GATE | 2 | VarT | 3.66 | - | - | 15.53 | -0.1 | 2.9 | 2.5 | 1.47 | -0.07 | 19 |
| GATE | 32 |  | 3.61 | 0.66 | 0.25 | 15.39 | -0.1 | 2.9 | 2.4 | 1.48 | -0.02 | 22 |
| IATE | 5000 |  | 4.52 | 6.87 | 2.81 | 36.76 | -0.1 | 2.9 | 6.7 | 2.13 | 0.23 | 51 |
| ATE | 1 | OneF. | 3.48 | - | - | 14.09 | -0.2 | 3.1 | 1.4 | 1.40 | -0.01 | 20 |
| GATE | 2 | MCE | 3.46 | - | - | 13.99 | -0.2 | 3.2 | 1.4 | 1.41 | 0.01 | 21 |
| GATE | 32 |  | 3.42 | 0.66 | 0.25 | 13.94 | -0.2 | 3.1 | 1.4 | 1.42 | 0.05 | 25 |
| IATE | 5000 |  | 4.26 | 6.87 | 2.56 | 37.01 | -0.1 | 3.0 | 2.0 | 2.04 | 0.35 | 57 |
| ATE | 1 | OneF. | 2.67 | - | - | 9.09 | 0.0 | 3.1 | 1.0 | 1.39 | 0.06 | 42 |
| GATE | 2 | MCE. | 2.65 | - | - | 9.03 | 0.0 | 3.1 | 1.3 | 1.41 | 0.07 | 44 |
| GATE | 32 | LC-2 | 2.59 | 0.66 | 0.13 | 9.08 | 0.0 | 3.2 | 1.7 | 1.44 | 0.08 | 47 |
| IATE | 5000 |  | 4.95 | 6.87 | 1.41 | 42.16 | 0.0 | 3.2 | 7.0 | 1.94 | 0.12 | 42 |
| ATE | 1 | OneF. | 2.29 | - | - | 8.01 | 0.0 | 2.8 | 1.8 | 1.66 | -0.03 | 60 |
| GATE | 2 | MCE. | 2.28 | - | - | 7.97 | 0.0 | 2.8 | 1.7 | 1.66 | -0.02 | 60 |
| GATE | 32 | Pen | 2.24 | 0.66 | 0.41 | 7.91 | 0.0 | 2.8 | 2.0 | 1.67 | 0.00 | 62 |
| IATE | 5000 |  | 2.96 | 6.87 | 4.42 | 19.76 | 0.1 | 2.9 | 9.6 | 2.47 | 0.30 | 72 |
| ATE | 1 | OneF. | 1.87 | - | - | 6.26 | -0.2 | 3.5 | 16 | 1.67 | -0.04 | 66 |
| GATE | 2 | MCE. | 1.85 | - | - | 6.23 | -0.2 | 3.5 | 15 | 1.67 | -0.04 | 67 |
| GATE | 32 | Pen | 1.80 | 0.66 | 0.29 | 6.25 | -0.2 | 3.5 | 14 | 1.69 | -0.03 | 68 |
| IATE | 5000 | LC-2 | 3.62 | 6.87 | 3.18 | 25.67 | -0.2 | 3.3 | 37 | 2.52 | -0.14 | 56 |
| ATE | 1 | OneF. | 1.78 | - | - | 5.08 | -0.2 | 3.3 | 11 | 1.38 | 0.00 | 62 |
| GATE | 2 | MCE. | 1.76 | - | - | 5.02 | -0.2 | 3.3 | 11 | 1.39 | 0.00 | 62 |
| GATE | 32 | Pen | 1.71 | 0.66 | 0.27 | 5.04 | -0.2 | 3.3 | 10 | 1.40 | 0.01 | 64 |
| IATE | 5000 | LC-5 | 3.68 | 6.87 | 3.03 | 24.58 | -0.2 | 3.2 | 37 | 2.00 | -0.15 | 49 |

Note: For GATE and IATE the *average bias* is the absolute value of the bias for the specific group (GATE) / observation (IATE) averaged over all groups / observation (each group / observation receives the same weight). *CovP (90%)* denotes the (average) probability that the true value is part of the 90% confidence interval. The simulation errors of the mean MSEs are around 0.15 for ATE/GATE and 0.3 for IATE.

*Table C.7: Simulation results for N=4'000, strong effect, and main estimators*

| | Groups # | Est. | True & estimated effects | | | Estimation error of effects (averages) | | | | | Estimation of std. error | |
|---|---|---|---|---|---|---|---|---|---|---|---|---|
| | | | Avg. bias | X-sectional std. dev. | | MSE | Skew ness | Kurt- osis | JB- Stat. | Std. err. | Avg. bias | CovP (90) in % |
| | | | | true | est. | | | | | | | |
| | (1) | | (2) | (3) | (4) | (5) | (6) | (7) | (8) | (9) | (10) | (11) |
| ATE | 1 | Basic | 2.19 | - | - | 5.21 | 0.1 | 3.0 | 0.8 | 0.64 | 0.04 | 5 |
| GATE | 2 | | 2.20 | - | - | 5.28 | 0.1 | 3.0 | 1.3 | 0.66 | 0.05 | 5 |
| GATE | 32 | | 2.16 | 0.66 | 0.52 | 5.28 | 0.1 | 3.2 | 2.4 | 0.74 | 0.08 | 15 |
| IATE | 5000 | | 2.38 | 6.87 | 5.12 | 14.78 | 0.0 | 3.0 | 2.4 | 1.69 | 0.30 | 67 |
| ATE | 1 | OneF. | 2.69 | - | - | 8.29 | -0.1 | 2.7 | 1.4 | 1.01 | -0.23 | 10 |
| GATE | 2 | VarT | 2.68 | - | - | 8.23 | -0.1 | 2.7 | 1.3 | 1.01 | -0.21 | 10 |
| GATE | 32 | | 2.64 | 0.66 | 0.38 | 8.13 | -0.1 | 2.7 | 1.5 | 1.03 | -0.17 | 14 |
| IATE | 5000 | | 3.38 | 6.87 | 4.49 | 19.78 | -0.1 | 2.9 | 3.5 | 1.74 | 0.06 | 51 |
| ATE | 1 | OneF. | 1.91 | - | - | 4.24 | 0.5 | 3.2 | 8.9 | 0.77 | 0.08 | 28 |
| GATE | 2 | MCE | 1.90 | - | - | 4.23 | 0.4 | 3.3 | 10.1 | 0.78 | 0.09 | 29 |
| GATE | 32 | | 1.87 | 0.66 | 0.44 | 4.21 | 0.4 | 3.2 | 8.3 | 0.80 | 0.13 | 37 |
| IATE | 5000 | | 2.38 | 6.87 | 4.49 | 13.54 | 0.2 | 3.3 | 7.3 | 1.45 | 0.42 | 68 |
| ATE | 1 | OneF. | 1.92 | - | - | 4.44 | -0.2 | 3.2 | 2.8 | 0.85 | -0.07 | 25 |
| GATE | 2 | MCE. | 1.91 | - | - | 4.41 | -0.2 | 3.2 | 2.6 | 0.86 | -0.06 | 27 |
| GATE | 32 | LC-2 | 1.87 | 0.66 | 0.26 | 4.46 | -0.2 | 3.1 | 2.5 | 0.89 | -0.05 | 31 |
| IATE | 5000 | | 3.69 | 6.87 | 2.79 | 24.46 | 0.0 | 3.2 | 5.0 | 1.06 | -0.11 | 41 |
| ATE | 1 | OneF. | 0.94 | - | - | 1.56 | 0.3 | 3.7 | 10.0 | 0.82 | 0.13 | 80 |
| GATE | 2 | MCE. | 0.95 | - | - | 1.56 | 0.3 | 3.7 | 9.7 | 0.83 | 0.14 | 80 |
| GATE | 32 | Pen | 0.92 | 0.66 | 0.55 | 1.58 | 0.3 | 3.6 | 8.5 | 0.84 | 0.17 | 82 |
| IATE | 5000 | | 1.50 | 6.87 | 5.74 | 6.39 | 0.1 | 3.1 | 4.8 | 1.59 | 0.47 | 85 |
| ATE | 1 | OneF. | 1.21 | - | - | 2.60 | -0.4 | 3.1 | 7.8 | 1.05 | -0.17 | 53 |
| GATE | 2 | MCE. | 1.20 | - | - | 2.58 | -0.4 | 3.1 | 7.5 | 1.06 | -0.16 | 56 |
| GATE | 32 | Pen | 1.18 | 0.66 | 0.41 | 2.62 | -0.4 | 3.1 | 7.8 | 1.08 | -0.15 | 58 |
| IATE | 5000 | LC-2 | 2.58 | 6.87 | 4.36 | 14.03 | -0.2 | 3.0 | 7.6 | 2.03 | -0.25 | 57 |
| ATE | 1 | OneF. | 1.24 | - | - | 2.21 | -0.5 | 3.9 | 20.6 | 0.74 | -0.08 | 45 |
| GATE | 2 | MCE. | 1.23 | - | - | 2.20 | -0.5 | 3.9 | 18.3 | 0.75 | -0.07 | 46 |
| GATE | 32 | Pen | 1.19 | 0.66 | 0.37 | 2.22 | -0.5 | 3.9 | 19.9 | 0.77 | -0.07 | 50 |
| IATE | 5000 | LC-5 | 2.76 | 6.87 | 4.17 | 13.92 | -0.1 | 3.3 | 11.8 | 1.48 | -0.21 | 47 |

Note: For GATE and IATE the *average bias* is the absolute value of the bias for the specific group (GATE) / observation (IATE) averaged over all groups / observation (each group / observation receives the same weight). *CovP (90%)* denotes the (average) probability that the true value is part of the 90% confidence interval. The simulation errors of the mean MSEs are around 0.1 for ATE/GATE and 0.15 for IATE.

*Table C.8: Simulation results for N=8'000, strong effect, and main estimators*

| | Groups # | Est. | True & estimated effects | | | Estimation error of effects (averages) | | | | | Estimation of std. error | |
|---|---|---|---|---|---|---|---|---|---|---|---|---|
| | | | Avg. bias | X-sectional std. dev. | | MSE | Skewness | Kurtosis | JB-Stat. | Std. err. | Avg. bias | CovP (90) in % |
| | | | | true | est. | | | | | | | |
| | (1) | | (2) | (3) | (4) | (5) | (6) | (7) | (8) | (9) | (10) | (11) |
| ATE | 1 | Basic | 1.82 | - | - | 3.52 | -0.6 | 3.4 | 7.2 | 0.44 | 0.05 | 2 |
| GATE | 2 | | 1.84 | - | - | 3.58 | -0.5 | 3.6 | 10.0 | 0.46 | 0.06 | 3 |
| GATE | 32 | | 1.82 | 0.66 | 0.60 | 3.66 | -0.3 | 2.9 | 2.8 | 0.55 | 0.11 | 12 |
| IATE | 5000 | | 2.07 | 6.87 | 5.58 | 11.06 | 0.0 | 2.9 | 2.0 | 1.52 | 0.26 | 68 |
| ATE | 1 | OneF. | 0.91 | - | - | 1.80 | 0.2 | 2.1 | 5.0 | 0.99 | -0.26 | 64 |
| GATE | 2 | VarT | 0.91 | - | - | 1.79 | 0.2 | 2.1 | 4.8 | 0.99 | -0.25 | 66 |
| GATE | 32 | | 0.88 | 0.66 | 0.55 | 1.80 | 0.1 | 2.2 | 4.5 | 1.00 | -0.22 | 67 |
| IATE | 5000 | | 1.39 | 6.87 | 5.90 | 6.94 | 0.1 | 2.7 | 4.1 | 1.86 | 0.12 | 82 |
| ATE | 1 | OneF. | 1.39 | - | - | 2.20 | -0.1 | 2.6 | 1.2 | 0.52 | 0.12 | 28 |
| GATE | 2 | MCE | 1.39 | - | - | 2.20 | -0.2 | 2.7 | 1.7 | 0.53 | 0.13 | 30 |
| GATE | 32 | | 1.35 | 0.66 | 0.49 | 2.21 | -0.1 | 2.6 | 1.3 | 0.57 | 0.17 | 39 |
| IATE | 5000 | | 1.86 | 6.87 | 5.05 | 8.76 | 0.0 | 2.9 | 1.7 | 1.23 | 0.40 | 70 |
| ATE | 1 | OneF. | 1.67 | - | - | 3.19 | -0.1 | 3.1 | 0.3 | 0.63 | -0.04 | 14 |
| GATE | 2 | MCE. | 1.66 | - | - | 3.16 | -0.1 | 3.0 | 0.5 | 0.64 | -0.04 | 15 |
| GATE | 32 | LC-2 | 1.62 | 0.66 | 0.33 | 3.19 | -0.1 | 3.0 | 0.7 | 0.66 | -0.03 | 24 |
| IATE | 5000 | | 3.16 | 6.87 | 3.46 | 18.23 | -0.1 | 2.8 | 2.6 | 1.47 | -0.15 | 43 |
| ATE | 1 | OneF. | 0.52 | - | - | 0.57 | 0.4 | 3.0 | 3.3 | 0.55 | 0.17 | 89 |
| GATE | 2 | MCE. | 0.52 | - | - | 0.58 | 0.4 | 2.9 | 3.1 | 0.56 | 0.17 | 88 |
| GATE | 32 | Pen | 0.50 | 0.66 | 0.58 | 0.62 | 0.3 | 3.0 | 2.6 | 0.59 | 0.20 | 89 |
| IATE | 5000 | | 1.01 | 6.87 | 6.14 | 4.17 | 0.1 | 3.1 | 2.0 | 1.40 | 0.48 | 90 |
| ATE | 1 | OneF. | 0.88 | - | - | 1.28 | -0.5 | 3.1 | 3.4 | 0.71 | -0.05 | 59 |
| GATE | 2 | MCE. | 0.87 | - | - | 1.27 | -0.4 | 3.1 | 2.9 | 0.72 | -0.05 | 60 |
| GATE | 32 | Pen | 0.85 | 0.66 | 0.48 | 1.32 | -0.4 | 3.1 | 3.8 | 0.75 | -0.05 | 63 |
| IATE | 5000 | LC-2 | 1.96 | 6.87 | 5.03 | 8.99 | 0.1 | 2.9 | 3.1 | 1.73 | -0.16 | 64 |
| ATE | 1 | OneF. | 0.97 | - | - | 1.30 | -0.3 | 3.4 | 3.0 | 0.60 | -0.07 | 38 |
| GATE | 2 | MCE. | 0.95 | - | - | 1.29 | -0.3 | 3.5 | 4.3 | 0.61 | -0.06 | 44 |
| GATE | 32 | Pen | 0.92 | 0.66 | 0.40 | 1.31 | -0.3 | 3.4 | 3.2 | 0.62 | -0.06 | 50 |
| IATE | 5000 | LC-5 | 2.55 | 6.87 | 4.33 | 11.6 | -0.1 | 3.1 | 5.8 | 1.46 | -0.18 | 44 |

Note: For GATE and IATE the *average bias* is the absolute value of the bias for the specific group (GATE) / observation (IATE) averaged over all groups / observation (each group / observation receives the same weight). *CovP (90%)* denotes the (average) probability that the true value is part of the 90% confidence interval. The simulation errors of the mean MSEs are around 0.1 (GATEs) -0.4 (IATEs).

Appendix C.2.3: Earnings dependent individual effects

The following tables contain the results for the IATEs that are dependent on insured earnings. A major difference to the other (non-zero) IATEs is that the earnings related IATEs have a much lower correlation with the propensity score. Thus, it should be 'easier' for the estimators to disentangle effect heterogeneity from selection bias.

*Table C.9: Simulation results for N=1'000, earnings dependent effect, and main estimators*

| | Groups # | Est. | True & estimated effects | | | Estimation error of effects (averages) | | | | | Estimation of std. error | |
|---|---|---|---|---|---|---|---|---|---|---|---|---|
| | | | Avg. bias | X-sectional std. dev. | | MSE | Skewness | Kurtosis | JB-Stat. | Std. err. | Avg. bias | CovP (90) in % |
| | | | | true | est. | | | | | | | |
| | (1) | | (2) | (3) | (4) | (5) | (6) | (7) | (8) | (9) | (10) | (11) |
| ATE | 1 | Basic | 1.20 | - | - | 2.87 | 0.0 | 3.1 | 0.3 | 1.21 | 0.01 | 74 |
| GATE | 2 | | 1.23 | - | - | 3.10 | 0.0 | 3.1 | 0.3 | 1.25 | 0.02 | 74 |
| GATE | 32 | | 1.14 | 0.35 | 0.20 | 2.98 | 0.0 | 3.1 | 0.6 | 1.28 | 0.72 | 79 |
| IATE | 5000 | | 1.28 | 1.71 | 1.39 | 6.38 | 0.0 | 3.1 | 3.6 | 2.00 | 0.34 | 87 |
| ATE | 1 | OneF. | 1.18 | - | - | 2.83 | -0.1 | 2.8 | 3.8 | 1.20 | 0.06 | 75 |
| GATE | 2 | VarT | 1.25 | - | - | 3.24 | -0.1 | 2.8 | 3.7 | 1.21 | 0.08 | 73 |
| GATE | 32 | | 1.11 | 0.35 | 0.08 | 2.81 | -0.1 | 2.8 | 3.5 | 1.22 | 0.10 | 78 |
| IATE | 5000 | | 1.53 | 1.71 | 0.67 | 5.88 | -0.1 | 3.0 | 2.4 | 1.58 | 0.50 | 83 |
| ATE | 1 | OneF. | 1.18 | - | - | 2.77 | 0.1 | 2.9 | 0.9 | 1.17 | 0.04 | 76 |
| GATE | 2 | MCE | 1.24 | - | - | 3.09 | 0.0 | 2.9 | 1.1 | 1.19 | 0.07 | 74 |
| GATE | 32 | | 1.13 | 0.35 | 0.13 | 2.83 | 0.1 | 3.0 | 1.5 | 1.22 | 0.11 | 80 |
| IATE | 5000 | | 1.41 | 1.71 | 0.80 | 5.48 | 0.0 | 3.1 | 2.8 | 1.65 | 0.54 | 87 |
| ATE | 1 | OneF. | 0.80 | - | - | 2.33 | -0.1 | 2.9 | 2.2 | 1.30 | 0.07 | 86 |
| GATE | 2 | MCE. | 0.86 | - | - | 2.69 | -0.1 | 2.7 | 2.2 | 1.32 | 0.08 | 84 |
| GATE | 32 | LC-2 | 0.74 | 0.35 | 0.10 | 2.46 | -0.1 | 2.9 | 1.8 | 1.36 | 0.08 | 86 |
| IATE | 5000 | | 1.24 | 1.71 | 0.58 | 5.54 | 0.0 | 3.0 | 2.0 | 1.77 | 0.18 | 82 |
| ATE | 1 | OneF. | 0.56 | - | - | 2.09 | -0.1 | 3.0 | 1.9 | 1.33 | 0.23 | 92 |
| GATE | 2 | MCE. | 0.64 | - | - | 2.44 | -0.1 | 3.0 | 1.8 | 1.34 | 0.24 | 89 |
| GATE | 32 | Pen | 0.50 | 0.35 | 0.11 | 2.15 | -0.1 | 2.9 | 1.8 | 1.35 | 0.25 | 92 |
| IATE | 5000 | | 1.37 | 1.71 | 1.05 | 6.70 | -0.1 | 3.0 | 4.1 | 1.94 | 0.67 | 89 |
| ATE | 1 | OneF. | 0.42 | - | - | 2.58 | 0.0 | 2.9 | 0.3 | 1.55 | 0.04 | 90 |
| GATE | 2 | MCE. | 0.51 | - | - | 2.93 | 0.0 | 3.0 | 0.3 | 1.56 | 0.05 | 87 |
| GATE | 32 | Pen | 0.37 | 0.35 | 0.09 | 2.70 | 0.0 | 3.0 | 0.5 | 1.58 | 0.04 | 90 |
| IATE | 5000 | LC-2 | 1.26 | 1.71 | 0.76 | 7.40 | 0.0 | 3.1 | 3.2 | 2.20 | 0.15 | 84 |
| ATE | 1 | OneF. | 0.36 | - | - | 1.93 | 0.1 | 3.1 | 2.7 | 1.34 | 0.05 | 90 |
| GATE | 2 | MCE. | 0.51 | - | - | 2.27 | 0.1 | 3.1 | 2.8 | 1.35 | 0.05 | 87 |
| GATE | 32 | Pen | 0.32 | 0.35 | 0.08 | 2.03 | 0.1 | 3.1 | 2.4 | 1.37 | 0.05 | 90 |
| IATE | 5000 | LC-5 | 1.22 | 1.71 | 0.69 | 6.15 | 0.0 | 3.2 | 6.0 | 1.91 | 0.12 | 82 |

Note: For GATE and IATE the *average bias* is the absolute value of the bias for the specific group (GATE) / observation (IATE) averaged over all groups / observation (each group / observation receives the same weight). *CovP (90%)* denotes the (average) probability that the true value is part of the 90% confidence interval. The simulation errors of the mean MSEs are around 0.1

*Table C.10: Simulation results for N=4'000, earnings dependent effect, and main estimators*

|  | | | True & estimated effects | | | Estimation error of effects (averages) | | | | | Estimation of std. error | |
|---|---|---|---|---|---|---|---|---|---|---|---|---|
|  | Groups # | Est. | Avg. bias | X-sectional std. dev. | | MSE | Skew ness | Kurt- osis | JB- Stat. | Std. err. | Avg. bias | CovP (90) in % |
|  | | | | true | est. | | | | | | | |
|  | (1) | | (2) | (3) | (4) | (5) | (6) | (7) | (8) | (9) | (10) | (11) |
| ATE | 1 | Basic | 0.94 | - | - | 1.20 | -0.3 | 3.2 | 4.8 | 0.56 | 0.07 | 54 |
| GATE | 2 | | 0.96 | - | - | 1.27 | -0.3 | 3.1 | 3.8 | 0.59 | 0.08 | 58 |
| GATE | 32 | | 0.91 | 0.35 | 0.28 | 1.30 | -0.2 | 3.1 | 4.0 | 0.67 | 0.12 | 69 |
| IATE | 5000 | | 1.09 | 1.71 | 1.82 | 4.36 | -0.1 | 3.0 | 2.2 | 1.61 | 0.23 | 85 |
| ATE | 1 | OneF. | 1.06 | - | - | 1.55 | 0.3 | 3.0 | 5.2 | 0.61 | 0.08 | 56 |
| GATE | 2 | VarT | 1.14 | - | - | 1.86 | 0.3 | 3.0 | 3.8 | 0.63 | 0.09 | 54 |
| GATE | 32 | | 1.02 | 0.35 | 0.13 | 1.53 | 0.3 | 3.0 | 3.5 | 0.65 | 0.11 | 64 |
| IATE | 5000 | | 1.40 | 1.71 | 0.93 | 4.24 | 0.0 | 3.3 | 11.2 | 1.15 | 0.48 | 79 |
| ATE | 1 | OneF. | 0.87 | - | - | 1.03 | -0.2 | 2.9 | 1.2 | 0.53 | 0.10 | 62 |
| GATE | 2 | MCE | 0.90 | - | - | 1.17 | -0.2 | 2.9 | 1.6 | 0.55 | 0.12 | 65 |
| GATE | 32 | | 0.83 | 0.35 | 0.20 | 1.08 | -0.1 | 2.8 | 1.5 | 0.60 | 0.17 | 74 |
| IATE | 5000 | | 1.03 | 1.71 | 1.22 | 2.78 | 0.0 | 3.0 | 2.0 | 1.13 | 0.54 | 90 |
| ATE | 1 | OneF. | 0.55 | - | - | 0.74 | 0.0 | 3.2 | 0.3 | 0.66 | 0.03 | 82 |
| GATE | 2 | MCE. | 0.60 | - | - | 0.96 | 0.0 | 3.2 | 0.4 | 0.68 | 0.03 | 76 |
| GATE | 32 | LC-2 | 0.50 | 0.35 | 0.13 | 0.81 | 0.0 | 3.2 | 1.9 | 0.72 | 0.04 | 84 |
| IATE | 5000 | | 0.97 | 1.71 | 0.78 | 3.06 | 0.1 | 3.1 | 3.7 | 1.22 | 0.13 | 80 |
| ATE | 1 | OneF. | 0.21 | - | - | 0.75 | -0.3 | 2.9 | 3.3 | 0.84 | 0.13 | 93 |
| GATE | 2 | MCE. | 0.41 | - | - | 0.96 | -0.3 | 2.9 | 3.1 | 0.85 | 0.14 | 89 |
| GATE | 32 | Pen | 0.21 | 0.35 | 0.15 | 0.83 | -0.3 | 2.9 | 3.3 | 0.87 | 0.16 | 93 |
| IATE | 5000 | | 1.10 | 1.71 | 1.05 | 4.33 | -0.1 | 3.0 | 2.2 | 1.54 | 0.56 | 89 |
| ATE | 1 | OneF. | 0.16 | - | - | 0.73 | 0.0 | 3.0 | 0.0 | 0.85 | 0.05 | 90 |
| GATE | 2 | MCE. | 0.45 | - | - | 0.98 | 0.0 | 3.0 | 0.0 | 0.86 | 0.05 | 87 |
| GATE | 32 | Pen | 0.19 | 0.35 | 0.12 | 0.83 | 0.0 | 3.0 | 0.6 | 0.87 | 0.06 | 90 |
| IATE | 5000 | LC-2 | 1.07 | 1.71 | 0.82 | 4.58 | 0.0 | 3.0 | 2.3 | 1.62 | 0.16 | 82 |
| ATE | 1 | OneF. | 0.12 | - | - | 0.54 | -0.2 | 3.0 | 1.1 | 0.73 | 0.05 | 92 |
| GATE | 2 | MCE. | 0.45 | - | - | 0.78 | -0.2 | 3.0 | 1.1 | 0.73 | 0.05 | 86 |
| GATE | 32 | Pen | 0.19 | 0.35 | 0.13 | 0.63 | -0.2 | 3.0 | 1.6 | 0.75 | 0.06 | 91 |
| IATE | 5000 | LC-5 | 1.06 | 1.71 | 0.74 | 4.04 | 0.0 | 3.1 | 2.5 | 1.43 | 0.10 | 79 |

Note: For GATE and IATE the *average bias* is the absolute value of the bias for the specific group (GATE) / observation (IATE) averaged over all groups / observation (each group / observation receives the same weight). *CovP (90%)* denotes the (average) probability that the true value is part of the 90% confidence interval. The simulation errors of the mean MSEs are around 0.07.

*Table C.11: Simulation results for N=8'000, earnings dependent effect, and main estimators*

| | | | True & estimated effects | | | Estimation error of effects (averages) | | | | | Estimation of std. error | |
|---|---|---|---|---|---|---|---|---|---|---|---|---|
| | Groups # | Est. | Avg. bias | X-sectional std. dev. | | MSE | Skewness | Kurtosis | JB-Stat. | Std. err. | Avg. bias | CovP (90) in % |
| | | | | true | est. | | | | | | | |
| | (1) | | (2) | (3) | (4) | (5) | (6) | (7) | (8) | (9) | (10) | (11) |
| ATE | 1 | Basic | 0.77 | - | - | 0.74 | -0.4 | 2.9 | 2.8 | 0.38 | 0.08 | 46 |
| GATE | 2 | | 0.78 | - | - | 0.77 | -0.4 | 2.9 | 2.7 | 0.40 | 0.09 | 50 |
| GATE | 32 | | 0.75 | 0.35 | 0.30 | 0.86 | 0.0 | 2.9 | 1.4 | 0.53 | 0.10 | 68 |
| IATE | 5000 | | 0.99 | 1.71 | 1.97 | 3.63 | 0.0 | 3.0 | 2.3 | 1.46 | 0.21 | 85 |
| ATE | 1 | OneF. | 0.69 | - | - | 0.73 | 0.4 | 3.1 | 2.9 | 0.51 | 0.12 | 78 |
| GATE | 2 | VarT | 0.75 | - | - | 1.00 | 0.3 | 3.1 | 2.6 | 0.52 | 0.13 | 66 |
| GATE | 32 | | 0.63 | 0.35 | 0.14 | 0.74 | 0.3 | 3.1 | 2.7 | 0.53 | 0.14 | 80 |
| IATE | 5000 | | 1.32 | 1.71 | 1.02 | 4.01 | 0.0 | 2.9 | 1.8 | 1.16 | 0.63 | 83 |
| ATE | 1 | OneF. | 0.73 | - | - | 0.68 | -0.2 | 3.0 | 1.2 | 0.39 | 0.06 | 51 |
| GATE | 2 | MCE | 0.76 | - | - | 0.77 | -0.2 | 3.0 | 1.0 | 0.41 | 0.08 | 54 |
| GATE | 32 | | 0.68 | 0.35 | 0.19 | 0.71 | -0.2 | 3.2 | 2.0 | 0.46 | 0.16 | 72 |
| IATE | 5000 | | 0.87 | 1.71 | 1.36 | 2.12 | 0.0 | 3.0 | 1.9 | 1.00 | 0.48 | 90 |
| ATE | 1 | OneF. | 0.51 | - | - | 0.50 | -0.1 | 2.8 | 0.4 | 0.50 | 0.00 | 73 |
| GATE | 2 | MCE. | 0.55 | - | - | 0.66 | -0.1 | 2.8 | 0.5 | 0.51 | 0.01 | 68 |
| GATE | 32 | LC-2 | 0.47 | 0.35 | 0.19 | 0.56 | -0.1 | 2.9 | 1.6 | 0.56 | 0.01 | 77 |
| IATE | 5000 | | 0.82 | 1.71 | 0.97 | 2.30 | 0.0 | 2.9 | 1.7 | 1.07 | 0.11 | 80 |
| ATE | 1 | OneF. | 0.06 | - | - | 0.28 | -0.3 | 3.7 | 4.5 | 0.53 | 0.24 | 98 |
| GATE | 2 | MCE. | 0.31 | - | - | 0.40 | -0.3 | 3.7 | 5.1 | 0.54 | 0.24 | 96 |
| GATE | 32 | Pen | 0.15 | 0.35 | 0.18 | 0.35 | -0.2 | 3.5 | 3.4 | 0.56 | 0.28 | 97 |
| IATE | 5000 | | 0.84 | 1.71 | 1.01 | 2.88 | 0.0 | 3.0 | 1.9 | 1.27 | 0.68 | 93 |
| ATE | 1 | OneF. | 0.00 | - | - | 0.44 | 0.0 | 2.7 | 0.6 | 0.67 | 0.01 | 90 |
| GATE | 2 | MCE. | 0.40 | - | - | 0.62 | 0.0 | 2.7 | 0.6 | 0.69 | 0.01 | 82 |
| GATE | 32 | Pen | 0.17 | 0.35 | 0.16 | 0.53 | 0.0 | 2.7 | 0.9 | 0.71 | 0.02 | 89 |
| IATE | 5000 | LC-2 | 0.94 | 1.71 | 0.72 | 3.75 | -0.1 | 2.9 | 1.7 | 1.57 | 0.10 | 82 |
| ATE | 1 | OneF. | 0.04 | - | - | 0.31 | 0.0 | 2.5 | 0.8 | 0.55 | 0.02 | 90 |
| GATE | 2 | MCE. | 0.40 | - | - | 0.48 | 0.0 | 2.6 | 0.6 | 0.56 | 0.02 | 80 |
| GATE | 32 | Pen | 0.17 | 0.35 | 0.14 | 0.39 | 0.0 | 2.7 | 0.8 | 0.58 | 0.02 | 89 |
| IATE | 5000 | LC-5 | 0.98 | 1.71 | 0.79 | 3.21 | 0.0 | 3.0 | 2.3 | 1.23 | 0.10 | 78 |

Note: For GATE and IATE the *average bias* is the absolute value of the bias for the specific group (GATE) / observation (IATE) averaged over all groups / observation (each group / observation receives the same weight). *CovP (90%)* denotes the (average) probability that the true value is part of the 90% confidence interval. The simulation errors of the mean MSEs are around 0.08.

## Appendix C.3 (Online): Further estimators

In this section, we present additional results for three further estimators using the DGPs with selectivity and the four specification of the IATEs. The first estimator is the *Basic* estimator in a configuration that might be considered standard: It uses the full sample for training (instead of the splitted half), but when building each tree, the respective subsample is splitted and 50% is used for building the tree, and the other 50% is used for computing the effects. This procedure has been termed 'honesty' in the literature (e.g. Wager and Athey, 2018).

*Table C.12: Simulation results for additional estimators: Main DGP with selectivity*

| | | | True & estimated effects | | | Estimation error of effects (averages) | | | | | Estimation of std. error | |
|---|---|---|---|---|---|---|---|---|---|---|---|---|
| | Groups # | Est. | Avg. bias | X-sectional std. dev. true | est. | MSE | Skewness | Kurtosis | JB-Stat. | Std. err. | Avg. bias | CovP (90) in % |
| | (1) | | (2) | (3) | (4) | (5) | (6) | (7) | (8) | (9) | (10) | (11) |
| | | | | | | N = 1'000 | | | | | | |
| ATE | 1 | Basic. | 1.78 | - | - | 3.87 | 0.0 | 2.8 | 1.2 | 0.84 | 0.02 | 33 |
| GATE | 2 | One | 1.78 | - | - | 3.92 | 0.0 | 2.9 | 0.5 | 0.87 | 0.02 | 37 |
| GATE | 32 | Sam | 1.74 | 0.17 | 0.11 | 3.93 | 0.0 | 3.0 | 1.3 | 0.97 | 0.01 | 44 |
| IATE | 5000 | | 1.80 | 1.72 | 1.08 | 8.00 | 0.0 | 3.1 | 3.2 | 3.27 | -0.13 | 67 |
| ATE | 1 | OneF | 1.91 | - | - | 5.03 | 0.1 | 2.7 | 3.6 | 1.18 | 0.04 | 50 |
| GATE | 2 | | 1.90 | - | - | 5.04 | 0.0 | 2.7 | 3.2 | 1.20 | 0.06 | 52 |
| GATE | 32 | | 1.88 | 0.17 | 0.07 | 5.12 | 0.0 | 2.8 | 3.9 | 1.24 | 0.13 | 57 |
| IATE | 5000 | | 1.94 | 1.72 | 0.59 | 8.64 | 0.0 | 3.0 | 1.8 | 1.71 | 0.63 | 76 |
| ATE | 1 | OneF. | 1.50 | - | - | 4.06 | 0.0 | 3.1 | 0.20 | 1.34 | 0.09 | 71 |
| GATE | 2 | VarT. | 1.50 | - | - | 4.09 | 0.0 | 3.1 | 0.21 | 1.35 | 0.10 | 72 |
| GATE | 32 | Pen | 1.49 | 0.17 | 0.16 | 4.06 | 0.0 | 3.1 | 0.53 | 1.36 | 0.12 | 73 |
| IATE | 5000 | | 1.51 | 1.72 | 1.26 | 6.40 | -0.1 | 3.5 | 42.5* | 1.89 | 0.53 | 89 |
| | | | | | | N = 4'000 | | | | | | |
| ATE | 1 | Basic. | 1.45 | - | - | 2.30 | 0.2 | 2.8 | 1.3 | 0.43 | 0.03 | 15 |
| GATE | 2 | One | 1.46 | - | - | 2.34 | 0.1 | 2.9 | 1.1 | 0.45 | 0.03 | 19 |
| GATE | 32 | Sam | 1.42 | 0.17 | 0.13 | 2.41 | 0.1 | 2.9 | 1.7 | 0.60 | -0.02 | 35 |
| IATE | 5000 | | 1.56 | 1.72 | 1.38 | 6.06 | 0.0 | 3.0 | 2.3 | 1.58 | -0.23 | 76 |
| ATE | 1 | OneF | 1.60 | - | - | 2.90 | -0.2 | 3.2 | 2.5 | 0.58 | 0.07 | 15 |
| GATE | 2 | | 1.60 | - | - | 2.90 | -0.2 | 3.2 | 2.1 | 0.60 | 0.08 | 19 |
| GATE | 32 | | 1.58 | 0.17 | 0.08 | 2.97 | -0.1 | 3.2 | 1.7 | 0.67 | 0.16 | 35 |
| IATE | 5000 | | 1.66 | 1.72 | 0.74 | 5.91 | 0.0 | 3.0 | 2.2 | 1.25 | 0.58 | 76 |
| ATE | 1 | OneF. | 1.21 | - | - | 1.93 | 0.2 | 3.6 | 5.6 | 0.70 | 0.12 | 58 |
| GATE | 2 | VarT. | 1.20 | - | - | 1.94 | 0.2 | 3.6 | 5.0 | 0.71 | 0.13 | 61 |
| GATE | 32 | Pen | 1.19 | 0.17 | 0.13 | 1.62 | 0.2 | 3.6 | 5.6 | 0.71 | 0.14 | 63 |
| IATE | 5000 | | 1.24 | 1.72 | 1.61 | 3.78 | 0.0 | 2.9 | 1.7 | 1.34 | 0.62 | 88 |
| | | | | | | N = 8'000 | | | | | | |
| ATE | 1 | Basic. | 1.34 | - | - | 1.92 | 0.2 | 2.7 | 1.5 | 0.37 | -0.04 | 0 |
| GATE | 2 | One | 1.35 | - | - | 1.96 | 0.2 | 2.9 | 0.9 | 0.39 | -0.04 | 3 |
| GATE | 32 | Sam | 1.30 | 0.17 | 0.14 | 1.97 | 0.1 | 3.0 | 1.5 | 0.49 | -0.04 | 14 |
| IATE | 5000 | | 1.47 | 1.72 | 1.42 | 4.84 | 0.0 | 2.9 | 1.8 | 1.31 | -0.18 | 59 |
| ATE | 1 | OneF | 0.92 | - | - | 1.08 | 0.1 | 3.4 | 0.8 | 0.48 | 0.02 | 43 |
| GATE | 2 | | 0.92 | - | - | 1.09 | 0.0 | 3.3 | 0.7 | 0.50 | 0.02 | 45 |
| GATE | 32 | | 0.90 | 0.17 | 0.06 | 1.09 | 0.0 | 2.0 | 2.0 | 0.57 | 0.05 | 52 |
| IATE | 5000 | | 1.18 | 1.72 | 0.62 | 3.56 | 0.0 | 1.1 | 1.9 | 1.19 | 0.13 | 70 |
| ATE | 1 | OneF. | 1.05 | - | - | 1.32 | 0.2 | 2.9 | 0.6 | 0.46 | 0.16 | 46 |
| GATE | 2 | VarT. | 1.04 | - | - | 1.32 | 0.2 | 2.9 | 0.6 | 0.47 | 0.16 | 52 |
| GATE | 32 | Pen | 1.04 | 0.17 | 0.14 | 1.32 | 0.1 | 2.9 | 0.7 | 0.48 | 0.17 | 55 |
| IATE | 5000 | | 1.10 | 1.72 | 1.77 | 3.07 | 0.0 | 2.9 | 1.7 | 1.18 | 0.61 | 89 |

Note: For GATE and IATE the *average bias* is the absolute value of the bias for the specific group (GATE) / observation (IATE) averaged over all groups / observation (each group / observation receives the same weight). *CovP (90%)* denotes the (average) probability that the true value is part of the 90% confidence interval. *OneF.VarT.Penalty:* Baseline penalty multiplied by 100.

The second estimator is *OneF*. It is identical to *OneF.MCE* when setting MCE to zero. Therefore, it has some small computational advantages compared to *OneF.MCE* and *OneF.MCE.Penalty*. The third estimator is *OneF.VarT* with and added penalty function. To economize on computation costs, for the latter two estimators we do report results for *N=8'000* only.

*Table C.13: Simulation results for additional estimators: DGP with no effect and selectivity*

| | Groups # | Est. | True & estimated effects | | | Estimation error of effects (averages) | | | | | Estimation of std. error | |
|---|---|---|---|---|---|---|---|---|---|---|---|---|
| | | | Avg. bias | X-sectional std. dev. | | MSE | Skewness | Kurtosis | JB-Stat. | Std. err. | Avg. bias | CovP (90) in % |
| | | | | true | est. | | | | | | | |
| | (1) | | (2) | (3) | (4) | (5) | (6) | (7) | (8) | (9) | (10) | (11) |
| | | | | | | N = 1'000 | | | | | | |
| ATE | 1 | Basic. | 1.01 | - | - | 1.68 | 0.0 | 3.2 | 0.7 | 0.81 | 0.05 | 69 |
| GATE | 2 | One | 1.02 | - | - | 1.75 | 0.0 | 3.2 | 1.6 | 0.84 | 0.04 | 69 |
| GATE | 32 | Sam | 0.97 | 0 | 0.21 | 1.89 | 0.0 | 3.1 | 0.9 | 0.96 | 0.01 | 74 |
| IATE | 5000 | | 1.06 | 0 | 0.93 | 4.71 | 0.0 | 3.0 | 2.5 | 1.79 | -0.09 | 79 |
| ATE | 1 | OneF | 1.08 | - | - | 2.34 | -0.1 | 2.9 | 2.3 | 1.07 | 0.12 | 78 |
| GATE | 2 | | 1.09 | - | - | 2.38 | -0.1 | 2.9 | 2.0 | 1.09 | 0.14 | 79 |
| GATE | 32 | | 1.08 | 0 | 0.07 | 2.49 | -0.1 | 2.9 | 1.6 | 1.15 | 0.21 | 83 |
| IATE | 5000 | | 1.09 | 0 | 0.43 | 4.04 | -0.1 | 3.0 | 2.8 | 1.63 | 0.68 | 94 |
| ATE | 1 | OneF. | 0.96 | - | - | 2.46 | -0.1 | 3.1 | 2.6 | 1.24 | 0.12 | 84 |
| GATE | 2 | VarT. | 0.96 | - | - | 2.48 | -0.1 | 3.1 | 2.8 | 1.25 | 0.13 | 84 |
| GATE | 32 | Pen | 0.96 | 0 | 0.04 | 2.51 | -0.1 | 3.1 | 2.9 | 1.26 | 0.14 | 85 |
| IATE | 5000 | | 0.97 | 0 | 0.56 | 4.07 | -0.1 | 3.0 | 11.3 | 1.68 | 0.57 | 93 |
| | | | | | | N = 4'000 | | | | | | |
| ATE | 1 | Basic. | 0.85 | - | - | 0.89 | 0.3 | 3.0 | 4.0 | 0.41 | 0.04 | 39 |
| GATE | 2 | One | 0.86 | - | - | 0.94 | 0.3 | 3.1 | 3.6 | 0.44 | 0.03 | 43 |
| GATE | 32 | Sam | 0.79 | 0 | 0.21 | 1.04 | 0.1 | 3.1 | 2.3 | 0.61 | -0.02 | 59 |
| IATE | 5000 | | 1.05 | 0 | 0.93 | 3.89 | 0.0 | 3.0 | 2.4 | 1.51 | -0.18 | 73 |
| ATE | 1 | OneF | 0.91 | - | - | 1.16 | -0.3 | 3.1 | 3.6 | 0.57 | 0.05 | 56 |
| GATE | 2 | | 0.90 | - | - | 1.18 | -0.3 | 3.0 | 3.7 | 0.59 | 0.06 | 56 |
| GATE | 32 | | 0.91 | - | - | 1.24 | -0.2 | 3.0 | 2.6 | 0.65 | 0.17 | 73 |
| IATE | 5000 | | 0.91 | - | 0.51 | 2.59 | -0.0 | 2.9 | 1.7 | 1.22 | 0.58 | 94 |
| ATE | 1 | OneF. | 0.86 | - | - | 1.22 | 0.0 | 3.0 | 0.0 | 0.69 | 0.13 | 77 |
| GATE | 2 | VarT. | 0.86 | - | - | 1.23 | 0.0 | 3.0 | 0.3 | 0.70 | 0.14 | 77 |
| GATE | 32 | Pen | 0.85 | 0 | 0.05 | 1.23 | 0.0 | 3.0 | 0.4 | 0.71 | 0.15 | 79 |
| IATE | 5000 | | 1.00 | 0 | 0.71 | 2.96 | 0.0 | 3.0 | 1.3 | 1.30 | 0.62 | 92 |

Note: For GATE and IATE the *average bias* is the absolute value of the bias for the specific group (GATE) / observation (IATE) averaged over all groups / observation (each group / observation receives the same weight). *CovP (90%)* denotes the (average) probability that the true value is part of the 90% confidence interval. *OneF.VarT.Penalty:* Baseline penalty multiplied by 100.

*Table C.14: Simulation results for additional estimators: DGP with strong effect and selectivity*

| | Groups # | Est. | True & estimated effects | | | Estimation error of effects (averages) | | | | | Estimation of std. error | |
|---|---|---|---|---|---|---|---|---|---|---|---|---|
| | | | Avg. bias | X-sectional std. dev. | | MSE | Skewness | Kurtosis | JB-Stat. | Std. err. | Avg. bias | CovP (90) in % |
| | | | | true | est. | | | | | | | |
| | (1) | | (2) | (3) | (4) | (5) | (6) | (7) | (8) | (9) | (10) | (11) |
| | | | | | | N = 1'000 | | | | | | |
| ATE | 1 | Basic. | 2.94 | - | - | 9.49 | 0.1 | 3.2 | 2.5 | 0.93 | -0.03 | 33 |
| GATE | 2 | One | 2.94 | - | - | 9.54 | 0.1 | 3.1 | 1.6 | 0.95 | -0.03 | 37 |
| GATE | 32 | Sam | 2.90 | 0.67 | 0.45 | 9.59 | 0.1 | 3.2 | 6.4 | 1.03 | -0.03 | 44 |
| IATE | 5000 | | 3.03 | 6.86 | 4.24 | 24.10 | 0.0 | 3.0 | 2.6 | 2.05 | -0.18 | 67 |
| ATE | 1 | OneF | 3.20 | - | - | 12.01 | 0.0 | 2.9 | 0.5 | 1.32 | 0.12 | 27 |
| GATE | 2 | | 3.19 | - | - | 11.93 | 0.0 | 2.9 | 0.5 | 1.33 | 0.14 | 29 |
| GATE | 32 | | 3.15 | 0.66 | 0.27 | 11.91 | 0.0 | 2.9 | 0.5 | 1.36 | 0.18 | 34 |
| IATE | 5000 | | 3.98 | 6.87 | 2.75 | 34.22 | 0.0 | 3.0 | 2.1 | 2.04 | 0.54 | 61 |
| ATE | 1 | OneF. | 1.79 | - | - | 5.64 | 0.0 | 2.8 | 2.2 | 1.56 | 0.21 | 74 |
| GATE | 2 | VarT. | 1.79 | - | - | 5.62 | 0.0 | 2.8 | 2.2 | 1.56 | 0.21 | 75 |
| GATE | 32 | Pen | 1.75 | 0.66 | 0.48 | 5.58 | 0.0 | 2.8 | 2.2 | 1.57 | 0.23 | 76 |
| IATE | 5000 | | 2.38 | 6.87 | 5.14 | 14.62 | 0.0 | 3.0 | 2.4 | 2.44 | 0.63 | 82 |
| | | | | | | N = 4'000 | | | | | | |
| ATE | 1 | Basic. | 2.10 | - | - | 4.65 | 0.0 | 3.1 | 0.1 | 0.49 | -0.01 | 1 |
| GATE | 2 | One | 2.11 | - | - | 4.72 | 0.0 | 3.1 | 0.6 | 0.51 | -0.01 | 1 |
| GATE | 32 | Sam | 2.10 | 0.67 | 0.56 | 4.81 | 0.0 | 3.0 | 1.5 | 0.61 | -0.01 | 4 |
| IATE | 5000 | | 2.31 | 6.86 | 5.27 | 13.74 | 0.0 | 3.0 | 2.1 | 1.62 | -0.18 | 57 |
| ATE | 1 | OneF | 2.02 | - | - | 4.61 | -0.1 | 2.9 | 0.7 | 0.71 | 0.14 | 18 |
| GATE | 2 | | 2.02 | - | - | 4.58 | -0.1 | 2.9 | 0.7 | 0.72 | 0.15 | 21 |
| GATE | 32 | | 1.98 | 0.66 | 0.42 | 4.61 | -0.1 | 2.9 | 1.4 | 0.76 | 0.19 | 29 |
| IATE | 5000 | | 2.56 | 6.87 | 4.13 | 16.44 | 0.0 | 2.9 | 1.8 | 1.49 | 0.52 | 67 |
| ATE | 1 | OneF. | 0.88 | - | - | 1.64 | 0.3 | 2.9 | 4.0 | 0.93 | 0.07 | 82 |
| GATE | 2 | VarT. | 0.87 | - | - | 1.64 | 0.3 | 2.9 | 4.0 | 0.93 | 0.09 | 82 |
| GATE | 32 | Pen | 0.85 | 0.66 | 0.56 | 1.64 | 0.3 | 2.9 | 3.7 | 0.95 | 0.11 | 82 |
| IATE | 5000 | | 1.37 | 6.87 | 5.91 | 6.61 | 0.2 | 3.2 | 4.8 | 1.83 | 0.46 | 89 |

Note: For GATE and IATE the *average bias* is the absolute value of the bias for the specific group (GATE) / observation (IATE) averaged over all groups / observation (each group / observation receives the same weight). *CovP (90%)* denotes the (average) probability that the true value is part of the 90% confidence interval. *OneF.VarT.Penalty:* Baseline penalty multiplied by 100.

*Table C.15: Simulation results for additional estimators: DGP with earnings dependent effect and selectivity*

|  | Groups # | Est. | True & estimated effects | | | Estimation error of effects (averages) | | | | | Estimation of std. error | |
|---|---|---|---|---|---|---|---|---|---|---|---|---|
|  |  |  | Avg. bias | X-sectional std. dev. | | MSE | Skew ness | Kurt- osis | JB- Stat. | Std. err. | Avg. bias | CovP (90) in % |
|  |  |  |  | true | est. |  |  |  |  |  |  |  |
|  | (1) |  | (2) | (3) | (4) | (5) | (6) | (7) | (8) | (9) | (10) | (11) |
| | | | | | | N = 1'000 | | | | | | |
| ATE | 1 | Basic. | 1.10 | - | - | 1.86 | 0.0 | 3.0 | 0.1 | 0.80 | -0.08 | 65 |
| GATE | 2 | One | 1.14 | - | - | 2.04 | 0.0 | 2.9 | 0.3 | 0.84 | -0.07 | 65 |
| GATE | 32 | Sam | 1.06 | 0.35 | 0.22 | 1.99 | 0.0 | 3.0 | 0.4 | 0.91 | -0.05 | 70 |
| IATE | 5000 |  | 1.20 | 1.71 | 1.45 | 4.83 | 0.0 | 3.0 | 2.1 | 1.65 | -0.10 | 76 |
| ATE | 1 | OneF | 1.09 | - | - | 2.44 | -0.1 | 2.9 | 1.2 | 1.11 | 0.10 | 78 |
| GATE | 2 |  | 1.14 | - | - | 2.70 | -0.1 | 2.9 | 1.2 | 1.14 | 0.12 | 80 |
| GATE | 32 |  | 1.05 | 0.35 | 0.15 | 2.54 | 0.0 | 2.9 | 0.6 | 1.18 | 0.19 | 83 |
| IATE | 5000 |  | 1.28 | 1.71 | 1.02 | 5.20 | 0.0 | 3.1 | 2.5 | 1.71 | 0.66 | 91 |
| ATE | 1 | OneF. | 0.97 | - | - | 2.47 | -0.1 | 2.7 | 5.9 | 1.23 | 0.14 | 83 |
| GATE | 2 | VarT. | 1.04 | - | - | 2.88 | -0.1 | 2.7 | 5.6 | 1.25 | 0.15 | 81 |
| GATE | 32 | Pen | 0.90 | 0.35 | 0.09 | 2.47 | -0.1 | 2.7 | 5.9 | 1.26 | 0.16 | 85 |
| IATE | 5000 |  | 1.47 | 1.71 | 0.78 | 6.17 | -0.1 | 3.1 | 7.5 | 1.71 | 0.58 | 85 |
| | | | | | | N = 4'000 | | | | | | |
| ATE | 1 | Basic. | 0.88 | - | - | 0.92 | -0.3 | 3.1 | 2.8 | 0.44 | 0.06 | 34 |
| GATE | 2 | One | 0.89 | - | - | 0.99 | -0.2 | 3.1 | 6.1 | 0.47 | 0.04 | 38 |
| GATE | 32 | Sam | 0.85 | 0.35 | 0.30 | 1.06 | -0.2 | 3.1 | 4.3 | 0.56 | 0.01 | 52 |
| IATE | 5000 |  | 1.05 | 1.71 | 1.84 | 3.79 | 0.0 | 3.0 | 2.1 | 1.25 | -0.20 | 71 |
| ATE | 1 | OneF | 0.81 | - | - | 0.99 | 0.0 | 3.6 | 3.8 | 0.57 | 0.05 | 68 |
| GATE | 2 |  | 0.84 | - | - | 1.11 | 0.0 | 3.5 | 3.2 | 0.60 | 0.07 | 66 |
| GATE | 32 |  | 0.77 | 0.35 | 0.21 | 1.06 | -0.1 | 3.4 | 3.4 | 0.66 | 0.16 | 80 |
| IATE | 5000 |  | 0.96 | 1.71 | 1.33 | 2.94 | 0.0 | 3.1 | 2.2 | 1.25 | 0.59 | 92 |
| ATE | 1 | OneF. | 0.75 | - | - | 1.00 | -0.3 | 2.9 | 4.1 | 0.82 | 0.16 | 79 |
| GATE | 2 | VarT. | 0.81 | - | - | 1.32 | -0.3 | 3.0 | 4.5 | 0.84 | 0.16 | 74 |
| GATE | 32 | Pen | 0.68 | 0.35 | 0.12 | 1.01 | -0.3 | 2.9 | 3.5 | 0.86 | 0.17 | 83 |
| IATE | 5000 |  | 1.40 | 1.71 | 1.04 | 4.63 | 0.0 | 3.0 | 1.6 | 1.93 | 0.65 | 83 |

Note: For GATE and IATE the *average bias* is the absolute value of the bias for the specific group (GATE) / observation (IATE) averaged over all groups / observation (each group / observation receives the same weight). *CovP (90%)* denotes the (average) probability that the true value is part of the 90% confidence interval. *OneF.VarT.Penalty:* Baseline penalty multiplied by 100.

# Appendix D (online): Simulation results for the experimental case

In this section, we consider the case without selection bias, i.e. where treatment is fully randomized with a propensity score of 0.5 for all observations. To save computation time, the cases of *N = 8'000* and LC-5 are omitted.

*Table D.1: Simulation results for N=1'000, main DGP, no selectivity, and main estimators*

| | Groups # | Est. | True & estimated effects | | | Estimation error of effects (averages) | | | | | Estimation of std. error | |
|---|---|---|---|---|---|---|---|---|---|---|---|---|
| | | | Avg. bias | X-sectional std. dev. | | MSE | Skewness | Kurtosis | JB-Stat. | Std. err. | Avg. bias | CovP (90) in % |
| | | | | true | est. | | | | | | | |
| | (1) | | (2) | (3) | (4) | (5) | (6) | (7) | (8) | (9) | (10) | (11) |
| ATE | 1 | Basic | -0.01 | - | - | 1.23 | -0.2 | 3.1 | 7.3 | 1.11 | 0.07 | 92 |
| GATE | 2 | | 0.03 | - | - | 1.29 | -0.2 | 3.1 | 7.1 | 1.13 | 0.09 | 92 |
| GATE | 32 | | 0.07 | 0.17 | 0.12 | 1.42 | -0.2 | 3.1 | 5.1 | 1.19 | 0.12 | 93 |
| IATE | 5000 | | 0.82 | 1.72 | 0.83 | 4.49 | -0.1 | 3.1 | 5.1 | 1.89 | 0.43 | 92 |
| ATE | 1 | OneF. | -0.02 | - | - | 1.28 | 0.1 | 2.9 | 2.7 | 1.13 | 0.05 | 92 |
| GATE | 2 | VarT | 0.05 | - | - | 1.31 | 0.1 | 2.9 | 2.5 | 1.15 | 0.07 | 92 |
| GATE | 32 | | 0.10 | 0.17 | 0.06 | 1.39 | 0.1 | 2.9 | 2.0 | 1.17 | 0.09 | 92 |
| IATE | 5000 | | 1.10 | 1.72 | 0.51 | 3.98 | 0.1 | 2.9 | 2.2 | 1.56 | 0.54 | 91 |
| ATE | 1 | OneF. | -0.05 | - | - | 1.41 | -0.1 | 3.1 | 1.4 | 1.19 | -0.01 | 91 |
| GATE | 2 | MCE | 0.06 | - | - | 1.45 | -0.1 | 3.1 | 1.3 | 1.20 | 0.01 | 91 |
| GATE | 32 | | 0.10 | 0.17 | 0.06 | 1.54 | -0.1 | 3.1 | 1.1 | 1.23 | 0.07 | 92 |
| IATE | 5000 | | 1.04 | 1.72 | 0.57 | 4.19 | 0.0 | 3.0 | 1.9 | 1.67 | 0.57 | 92 |
| ATE | 1 | OneF. | -0.05 | - | - | 1.68 | 0.0 | 2.8 | 2.8 | 1.29 | 0.06 | 92 |
| GATE | 2 | MCE. | 0.06 | - | - | 1.72 | 0.0 | 2.8 | 2.1 | 1.31 | 0.07 | 92 |
| GATE | 32 | LC-2 | 0.11 | 0.17 | 0.05 | 1.82 | 0.0 | 2.9 | 1.5 | 1.34 | 0.08 | 92 |
| IATE | 5000 | | 1.08 | 1.72 | 0.53 | 4.83 | 0.0 | 3.0 | 1.7 | 1.82 | 0.17 | 86 |
| ATE | 1 | OneF. | 0.02 | - | - | 1.28 | -0.1 | 2.8 | 4.2 | 1.13 | 0.05 | 91 |
| GATE | 2 | MCE. | 0.05 | - | - | 1.32 | -0.1 | 2.8 | 3.6 | 1.15 | 0.07 | 91 |
| GATE | 32 | Pen | 0.10 | 0.17 | 0.06 | 1.41 | -0.1 | 2.9 | 4.0 | 1.18 | 0.12 | 93 |
| IATE | 5000 | | 1.06 | 1.72 | 0.54 | 4.03 | 0.0 | 3.0 | 2.2 | 1.60 | 0.58 | 92 |
| ATE | 1 | OneF. | -0.08 | - | - | 1.85 | -0.1 | 3.0 | 1.0 | 1.36 | 0.00 | 90 |
| GATE | 2 | MCE. | 0.08 | - | - | 1.89 | -0.1 | 3.0 | 0.9 | 1.37 | 0.00 | 90 |
| GATE | 32 | Pen | 0.12 | 0.17 | 0.05 | 2.01 | -0.1 | 2.9 | 1.3 | 1.42 | 0.01 | 89 |
| IATE | 5000 | LC-2 | 1.13 | 1.72 | 0.48 | 5.05 | 0.0 | 3.0 | 1.9 | 1.97 | 0.12 | 85 |

Note: For GATE and IATE the *average bias* is the absolute value of the bias for the specific group (GATE) / observation (IATE) averaged over all groups / observation (each group / observation receives the same weight). *CovP (90%)* denotes the (average) probability that the true value is part of the 90% confidence interval. The simulation errors of the mean MSEs are around 0.06.

*Table D.2: Simulation results for N=4'000, main DGP, no selectivity, and main estimators*

| | Groups # | Est. | True & estimated effects | | | Estimation error of effects (averages) | | | | | Estimation of std. error | |
|---|---|---|---|---|---|---|---|---|---|---|---|---|
| | | | Avg. bias | X-sectional std. dev. | | MSE | Skew ness | Kurt- osis | JB- Stat. | Std. err. | Avg. bias | CovP (90) in % |
| | | | | true | est. | | | | | | | |
| | (1) | | (2) | (3) | (4) | (5) | (6) | (7) | (8) | (9) | (10) | (11) |
| ATE | 1 | Basic | 0.06 | - | - | 0.29 | 0.2 | 3.0 | 1.9 | 0.53 | 0.06 | 94 |
| GATE | 2 | | 0.06 | - | - | 0.31 | 0.2 | 3.0 | 2.0 | 0.55 | 0.07 | 94 |
| GATE | 32 | | 0.07 | 0.17 | 0.15 | 0.43 | 0.1 | 3.0 | 1.0 | 0.65 | 0.10 | 95 |
| IATE | 5000 | | 0.44 | 1.72 | 1.34 | 2.81 | 0.0 | 3.0 | 2.2 | 1.56 | 0.28 | 92 |
| ATE | 1 | OneF. | -0.1 | - | - | 0.34 | -0.1 | 2.7 | 1.5 | 0.58 | 0.01 | 90 |
| GATE | 2 | VarT | 0.03 | - | - | 0.36 | -0.2 | 2.7 | 1.7 | 0.60 | 0.03 | 91 |
| GATE | 32 | | 0.08 | 0.17 | 0.08 | 0.39 | -0.1 | 2.7 | 2.6 | 0.62 | 0.05 | 92 |
| IATE | 5000 | | 0.82 | 1.72 | 0.84 | 2.10 | 0.0 | 3.0 | 1.9 | 1.10 | 0.51 | 93 |
| ATE | 1 | OneF. | -0.01 | - | - | 0.32 | 0.1 | 2.9 | 0.3 | 0.57 | 0.02 | 92 |
| GATE | 2 | MCE | 0.05 | - | - | 0.34 | 0.1 | 3.0 | 0.6 | 0.58 | 0.03 | 92 |
| GATE | 32 | | 0.06 | 0.17 | 0.09 | 0.41 | 0.1 | 2.9 | 0.9 | 0.63 | 0.12 | 94 |
| IATE | 5000 | | 0.72 | 1.72 | 0.93 | 2.15 | 0.0 | 3.0 | 2.1 | 1.19 | 0.54 | 94 |
| ATE | 1 | OneF. | 0.02 | - | - | 0.46 | 0.1 | 2.9 | 0.3 | 0.68 | -0.01 | 91 |
| GATE | 2 | MCE. | 0.04 | - | - | 0.48 | 0.1 | 2.9 | 0.8 | 0.69 | -0.01 | 90 |
| GATE | 32 | LC-2 | 0.09 | 0.17 | 0.07 | 0.55 | 0.0 | 2.9 | 0.6 | 0.73 | 0.00 | 91 |
| IATE | 5000 | | 0.88 | 1.72 | 0.76 | 2.64 | 0.0 | 3.0 | 2.2 | 1.27 | 0.13 | 84 |
| ATE | 1 | OneF. | 0.03 | - | - | 0.32 | 0.0 | 2.8 | 0.4 | 0.57 | 0.02 | 90 |
| GATE | 2 | MCE. | 0.04 | - | - | 0.34 | 0.0 | 2.9 | 0.6 | 0.58 | 0.04 | 91 |
| GATE | 32 | Pen | 0.08 | 0.17 | 0.08 | 0.41 | -0.1 | 2.9 | 0.9 | 0.63 | 0.11 | 93 |
| IATE | 5000 | | 0.78 | 1.72 | 0.86 | 2.17 | 0.0 | 3.0 | 2.0 | 1.15 | 0.52 | 93 |
| ATE | 1 | OneF. | 0.00 | - | - | 0.44 | -0.1 | 2.8 | 0.7 | 0.67 | 0.01 | 88 |
| GATE | 2 | MCE. | 0.03 | - | - | 0.46 | -0.1 | 2.8 | 0.7 | 0.69 | 0.01 | 89 |
| GATE | 32 | Pen | 0.08 | 0.16 | 0.08 | 0.54 | -0.1 | 2.9 | 1.0 | 0.74 | 0.02 | 89 |
| IATE | 5000 | LC-2 | 0.90 | 1.72 | 0.76 | 2.68 | 0.0 | 3.0 | 2.3 | 1.42 | 0.14 | 83 |

Note: For GATE and IATE the *average bias* is the absolute value of the bias for the specific group (GATE) / observation (IATE) averaged over all groups / observation (each group / observation receives the same weight). *CovP (90%)* denotes the (average) probability that the true value is part of the 90% confidence interval. The simulation errors of the mean MSEs are around 0.06.

*Table D.3: Simulation results for N=1'000, no effect, no selectivity, and main estimators*

| | Groups # | Est. | True & estimated effects | | | Estimation error of effects (averages) | | | | | Estimation of std. error | |
|---|---|---|---|---|---|---|---|---|---|---|---|---|
| | | | Avg. bias | X-sectional std. dev. | | MSE | Skewness | Kurtosis | JB-Stat. | Std. err. | Avg. bias | CovP (90) in % |
| | | | | true | est. | | | | | | | |
| | (1) | | (2) | (3) | (4) | (5) | (6) | (7) | (8) | (9) | (10) | (11) |
| ATE | 1 | Basic | 0.02 | - | - | 1.27 | -0.1 | 3.0 | 1.8 | 1.13 | 0.04 | 91 |
| GATE | 2 | | 0.02 | - | - | 1.32 | -0.1 | 3.0 | 1.7 | 1.15 | 0.05 | 91 |
| GATE | 32 | | 0.02 | 0 | 0.01 | 1.47 | -0.1 | 3.0 | 2.8 | 1.21 | 0.09 | 92 |
| IATE | 5000 | | 0.04 | 0 | 0.04 | 3.36 | -0.1 | 3.1 | 3.1 | 1.83 | 0.45 | 96 |
| ATE | 1 | OneF. | 0.00 | - | - | 1.26 | 0.0 | 3.1 | 0.7 | 1.12 | 0.04 | 92 |
| GATE | 2 | VarT | 0.00 | - | - | 1.29 | 0.0 | 3.1 | 1.1 | 1.14 | 0.07 | 92 |
| GATE | 32 | | 0.02 | 0 | 0.01 | 1.34 | 0.0 | 3.1 | 0.9 | 1.16 | 0.09 | 92 |
| IATE | 5000 | | 0.02 | 0 | 0.03 | 2.36 | 0.0 | 3.0 | 1.9 | 1.53 | 0.53 | 97 |
| ATE | 1 | OneF. | -0.04 | - | - | 1.24 | 0.0 | 3.2 | 2.3 | 1.11 | 0.05 | 92 |
| GATE | 2 | MCE | 0.04 | - | - | 1.28 | 0.0 | 3.2 | 2.0 | 1.13 | 0.07 | 92 |
| GATE | 32 | | 0.04 | 0 | 0.01 | 1.38 | 0.0 | 3.2 | 2.8 | 1.18 | 0.13 | 93 |
| IATE | 5000 | | 0.04 | 0 | 0.03 | 2.56 | 0.0 | 3.1 | 2.0 | 1.59 | 0.60 | 97 |
| ATE | 1 | OneF. | -0.08 | - | - | 1.73 | 0.1 | 3.1 | 2.3 | 1.31 | 0.03 | 91 |
| GATE | 2 | MCE. | 0.08 | - | - | 1.78 | 0.1 | 3.2 | 2.7 | 1.33 | 0.04 | 91 |
| GATE | 32 | LC-2 | 0.08 | 0 | 0.00 | 1.88 | 0.1 | 3.1 | 2.6 | 1.37 | 0.05 | 92 |
| IATE | 5000 | | 0.08 | 0 | 0.04 | 3.31 | 0.0 | 3.8 | 3.8 | 1.82 | 0.16 | 93 |
| ATE | 1 | OneF. | 0.03 | - | - | 1.32 | 0.0 | 2.9 | 0.9 | 1.15 | 0.01 | 91 |
| GATE | 2 | MCE. | 0.03 | - | - | 1.35 | 0.0 | 2.9 | 1.0 | 1.16 | 0.03 | 91 |
| GATE | 32 | Pen | 0.03 | 0 | 0.05 | 1.43 | 0.0 | 2.9 | 0.6 | 1.20 | 0.09 | 93 |
| IATE | 5000 | | 0.03 | 0 | 0.03 | 2.54 | 0.0 | 2.9 | 1.9 | 1.59 | 0.57 | 97 |
| ATE | 1 | OneF. | 0.00 | - | - | 1.73 | -0.1 | 3.0 | 3.0 | 1.32 | 0.03 | 91 |
| GATE | 2 | MCE. | 0.01 | - | - | 1.78 | -0.1 | 3.0 | 2.8 | 1.34 | 0.03 | 91 |
| GATE | 32 | Pen | 0.01 | 0 | 0.01 | 1.89 | -0.1 | 3.0 | 3.1 | 1.37 | 0.04 | 90 |
| IATE | 5000 | LC-2 | 0.03 | 0 | 0.04 | 3.24 | -0.1 | 3.0 | 2.6 | 1.80 | 0.15 | 93 |

Note: For GATE and IATE the *average bias* is the absolute value of the bias for the specific group (GATE) / observation (IATE) averaged over all groups / observation (each group / observation receives the same weight). *CovP (90%)* denotes the (average) probability that the true value is part of the 90% confidence interval. The simulation errors of the mean MSEs are around 0.06.

*Table D.4: Simulation results for N=4'000, no effect, no selectivity, and main estimators*

| | Groups # | Est. | True & estimated effects | | | Estimation error of effects (averages) | | | | | Estimation of std. error | |
|---|---|---|---|---|---|---|---|---|---|---|---|---|
| | | | Avg. bias | X-sectional std. dev. | | MSE | Skew ness | Kurt- osis | JB- Stat. | Std. err. | Avg. bias | CovP (90) in % |
| | | | | true | est. | | | | | | | |
| | (1) | | (2) | (3) | (4) | (5) | (6) | (7) | (8) | (9) | (10) | (11) |
| ATE | 1 | Basic | -0.01 | - | - | 0.28 | 0.1 | 2.8 | 1.4 | 0.53 | 0.05 | 93 |
| GATE | 2 | | 0.01 | - | - | 0.30 | 0.1 | 2.8 | 1.4 | 0.55 | 0.07 | 92 |
| GATE | 32 | | 0.02 | 0 | 0.01 | 0.47 | 0.1 | 2.7 | 1.9 | 0.67 | 0.08 | 94 |
| IATE | 5000 | | 0.07 | 0 | 0.09 | 2.28 | 0.0 | 3.0 | 2.1 | 1.50 | 0.29 | 94 |
| ATE | 1 | OneF. | -0.02 | - | - | 0.28 | 0.2 | 3.1 | 2.0 | 0.58 | 0.05 | 94 |
| GATE | 2 | VarT | 0.02 | - | - | 0.30 | 0.2 | 3.0 | 1.5 | 0.62 | 0.08 | 94 |
| GATE | 32 | | 0.02 | 0 | 0.01 | 0.32 | 0.2 | 3.1 | 1.8 | 0.66 | 0.10 | 95 |
| IATE | 5000 | | 0.05 | 0 | 0.05 | 1.06 | 0.1 | 3.0 | 1.9 | 1.55 | 0.53 | 98 |
| ATE | 1 | OneF. | 0.07 | - | - | 0.30 | -0.2 | 2.7 | 2.0 | 0.54 | 0.04 | 94 |
| GATE | 2 | MCE | 0.07 | - | - | 0.32 | -0.2 | 2.8 | 2.1 | 0.56 | 0.06 | 94 |
| GATE | 32 | | 0.07 | 0 | 0.02 | 0.40 | -0.2 | 2.9 | 2.4 | 0.63 | 0.13 | 95 |
| IATE | 5000 | | 0.08 | 0 | 0.06 | 1.26 | 0.0 | 3.0 | 1.8 | 1.11 | 0.57 | 98 |
| ATE | 1 | OneF. | 0.07 | - | | 0.44 | 0.1 | 2.7 | 1.1 | 0.66 | 0.00 | 89 |
| GATE | 2 | MCE. | 0.07 | - | | 0.46 | 0.1 | 2.7 | 1.2 | 0.68 | 0.01 | 90 |
| GATE | 32 | LC-2 | 0.08 | 0 | 0.02 | 0.52 | 0.1 | 3.0 | 1.2 | 0.72 | 0.02 | 91 |
| IATE | 5000 | | 0.08 | 0 | 0.07 | 1.57 | 0.0 | 3.0 | 2.3 | 1.25 | 0.13 | 93 |
| ATE | 1 | OneF. | -0.02 | - | - | 0.28 | 0.1 | 2.7 | 0.9 | 0.53 | 0.05 | 93 |
| GATE | 2 | MCE. | 0.02 | - | - | 0.30 | 0.1 | 2.8 | 1.0 | 0.55 | 0.07 | 93 |
| GATE | 32 | Pen | 0.02 | 0 | 0.01 | 0.37 | 0.1 | 2.8 | 0.8 | 0.61 | 0.14 | 96 |
| IATE | 5000 | | 0.05 | 0 | 0.06 | 1.16 | -0.1 | 3.0 | 2.9 | 1.07 | 0.57 | 98 |
| ATE | 1 | OneF. | 0.01 | - | - | 0.38 | 0.0 | 3.0 | 0.0 | 0.62 | 0.05 | 92 |
| GATE | 2 | MCE. | 0.01 | - | - | 0.40 | 0.0 | 3.0 | 0.7 | 0.63 | 0.05 | 92 |
| GATE | 32 | Pen | 0.02 | 0 | 0.01 | 0.47 | 0.0 | 3.0 | 0.9 | 0.68 | 0.06 | 93 |
| IATE | 5000 | LC-2 | 0.05 | 0 | 0.06 | 1.57 | 0.0 | 3.1 | 3.4 | 1.25 | 0.15 | 94 |

Note: For GATE and IATE the *average bias* is the absolute value of the bias for the specific group (GATE) / observation (IATE) averaged over all groups / observation (each group / observation receives the same weight). *CovP (90%)* denotes the (average) probability that the true value is part of the 90% confidence interval. The simulation errors of the mean MSEs are around 0.03.

*Table D.5: Simulation results for N=1'000, strong effect, no selectivity, and main estimators*

| | Groups # | Est. | True & estimated effects | | | Estimation error of effects (averages) | | | | | Estimation of std. error | |
|---|---|---|---|---|---|---|---|---|---|---|---|---|
| | | | Avg. bias | X-sectional std. dev. | | MSE | Skewness | Kurtosis | JB-Stat. | Std. err. | Avg. bias | CovP (90) in % |
| | | | | true | est. | | | | | | | |
| | (1) | | (2) | (3) | (4) | (5) | (6) | (7) | (8) | (9) | (10) | (11) |
| ATE | 1 | Basic | 0.04 | - | - | 1.27 | 0.1 | 3.1 | 1.8 | 1.12 | 0.15 | 93 |
| GATE | 2 | | 0.04 | - | - | 1.31 | 0.1 | 3.1 | 2.0 | 1.14 | 0.17 | 93 |
| GATE | 32 | | 0.12 | 0.67 | 0.62 | 1.45 | 0.1 | 3.1 | 3.1 | 1.19 | 0.21 | 94 |
| IATE | 5000 | | 1.27 | 6.87 | 5.93 | 8.04 | 0.0 | 3.1 | 3.1 | 2.20 | 0.41 | 86 |
| ATE | 1 | OneF. | -0.09 | - | - | 1.30 | -0.1 | 3.3 | 4.3 | 1.14 | 0.14 | 93 |
| GATE | 2 | VarT | 0.10 | - | - | 1.34 | -0.1 | 3.3 | 4.2 | 1.15 | 0.16 | 94 |
| GATE | 32 | | 0.23 | 0.67 | 0.43 | 1.46 | -0.1 | 3.2 | 3.4 | 1.18 | 0.19 | 93 |
| IATE | 5000 | | 2.23 | 6.87 | 4.50 | 11.25 | 0.0 | 3.0 | 1.8 | 2.00 | 0.50 | 77 |
| ATE | 1 | OneF. | -0.02 | - | - | 1.22 | 0.0 | 2.9 | 0.7 | 1.11 | 0.18 | 95 |
| GATE | 2 | MCE | 0.09 | - | - | 1.25 | 0.0 | 2.9 | 0.9 | 1.12 | 0.19 | 95 |
| GATE | 32 | | 0.21 | 0.67 | 0.45 | 1.39 | 0.0 | 2.9 | 0.8 | 1.15 | 0.23 | 95 |
| IATE | 5000 | | 2.01 | 6.87 | 4.67 | 10.77 | 0.0 | 2.9 | 2.3 | 2.08 | 0.51 | 79 |
| ATE | 1 | OneF. | -0.02 | - | - | 1.83 | 0.0 | 3.0 | 0.4 | 1.35 | 0.06 | 91 |
| GATE | 2 | MCE. | 0.15 | - | - | 1.91 | 0.0 | 3.0 | 0.3 | 1.37 | 0.06 | 91 |
| GATE | 32 | LC-2 | 0.31 | 0.67 | 0.32 | 2.09 | 0.1 | 3.0 | 0.9 | 1.40 | 0.07 | 90 |
| IATE | 5000 | | 3.25 | 6.87 | 3.38 | 18.63 | 0.0 | 3.2 | 8.9 | 2.12 | 0.03 | 54 |
| ATE | 1 | OneF. | 0.00 | - | - | 1.33 | 0.0 | 2.9 | 0.8 | 1.28 | 0.13 | 94 |
| GATE | 2 | MCE. | 0.09 | - | - | 1.37 | 0.0 | 2.9 | 1.1 | 1.31 | 0.14 | 93 |
| GATE | 32 | Pen | 0.21 | 0.67 | 0.45 | 1.49 | 0.0 | 2.9 | 0.8 | 1.37 | 0.18 | 94 |
| IATE | 5000 | | 1.97 | 6.87 | 4.47 | 10.59 | 0.0 | 2.9 | 3.1 | 2.54 | 0.46 | 79 |
| ATE | 1 | OneF. | 0.00 | - | - | 1.94 | -0.1 | 3.0 | 0.60 | 1.39 | 0.02 | 91 |
| GATE | 2 | MCE. | 0.14 | - | - | 2.00 | -0.1 | 3.0 | 0.68 | 1.41 | 0.02 | 91 |
| GATE | 32 | Pen | 0.32 | 0.67 | 0.31 | 2.18 | -0.1 | 3.0 | 0.57 | 1.46 | 0.03 | 90 |
| IATE | 5000 | LC-2 | 3.29 | 6.87 | 3.34 | 18.85 | 0.0 | 3.1 | 6.18 | 2.10 | 0.02 | 53 |

Note: For GATE and IATE the *average bias* is the absolute value of the bias for the specific group (GATE) / observation (IATE) averaged over all groups / observation (each group / observation receives the same weight). *CovP (90%)* denotes the (average) probability that the true value is part of the 90% confidence interval. The simulation errors of the mean MSEs are around 0.15.

*Table D.6: Simulation results for N=4'000, strong effect, no selectivity, and main estimators*

| | Groups # | Est. | True & estimated effects | | | Estimation error of effects (averages) | | | | | Estimation of std. error | |
|---|---|---|---|---|---|---|---|---|---|---|---|---|
| | | | Avg. bias | X-sectional std. dev. | | MSE | Skew ness | Kurt-osis | JB-Stat. | Std. err. | Avg. bias | CovP (90) in % |
| | | | | true | est. | | | | | | | |
| | (1) | | (2) | (3) | (4) | (5) | (6) | (7) | (8) | (9) | (10) | (11) |
| ATE | 1 | Basic | 0.01 | - | - | 0.29 | -0.1 | 2.6 | 1.6 | 0.54 | 0.10 | 95 |
| GATE | 2 | | 0.06 | - | - | 0.32 | -0.1 | 2.6 | 1.9 | 0.56 | 0.11 | 97 |
| GATE | 32 | | 0.16 | 0.67 | 0.74 | 0.48 | -0.1 | 2.7 | 1.7 | 0.66 | 0.13 | 94 |
| IATE | 5000 | | 1.10 | 6.87 | 6.53 | 5.24 | 0.0 | 3.0 | 2.2 | 1.64 | 0.33 | 84 |
| ATE | 1 | OneF. | 0.00 | - | - | 0.24 | -0.1 | 2.9 | 0.5 | 0.50 | 0.14 | 98 |
| GATE | 2 | VarT | 0.05 | - | - | 0.27 | -0.1 | 2.8 | 0.6 | 0.51 | 0.15 | 98 |
| GATE | 32 | | 0.13 | 0.67 | 0.56 | 0.32 | 0.0 | 2.8 | 1.0 | 0.55 | 0.18 | 97 |
| IATE | 5000 | | 1.23 | 6.87 | 5.77 | 4.62 | 0.0 | 2.9 | 1.9 | 1.35 | 0.50 | 86 |
| ATE | 1 | OneF. | 0.05 | - | - | 0.29 | -0.1 | 3.1 | 0.7 | 0.54 | 0.10 | 94 |
| GATE | 2 | MCE | 0.05 | - | - | 0.31 | -0.1 | 3.1 | 0.8 | 0.55 | 0.11 | 94 |
| GATE | 32 | | 0.13 | 0.67 | 0.56 | 0.39 | -0.1 | 3.1 | 2.0 | 0.60 | 0.17 | 95 |
| IATE | 5000 | | 1.11 | 6.87 | 5.93 | 4.36 | 0.0 | 3.0 | 1.9 | 1.35 | 0.60 | 89 |
| ATE | 1 | OneF. | 0.03 | - | - | 0.48 | -0.1 | 2.8 | 0.4 | 0.70 | 0.00 | 90 |
| GATE | 2 | MCE. | 0.16 | - | - | 0.53 | -0.1 | 2.9 | 0.9 | 0.72 | 0.01 | 90 |
| GATE | 32 | LC-2 | 0.22 | 0.67 | 0.44 | 0.61 | -0.1 | 2.9 | 0.7 | 0.75 | 0.01 | 88 |
| IATE | 5000 | | 2.21 | 6.87 | 4.78 | 9.40 | 0.0 | 3.1 | 5.0 | 1.52 | 0.04 | 58 |
| ATE | 1 | OneF. | 0.06 | - | - | 0.29 | 0.0 | 2.8 | 0.3 | 0.53 | 0.10 | 95 |
| GATE | 2 | MCE. | 0.06 | - | - | 0.30 | 0.0 | 2.8 | 0.4 | 0.54 | 0.12 | 95 |
| GATE | 32 | Pen | 0.13 | 0.67 | 0.57 | 0.36 | 0.0 | 2.7 | 1.3 | 0.58 | 0.18 | 96 |
| IATE | 5000 | | 1.14 | 6.87 | 5.89 | 4.44 | 0.0 | 3.0 | 1.7 | 1.34 | 0.55 | 87 |
| ATE | 1 | OneF. | -0.05 | - | - | 0.41 | -0.1 | 3.0 | 0.6 | 0.64 | 0.05 | 91 |
| GATE | 2 | MCE. | 0.12 | - | - | 0.45 | -0.1 | 3.0 | 1.1 | 0.66 | 0.06 | 92 |
| GATE | 32 | Pen | 0.23 | 0.67 | 0.45 | 0.55 | -0.1 | 3.0 | 0.8 | 0.69 | 0.06 | 91 |
| IATE | 5000 | LC-2 | 2.12 | 6.87 | 4.90 | 8.93 | -0.0 | 3.0 | 3.6 | 1.51 | 0.05 | 60 |

Note: For GATE and IATE the *average bias* is the absolute value of the bias for the specific group (GATE) / observation (IATE) averaged over all groups / observation (each group / observation receives the same weight). *CovP (90%)* denotes the (average) probability that the true value is part of the 90% confidence interval. The simulation errors of the mean MSEs are around 0.06.

*Table D.7: Simulation results for N=1'000, earnings dependent effect, no selectivity, and main estimators*

| | Groups # | Est. | True & estimated effects | | | Estimation error of effects (averages) | | | | | Estimation of std. error | |
|---|---|---|---|---|---|---|---|---|---|---|---|---|
| | | | Avg. bias | X-sectional std. dev. | | MSE | Skewness | Kurtosis | JB-Stat. | Std. err. | Avg. bias | CovP (90) in % |
| | | | | true | est. | | | | | | | |
| | (1) | | (2) | (3) | (4) | (5) | (6) | (7) | (8) | (9) | (10) | (11) |
| ATE | 1 | Basic | -0.08 | - | - | 1.27 | 0.0 | 3.3 | 4.2 | 1.13 | 0.06 | 91 |
| GATE | 2 | | 0.19 | - | - | 1.39 | 0.0 | 3.3 | 4.7 | 1.16 | 0.07 | 91 |
| GATE | 32 | | 0.14 | 0.35 | 0.24 | 1.46 | 0.0 | 3.3 | 3.8 | 1.20 | 0.13 | 92 |
| IATE | 5000 | | 0.50 | 1.71 | 1.02 | 4.47 | 0.0 | 3.1 | 3.1 | 1.94 | 0.41 | 93 |
| ATE | 1 | OneF. | 0.03 | - | - | 1.26 | 0.1 | 3.0 | 0.6 | 1.12 | 0.06 | 92 |
| GATE | 2 | VarT | 0.42 | - | - | 1.48 | 0.1 | 3.0 | 0.8 | 1.14 | 0.09 | 90 |
| GATE | 32 | | 0.20 | 0.35 | 0.09 | 1.41 | 0.1 | 3.0 | 0.7 | 1.16 | 0.12 | 92 |
| IATE | 5000 | | 1.01 | 1.71 | 0.40 | 4.24 | 0.0 | 3.0 | 2.0 | 1.54 | 0.56 | 90 |
| ATE | 1 | OneF. | -0.03 | - | - | 1.20 | 0.1 | 2.9 | 1.1 | 1.09 | 0.09 | 92 |
| GATE | 2 | MCE | 0.35 | - | - | 1.37 | 0.1 | 2.9 | 1.1 | 1.12 | 0.11 | 92 |
| GATE | 32 | | 0.19 | 0.35 | 0.14 | 1.39 | 0.1 | 2.9 | 1.3 | 1.16 | 0.16 | 94 |
| IATE | 5000 | | 0.82 | 1.71 | 0.61 | 4.01 | 0.0 | 3.0 | 2.2 | 1.62 | 0.63 | 93 |
| ATE | 1 | OneF. | -0.05 | - | - | 1.85 | 0.0 | 2.7 | 3.9 | 1.36 | 0.00 | 90 |
| GATE | 2 | MCE. | 0.42 | - | - | 2.07 | 0.0 | 2.7 | 3.4 | 1.38 | 0.00 | 88 |
| GATE | 32 | LC-2 | 0.22 | 0.16 | 0.11 | 2.08 | 0.0 | 2.7 | 3.6 | 1.42 | 0.01 | 90 |
| IATE | 5000 | | 0.97 | 1.71 | 0.45 | 5.24 | 0.0 | 3.0 | 2.1 | 1.87 | 0.13 | 85 |
| ATE | 1 | OneF. | 0.06 | - | - | 1.34 | 0.0 | 2.9 | 0.3 | 1.16 | 0.03 | 91 |
| GATE | 2 | MCE. | 0.36 | - | - | 1.52 | 0.0 | 3.0 | 0.3 | 1.18 | 0.05 | 90 |
| GATE | 32 | Pen | 0.15 | 0.35 | 0.15 | 1.50 | 0.0 | 2.9 | 0.6 | 1.21 | 0.10 | 92 |
| IATE | 5000 | | 0.83 | 1.71 | 0.62 | 4.06 | 0.0 | 3.0 | 2.7 | 1.64 | 0.56 | 92 |
| ATE | 1 | OneF. | 0.01 | - | - | 1.68 | 0.0 | 2.8 | 1.6 | 1.30 | 0.06 | 92 |
| GATE | 2 | MCE. | 0.40 | - | - | 1.89 | 0.0 | 2.8 | 1.5 | 1.32 | 0.06 | 90 |
| GATE | 32 | Pen. | 0.19 | 0.16 | 0.12 | 1.89 | 0.0 | 2.9 | 1.6 | 1.35 | 0.07 | 91 |
| IATE | 5000 | LC-2 | 0.96 | 1.71 | 0.47 | 4.92 | 0.0 | 2.2 | 2.2 | 1.80 | 0.17 | 86 |

Note: For GATE and IATE the *average bias* is the absolute value of the bias for the specific group (GATE) / observation (IATE) averaged over all groups / observation (each group / observation receives the same weight). *CovP (90%)* denotes the (average) probability that the true value is part of the 90% confidence interval. The simulation errors of the mean MSEs are around 0.06

*Table D.8: Simulation results for N=4'000, earnings dependent effect, no selectivity, and main estimators*

| | Groups # | Est. | True & estimated effects | | | Estimation error of effects (averages) | | | | | Estimation of std. error | |
|---|---|---|---|---|---|---|---|---|---|---|---|---|
| | | | Avg. bias | X-sectional std. dev. | | MSE | Skewness | Kurtosis | JB-Stat. | Std. err. | Avg. bias | CovP (90) in % |
| | | | | true | est. | | | | | | | |
| | (1) | | (2) | (3) | (4) | (5) | (6) | (7) | (8) | (9) | (10) | (11) |
| **ATE** | *1* | Basic | 0.02 | - | - | 0.29 | -0.1 | 2.9 | 0.5 | 0.59 | 0.05 | 92 |
| **GATE** | *2* | | 0.03 | - | - | 0.34 | -0.5 | 2.9 | 1.1 | 0.64 | 0.06 | 92 |
| **GATE** | *32* | | 0.04 | 0.35 | 0.34 | 0.43 | -0.5 | 2.9 | 0.7 | 0.76 | 0.11 | 95 |
| **IATE** | *5000* | | 0.40 | 1.71 | 1.50 | 2.78 | 0.0 | 3.0 | 1.6 | 1.85 | 0.28 | 92 |
| **ATE** | *1* | OneF. | 0.00 | - | - | 0.31 | 0.0 | 2.8 | 0.3 | 0.56 | 0.03 | 91 |
| **GATE** | *2* | VarT | 0.33 | - | - | 0.45 | 0.0 | 2.9 | 1.3 | 0.58 | 0.05 | 87 |
| **GATE** | *32* | | 0.16 | 0.35 | 0.16 | 0.40 | 0.0 | 2.8 | 0.8 | 0.60 | 0.08 | 93 |
| **IATE** | *5000* | | 0.80 | 1.71 | 0.66 | 2.39 | 0.0 | 3.0 | 2.3 | 1.07 | 0.50 | 90 |
| **ATE** | *1* | OneF. | 0.05 | - | - | 0.31 | -0.2 | 3.2 | 3.1 | 0.56 | 0.03 | 92 |
| **GATE** | *2* | MCE | 0.24 | - | - | 0.40 | -0.2 | 3.3 | 2.9 | 0.58 | 0.06 | 90 |
| **GATE** | *32* | | 0.10 | 0.35 | 0.23 | 0.42 | -0.2 | 3.2 | 2.6 | 0.63 | 0.13 | 94 |
| **IATE** | *5000* | | 0.57 | 1.71 | 0.95 | 2.15 | -0.1 | 3.0 | 2.5 | 1.18 | 0.55 | 94 |
| **ATE** | *1* | OneF. | -0.01 | - | - | 0.45 | 0.0 | 3.0 | 0.0 | 0.67 | 0.00 | 91 |
| **GATE** | *2* | MCE. | 0.33 | - | - | 0.58 | 0.0 | 3.1 | 0.2 | 0.70 | 0.01 | 86 |
| **GATE** | *32* | LC-2 | 0.17 | 0.35 | 0.15 | 0.57 | 0.0 | 2.9 | 0.9 | 0.73 | 0.01 | 89 |
| **IATE** | *5000* | | 0.80 | 1.71 | 0.66 | 2.85 | 0.0 | 3.1 | 1.3 | 1.26 | 0.13 | 83 |
| **ATE** | *1* | OneF. | 0.01 | - | - | 0.29 | -0.1 | 2.8 | 1.2 | 0.59 | 0.05 | 93 |
| **GATE** | *2* | MCE. | 0.23 | - | - | 0.37 | -0.2 | 2.9 | 2.0 | 0.64 | 0.07 | 91 |
| **GATE** | *32* | Pen | 0.10 | 0.35 | 0.23 | 0.39 | -0.1 | 2.8 | 1.0 | 0.75 | 0.14 | 95 |
| **IATE** | *5000* | | 0.54 | 1.71 | 0.97 | 1.98 | -0.1 | 3.0 | 2.3 | 1.68 | 0.56 | 95 |
| **ATE** | *1* | OneF. | -0.05 | - | - | 0.40 | -0.1 | 3.3 | 1.2 | 0.63 | 0.04 | 92 |
| **GATE** | *2* | MCE. | 0.32 | - | - | 0.53 | 0.0 | 3.3 | 1.7 | 0.65 | 0.04 | 89 |
| **GATE** | *32* | Pen | 0.17 | 0.16 | 0.17 | 0.53 | -0.1 | 3.3 | 1.9 | 0.70 | 0.05 | 92 |
| **IATE** | *5000* | LC-2 | 0.77 | 1.72 | 0.71 | 2.73 | 0.0 | 3.1 | 2.8 | 1.26 | 0.15 | 84 |

Note: For GATE and IATE the *average bias* is the absolute value of the bias for the specific group (GATE) / observation (IATE) averaged over all groups / observation (each group / observation receives the same weight). *CovP (90%)* denotes the (average) probability that the true value is part of the 90% confidence interval. The simulation errors of the mean MSEs are around 0.04.